\documentclass[twocolumn,amssymb,amsmath,rmp,superscriptaddress,%
unsortedaddress,floatfix]{revtex4}
\usepackage{graphicx,bm}
\usepackage{type1cm}
\def\hw{$\hbar\omega\,$}
\def\H{${\cal H}$}
\def\Hm{${\cal H}_m$}

\def\HM{${\cal H}_M$}
\begin{document}
\title{The Shell Model as Unified View of Nuclear Structure}

\author{E. Caurier}
\email{etienne.caurier@ires.in2p3.fr}
\affiliation{Institut de Recherches Subatomiques, IN2P3-CNRS,
  Universit\'e Louis Pasteur, F-67037 Strasbourg, France}

\author{G. Mart\'{\i}nez-Pinedo}
\email{martinez@ieec.fcr.es}
\affiliation{Institut d'Estudis Espacials de Catalunya,
Edifici Nexus, Gran Capit\`a 2,
E-08034 Barcelona, Spain}
\affiliation{Instituci\'o Catalana de Recerca i Estudis Avan\c{c}ats,
Llu\'{\i}s Companys 23,
E-08010 Barcelona, Spain}

\author{F. Nowacki}
\email{frederic.nowacki@ires.in2p3.fr}
\affiliation{Institut de Recherches Subatomiques, IN2P3-CNRS,
  Universit\'e Louis Pasteur, F-67037 Strasbourg, France}

\author{A. Poves}
\email{alfredo.poves@uam.es}
\affiliation{Departamento de F\'{\i}sica Te\'orica, Universidad
  Aut\'onoma, Cantoblanco, 28049, Madrid, Spain}

\author{A. P. Zuker}
\email{andres.zuker@ires.in2p3.fr}
\affiliation{Institut de Recherches Subatomiques, IN2P3-CNRS,
  Universit\'e Louis Pasteur, F-67037 Strasbourg, France}

\date{\today}

\begin{abstract}
  The last decade has witnessed both quantitative and qualitative
  progresses in  Shell Model studies, which have resulted in remarkable
  gains in our
  understanding of the structure of the nucleus. Indeed,  it is now possible
  to diagonalize matrices in determinantal spaces of dimensionality up to
  $10^9$ using the Lanczos tridiagonal construction, whose formal and
  numerical aspects we will analyze. Besides, many new 
  approximation methods have been developed in order to overcome
  the dimensionality limitations. Furthermore, new effective
  nucleon-nucleon interactions have been constructed that contain both
  two and
  three-body contributions. The former are derived from realistic
  potentials ({\it i.e.}, consistent with two nucleon data). The latter
  incorporate the pure monopole terms necessary to
  correct the bad saturation and shell-formation properties of the
  realistic two-body forces. This combination appears to solve a number of
  hitherto puzzling problems. In the present review we will
  concentrate on those results which illustrate the global features
  of the approach:  the universality of
  the effective interaction and the capacity of the Shell Model to
  describe simultaneously all the manifestations of the nuclear
  dynamics either of single particle or collective nature.  We will
  also treat in some detail the problems associated with rotational motion, the
  origin of quenching of the Gamow Teller transitions, the double
  $\beta$-decays, the effect of isospin non conserving nuclear forces,
  and the specificities of the very neutron rich nuclei.  Many other
  calculations---that appear to have ``merely'' spectroscopic
  interest---are touched upon briefly, although we are fully aware that
  much of the credibility of the Shell Model rests on them.

\end{abstract}

\maketitle

\tableofcontents

\section{Introduction}
\label{sec:introduction}
In the early days of nuclear physics, the  nucleus, 
composed of strongly interacting neutrons and
protons confined in a very small volume, didn't appear as a system
to which  the shell model, so successful in the atoms, could be of
much relevance. Other descriptions---based on the analogy with a
charged liquid drop---seemed more natural. However, experimental
evidence of independent particle behavior in nuclei soon began to accumulate,
such as the extra binding related to some precise values of the number
of neutrons and protons (magic numbers) and the systematics of spins
and parities.
 
The existence of shell structure in nuclei had already been noticed in
the thirties but it took more than a decade and numerous
papers~\cite[see][for the early history]{Elliott.Lane:1957} before the
correct prescription was found by~\citet*{Mayer:1949}
and~\citet*{Haxel.Jensen.Suess:1949}. To explain the regularities of
the nuclear properties associated to ``magic numbers''---i.e.,
specific values of the number of protons $Z$ and neutrons $N$---the
authors proposed a model of independent nucleons confined by a surface
corrected, isotropic harmonic oscillator, plus a strong attractive
spin-orbit term\footnote{We assume throughout adimensional oscillator
  coordinates, i.e., $r\longrightarrow (m\omega/\hbar)^{1/2} \; r$.},
\begin{equation}  
 U(r) = \frac{1}{2} \; \hbar\omega \;r^2 + D \; \vec{l} \;^2
+ C \; \vec{l} \cdot \vec{s}.
\label{ho+ls} 
\end{equation}
In modern language this proposal amounts to assume that the main
effect of the two-body nucleon-nucleon interactions is to generate a
spherical mean field. The wave function of the ground state of a given
nucleus is then the product of one Slater determinant for the protons
and another for the neutrons, obtained by filling the lowest subshells
(or ``orbits'') of the potential.  This primordial shell model is
nowadays called ``independent particle model'' (IPM) or, ``naive''
shell model. Its foundation was pioneered by \citet{Brueckner:1954}
who showed how the short range nucleon-nucleon repulsion combined with
the Pauli principle could lead to nearly independent particle motion.

As the number of protons and neutrons depart from the
magic numbers it becomes indispensable to include in some way the
``residual'' two-body interaction, to break the degeneracies inherent
to the filling of orbits with two or more nucleons. 
At this point difficulties accumulate:
``Jensen himself never lost his skeptical attitude
towards the extension of the single-particle model to include the
dynamics of several nucleons outside closed shells in terms of a
residual interaction'' \citep{Weidenmuller:1990}. Nonetheless, some
physicists chose to persist.
One of our purposes  is to explain and illustrate why it was worth
persisting. 
The keypoint is that passage from the
IPM to the interacting shell model, the shell model (SM) for short, is
conceptually simple but difficult in practice. In the next section 
we succinctly  review  the steps involved. Our aim
is to give the  reader an overall view of the Shell Model as a
sub-discipline of the Many Body problem. ``Inverted commas'' are 
used for terms of the nuclear jargon when they  appear for the first time.
The rest of the introduction will sketch  the competing
views of nuclear structure and establish their connections with the
shell model. Throughout, the reader will be directed to the
sections of this review where specific topics are discussed.

\subsection{The three pillars of the shell
  model}\label{sec:three-pillars-shell} The strict validity of the IPM
may be limited to closed shells (and single particle --or hole--
states built on them), but it provides a framework with two important
components that help in dealing with more complex situations.  One is
of mathematical nature: the oscillator orbits define a basis of Slater
determinants, the $m$-scheme, in which to formulate the Schr\"odinger
problem in the occupation number representation (Fock space). It is
important to realize that the oscillator basis is relevant, not so
much because it provides an approximation to the individual nucleon
wave-functions, but because it provides the natural quantization
condition for self-bound systems.

Then, the many-body problem becomes one of diagonalizing a simple
matrix. We have a set of determinantal states for $A$ particles,
$a^{\dagger}_{i1}\ldots a^{\dagger}_{iA}|0\rangle=|\phi_{I}\rangle$,
and a Hamiltonian containing kinetic ${\cal K}$ and potential energies
${\cal V}$
\begin{gather}
  \label{Hec}
  {\cal H}=\sum_{ij} {\cal K}_{ij}\, a^{\dagger}_ia_j- \sum_{i\le j\, k\le
    l}{\cal V}_{ijkl}\,a^{\dagger}_ia^{\dagger}_ja_ka_l
\end{gather}
that adds one or two particles in orbits $i,\, j$ and removes one or
two from orbits $k,\, l$, subject to the Pauli principle
($\{a_i^{\dagger}a_j\}=\delta_{ij}$). The eigensolutions of the
problem $|\Phi_{\alpha}\rangle = \sum_I\,
c_{I,\alpha}|\phi_{I}\rangle$, are the result of diagonalizing the
matrix $\langle \phi_{I}|{\cal H}|\phi_{I'}\rangle$ whose off-diagonal
elements are either 0 or $\pm {\cal V}_{ijkl}$.  However, the
dimensionalities of the matrices---though not infinite, since a cutoff
is inherent to a non-relativistic approach---are so large as to make
the problem intractable, except for the lightest nuclei. Stated in
these terms, the nuclear many body problem does not differ much from
other many fermion problems in condensed matter, quantum liquids or
cluster physics; with which it shares quite often concepts
and techniques.  The differences come from the interactions, which in
the nuclear case are particularly complicated, but paradoxically quite
weak, in the sense that they produce sufficiently little mixing of
basic states so as to make the zeroth order approximations bear
already a significant resemblance with reality.

This is the second far-reaching --physical-- component of the IPM: the
basis can be taken to be small enough as to be, often, tractable. Here
some elementary definitions are needed: The (neutron and proton) major
oscillator shells of principal quantum number $p=0,\, 1,\, 2,\,
3\ldots$, called $s,\, p,\, sd,\, pf\ldots$ respectively, of energy
$\hbar \omega (p+3/2)$, contain orbits of total angular momentum
$j=1/2,\, 3/2\ldots p+1/2$, each with its possible $j_z$ projections,
for a total degeneracy $D_j=2j+1$ for each subshell, and
$D_p=(p+1)(p+2)$ for the major shells (One should not confuse the
$p=1$ shell ---$p$-shell in the old spectroscopic notation---with the
generic harmonic oscillator shell of energy $\hbar \omega (p+3/2)$).

 When for a given nucleus the particles are restricted to have the
lowest possible values of $p$ compatible with the Pauli principle,
 we speak of a 0$\hbar \omega$ space.
When the many-body states are allowed to have components involving basis
states with up to $N$ oscillator quanta more than
those pertaining to the 0$\hbar \omega$ space we speak of a $N$$\hbar \omega$
space (often refered as ``no core'')

For nuclei up to $A\approx 60$, the major oscillator shells provide
physically meaningful 0$\hbar \omega$ spaces. The simplest possible
example will suggest why this is so.  Start from the first of the
magic numbers $N=Z=2$; $^4$He is no doubt a closed shell. Adding one
particle should produce a pair of single-particle levels: $p_{3/2}$
and $p_{1/2}$.  Indeed, this is what is found in $^5$He and $^5$Li.
What about $^6$Li?  The lowest configuration, $p_{3/2}^2$, should
produce four states of angular momentum and isospin $JT=01,10,21,30$.
They are experimentaly observed. There is also a $JT=20$ level that
requires the $p_{3/2}\, p_{1/2}$ configuration. Hence the idea of
choosing as basis $p^m$ ($p$ stands generically for $p_{3/2}\,
p_{1/2}$) for nuclei up to the next closure $N=Z=8$, i.e., $^{16}$O.
Obviously, the general Hamiltonian in~\eqref{Hec} must be transformed
into an ``effective'' one adapted to the restricted basis (the
``valence space''). When this is done, the results are very
satisfactory~\cite{Cohen.Kurath:1965}.  The argument extends to the
$sd$ and $pf$ shells.

By now, we have identified two of the three ``pillars'' of the Shell
Model: a good valence space and an effective interaction adapted to
it. The third is a shell model code capable of coping with the
secular problem. In Section~\ref{sec:valence-space} we shall examine
the reasons for the success of the classical 0\hw\ SM spaces and
propose extensions capable of dealing with more general cases. The
interaction will touched upon in Section~\ref{sec:realistic-it-vs} and
discussed at length in Section~\ref{sec:interaction}. The codes will
be the subject of Section~\ref{sec:solut-secul-probl}.

\subsection{The competing views of nuclear
  structure}\label{sec:comp-views-nucl} Because Shell Model work is so
computer-intensive, it is instructive to compare its history and
recent developments with the competing---or alternative---views of
nuclear structure that demand less (or no) computing power.

\subsubsection{Collective {\it vs.} Microscopic}
\label{sec:collective-it-vs}
The early SM was hard to reconcile with the idea of the compound
nucleus and the success of the liquid drop model. With the discovery
of rotational motion~\cite{Bohr:1952,Bohr.Mottelson:1953} which was at
first as surprising as the IPM, the reconciliation became apparently
harder. It came with the realization that collective rotors are
associated with ``intrinsic states'' very well approximated by
deformed mean field determinants~\cite{Nilsson:1955}, from which the
exact eigenstates can be extracted by projection to good angular
momentum~\cite{Peierls.Yoccoz:1957}; an early and spectacular example
of spontaneous symmetry breaking. By further introducing nuclear
superfluidity \cite{Bohr.Mottelson.Pines:1958}, the unified model was
born, a basic paradigm that remains valid.

Compared with the impressive architecture of the unified model, what
the SM could offer were some striking but isolated examples which
pointed to the soundness of a many-particle description. Among them,
let us mention the elegant work of~\citet{Talmi.Unna:1960}, the
$f_{7/2}^n$ model of~\citet*{McCullen.Bayman.Zamick:1964}---probably
the first successful diagonalization involving both neutrons and
protons---and the~\citet[]{Cohen.Kurath:1965} fit to the $p$ shell,
the first of the classical 0$\hbar \omega$ regions. It is worth noting
that this calculation involved spaces of $m$-scheme dimensionalities,
$d_m$, of the order of 100, while at fixed total angular momentum and
isospin $d_{JT}\approx 10$.

The microscopic origin of rotational motion was found by 
\citet{Elliott:1958a,Elliott:1958b}. The interest of this
contribution was immediately recognized but it took quite a few years
to realize that Elliott's quadrupole force and the underlying SU(3)
symmetry were the foundation, rather than an example, of rotational
motion (see Section~\ref{sec:spher-shell-model}).

It is fair to say that for almost twenty years after its inception, in
the mind of many physicists, the shell model still suffered from an
implicit separation of roles, which assigned it the task of accurately
describing a few specially important nuclei, while the overall
coverage of the nuclear chart was understood to be the domain of the
unified model.

 A somehow intermediate path was opened by the well known Interacting Boson
 Model of \citet{Arima.Iachello:1975} and its developments,
 that we do not touch here. The interested reader can find the
 details in \cite{Iachello.Arima:1987}.

\subsubsection{Mean field {\it vs} Diagonalizations}
\label{sec:mean-field-it}

The first realistic\footnote{Realistic interactions are those
  consistent with data obtained in two (and nowadays three) nucleon
  systems.} matrix elements of~\citet{Kuo.Brown:1966} and the first
modern shell model code by~\citet*{French.Halbert.ea:1969} came almost
simultaneously, and opened the way for the first generation of ``large
scale'' calculations, which at the time meant $d_{JT}\approx 100$-600.
They made it possible to describe the neighborhood of
$^{16}$O~\cite*{Zuker.Buck.McGrory:1968} and the lower part of the
$sd$ shell~\cite{Halbert.Mcgrory.ea:1971}. However, the increases in
tractable dimensionalities were insufficient to promote the shell
model to the status of a general description, and the role-separation
mentioned above persisted. Moreover the work done exhibited serious
problems which could be traced to the realistic matrix elements
themselves (see next Section~\ref{sec:realistic-it-vs}).

However, a fundamental idea emerged at the time: 
 the existence of an underlying universal two-body interaction, which
 could allow a replacement of 
the unified model description by a fully microscopic one,
based on mean-field theory. The first breakthrough came
when~\citet{Baranger.Kumar:1968} proposed a new form of the unified
model by showing that Elliott's quadrupole force could be somehow
derived from the Kuo-Brown matrix elements.  Adding a pairing
interaction and a spherical mean field they proceeded to perform Hartree
Fock Bogoliubov calculations in the first of a successful
series of papers~\cite{Kumar.Baranger:1968}. Their work could be
described as shell model by other means, as it was restricted to
valence spaces of two contiguous major shells. 

This limitation was overcome when~\citet{Vautherin.Brink:1972}
and~\citet{Decharge.Gogny:1980} initiated the two families of Hartree
Fock (HF) calculations (Skyrme and Gogny for short) that remain to
this day the only tools capable of giving microscopic descriptions
throughout the periodic table. They were later joined by the
relativistic HF approach~\cite{Serot.Walecka:1986}.  \cite*[For a
review of the three variants see][]{Bender.Heenen.Reinhard:2003}. See
also \citet*{Peru.Girod.Berger:2000}
and~\citet*{Rodriguez-Guzman.Egido.Robledo:2000} and references
therein for recent work with the Gogny force, the one for which
closest contact with realistic interactions and the shell model can be
established
(Sections~\ref{sec:rotors-pf-shell} and~\ref{sec:univ-real-inter}).

Since single determinantal states can hardly be expected accurately to
describe many body solutions, everybody admits the need of going
beyond the mean field. Nevertheless, as it provides such a good
approximation to the wavefunctions (typically about 50\%) it suggests
efficient truncation schemes, as will be explained in
Sections~\ref{sec:the-exps-method} and~\ref{sec:approx}.

\subsubsection{Realistic {\it vs} Phenomenological}
\label{sec:realistic-it-vs}
Nowadays, many regions remain outside the direct reach of the Shell
Model, but enough has happened in the last decade to transform it into
a unified view. Many steps---outlined in the next
Section~\ref{sec:about-this-review}---are needed to substantiate this
claim. Here we introduce the first: A unified view requires a unique
interaction. Its free parameters must be few and well defined, so as
to make the calculations independent of the quantities they are meant
to explain.  Here, because we touch upon a different competing view,
of special interest for shell model experts, we have to recall the
remarks at the end of the first paragraph of the preceding
Section~\ref{sec:mean-field-it}:

The exciting prospect that realistic matrix elements could lead to
parameter-free spectroscopy did not materialize. As the growing
sophistication of numerical methods allowed to treat an increasing
number of particles in the $sd$ shell, the results became disastrous.
Then, two schools of thought on the status of phenomenological
corrections emerged: In one of them, all matrix elements were
considered to be free parameters. In the other, only average matrix
elements (``centroids'')---related to the bad saturation and shell
formation properties of the two-body potentials---needed to be fitted
(the monopole way).  The former lead to the ``Universal $sd$'' (USD)
interaction~\cite{Wildenthal:84}, that enjoyed an immense success and
for ten years set the standard for shell model calculations in a large
valence space~\cite{Brown.Wildenthal:1988},~\cite[see][for a recent
review]{Brown:2001}. The second, which we adopt here, was initiated in
Eduardo Pasquini PhD thesis (1976), where the first calculations in
the full $pf$ shell, involving both neutrons and protons were
done\footnote{Two months after his PhD, Pasquini and his wife
  ``disappeared'' in Argentina. This dramatic event explains why his
  work is only publicly available in condensed
  form~\cite{Pasquini.Zuker:1978}.}. Twenty years later, we know that
the minimal monopole corrections proposed in his work are sufficient
to provide results of a quality comparable to those of USD for nuclei
in $f_{7/2}$ region.

However, there were still problems with the monopole way: they showed
around and beyond $^{56}$Ni, as well as in the impossibility to be
competitive with USD in the $sd$ shell (note that USD contains
non-monopole two-body corrections to the realistic matrix elements).
The solution came about only recently with the introduction of
three-body forces: This far reaching development will be explained in
detail in
Sections~\ref{sec:interaction},~\ref{sec:monopole-hamiltonian}
and~\ref{sec:monop-probl-three}. We will only anticipate on the
conclusions of these sections through two syllogisms. The first is:
The case for realistic two-body interactions is so strong that we have
to accept them as they are. Little is known about three-body forces
except that they exist.  Therefore problems with calculations
involving only two body forces must be blamed on the absence of three
body forces.  This argument raises (at least) one question: What to
make of all calculations, using only two-body forces, that give
satisfactory results? The answer lies in the second syllogism: All
clearly identifiable problems are of monopole origin.  They can be
solved reasonably well by phenomenological changes of the monopole
matrix elements. Therefore, fitted interactions can differ from the
realistic ones basically through monopole matrix elements. (We speak
of R-compatibility in this case). The two syllogisms become fully
consistent by noting that the inclusion of three-body monopole terms
always improves the performance of the forces that adopt the monopole
way.

For the 0$\hbar \omega$ spaces, the Cohen
Kurath~\cite{Cohen.Kurath:1965}, the Chung
Wildenthal~\cite{Chung.Wildenthal:1979} and the
FPD6~\cite{Richter.Brown:1991} interactions turn out to be
R-compatible. The USD~\cite{Wildenthal:84} interaction is
R-incompatible.  It will be of interest to analyze the recent set of
$pf$-shell matrix elements (GXPF1), proposed
by~\citet{Honma.Otsuka.ea:2002}, but not yet released.

\subsection{The valence space}\label{sec:valence-space}

The choice of the valence space should reflect a basic physical fact:
that the most significant components of the low lying states of nuclei
can be accounted for by many-body states involving the excitation of
particles in a few orbitals around the Fermi level.  The history of
the Shell Model is that of the interplay between experiment and theory
to establish the validity of this concept. Our present understanding
can be roughly summed up by saying that ``few'' mean essentially one
or two contiguous major shells\footnote{Three major shells are
  necessary to deal with super-deformation.}.

For a single major shell, the classical 0$\hbar \omega$ spaces, exact
solutions are now available from which instructive conclusions can be
drawn.  Consider, for instance,the spectrum of $^{41}$Ca from which we
want to extract the single-particle states of the $pf$ shell.
Remember that in $^5$Li we expected to find two states, and indeed
found two.  Now, in principle, we expect four states. They are
certainly in the data. However, they must be retrieved from a jungle
of some 140 levels seen below 7 MeV. Moreover, the $f_{7/2}$ ground
state is the only one that is a pure single particle state, the other
ones are split, even severely in the case of the highest ($f_{5/2}$)
~\cite{Endt.van-der-Leun:1990}.  (Section~\ref{sec:spec} contains an
interesting example of the splitting mechanism).  Thus, how can we
expect the $pf$ shell to be a good valence space? For few particles
above $^{40}$Ca it is certainly not. However, as we report in this
work, it turns out that above $A=46$, \emph{the lowest $(pf)^m$
  configurations}~\footnote{A configuration is a set of states having
  fixed number of particles in each orbit.} \emph{are sufficiently
  detached from all others so as to generate wavefunctions that can
  evolve to the exact ones through low order perturbation theory.}

The ultimate ambition of Shell Model theory is to get exact solutions.
The ones provided by the sole valence space are, so to speak, the tip
of the iceberg in terms of number of basic states involved. The rest
may be so well hidden as to make us believe in the literal validity of
the shell model description in a restricted valence space.  For
instance, the magic closed shells are good valence spaces consisting
of a single state. This does not mean that a magic nucleus is 100\%
closed shell: 50 or 60\% should be enough, as we shall see in
Section~\ref{sec:the-exps-method}. In section~\ref{meaning-val-space}
it will be argued that the 0\hw\ valence spaces account for basically
the same percentage of the full wavefunctions.  Conceptually the
valence space may be thought as defining a representation intermediate
between the Schr\"{o}dinger one (the operators are fixed and the
wavefunction contains all the information) and the Heisenberg one (the
reverse is true).

At best, 0$\hbar \omega$ spaces can describe only a limited number of
low-lying states of the same (``natural'') parity. Two contiguous
major shells can most certainly cope with all levels of interest but
they lead to intractably large spaces and suffer from a ``Center of
Mass problem'' (analyzed in Appendix~\ref{sec:center-mass-problem}).
Here, a physically sound pruning of the space is suggested by the IPM.
What made the success of the model is the explanation of the observed
magic numbers generated by the spin-orbit term, as opposed to the HO
(harmonic oscillator) ones (for shell formation see
Section~\ref{sec:shell-formation}).  If we separate a major shell as
HO($p$)$=p_> \oplus r_p$, where $p_>$ is the largest subshell having
$j=p+1/2$, and $r_p$ is the ``rest'' of the HO $p$-shell, we can
define the EI (extruder-intruder) spaces as EI($p$)$=r_p \oplus
(p+1)_>$: The $(p+1)_>$ orbit is expelled from HO($p+1$) by the
spin-orbit interaction and intrudes into HO($p$). The EI spaces are
well established standards when only one ``fluid'' (proton or neutron)
is active: The $Z$ or $N=28,\, 50,\, 82$ and 126 isotopes or isotones.
These nuclei are ``spherical'', amenable to exact diagonalizations and
fairly well understood~\cite[see for
example][]{Abzouzi.Caurier.Zuker:1991}.

As soon as both fluids are active, ``deformation'' effects become
appreciable, leading to ``coexistence'' between spherical and deformed
states, and eventually to dominance of the latter. To cope with this
situation we propose ``Extended EI spaces'' defined as EEI($p$)$=r_p
\; \oplus \Delta_{p+1}$, where $\Delta_p=p_> \oplus (p_>-2) \oplus
\ldots$, {\em i.e.}, the $\Delta j=2$ sequence of orbits that contain
$p_>$, which are needed to account for rotational motion as explained
in Section~\ref{sec:spher-shell-model}.  The EEI(1) ($p_{1/2},\,
d_{5/2},\, s_{1/2}$) space was successful in describing the full low
lying spectra for
$A=15$-18~\cite{Zuker.Buck.McGrory:1968,Zuker.Buck.McGrory:1969}.  It
is only recently that the EEI(2) ($s_{1/2},\, d_{3/2},\, f_{7/2},\,
p_{3/2}$) space (region around $^{40}$Ca) has become tractable (see
Sections~\ref{sec:36ar-40ca-super} and~\ref{sec:is}). EEI(3) is the
natural space for the proton rich region centered in $^{80}$Zr.  For
heavier nuclei exact diagonalizations are not possible but the EEI
spaces provide simple (and excellent) estimates of quadrupole moments
at the beginning of the well deformed regions
(Section~\ref{sec:heavier-nuclei}).

As we have presented it, the choice of valence space is primarily a
matter of physics. In practice, when exact diagonalizations are
impossible, truncations are introduced. They may be based on systematic
approaches, such as approximation schemes discussed in
Sections~\ref{sec:the-exps-method} and~\ref{sec:approx} or mean field
methods mentioned in Section~\ref{sec:mean-field-it}, and analyzed in
Section~\ref{sec:rotors-pf-shell}.

\subsection{About this review: the unified view}
\label{sec:about-this-review}

 We expect this review to highlight several unifying aspects common to
 the most recent successful shell model calculations:

 a) An effective interaction connected with both
 the two and three nucleon bare forces.

 b) The explanation of the global properties of
   nuclei via the monopole Hamiltonian.

 c) The universality of the multipole Hamiltonian.  

 d) The description of the collective behavior in the laboratory
    frame, by means of the  spherical shell model.

 e) The description of resonances using the Lanczos strength function method.

Let us see now how this review is organized along these lines. References
are only given for work that will not be mentioned later.

Section II. The basic tool in analyzing the interaction is the
monopole-multipole separation. The former is in charge of saturation
and shell properties; it can be thought as the correct generalization
of Eq.~(\ref{ho+ls}). Monopole theory is scattered in many references.
Only by the inclusion of three-body forces could a satisfactory
formulation be achieved. As this is a very recent and fundamental
development which makes possible a unified viewpoint of the monopole
field concept, this section is largely devoted to it. The ``residual''
multipole force has been extensively described in a single reference
which will be reviewed briefly and updated.  The aim of the section is
to show how the realistic interactions can be characterized by a small
number of parameters.

Section III. The ANTOINE and NATHAN codes have made it possible to
evolve from dimensionalities $d_m\sim 5\times 10^4$ in 1989 to
$d_m\sim 10^9,\, d_J\sim 10^7$ nowadays. Roughly half of the--order
of magnitude--gains are due to increases in computing power. The rest
comes from algorithmic advances in the construction of tridiagonal
Lanczos matrices that will be described in this section.

Section IV. The Lanczos construction can be used to eliminate the
``black box'' aspect of the diagonalizations, to a large extent. We
show how it can be related to the notions of partition function,
evolution operator and level densities. Furthermore, it can be turned
into a powerful truncation method by combining it to coupled cluster
theory. Finally, it describes strength functions with maximal efficiency.

Section V. After describing the three body mechanism which solves 
the monopole problem that had plagued the classical 0$\hbar
\omega$ calculations, some selected examples of $pf$ shell
spectroscopy are presented. Special attention is given to Gamow Teller
transitions, one of the main achievements of modern SM work.

Section VI. Another major recent achievement is the shell model
description of rotational nuclei.  The new generation of gamma
detectors, Euroball and Gammasphere has made it possible to access
high spin states in medium mass nuclei for which full 0$\hbar \omega$
calculations are available. Their remarkable harvest includes a large
spectrum of collective manifestations which the spherical shell model
can predict or explain as, for instance, deformed rotors, backbending,
band terminations, yrast traps, etc. Configurations involving two
major oscillator shells are also shown to account well for the
appearance of superdeformed excited bands.

Section VII. A conjunction of factors make light and medium-light
nuclei near the neutron drip line specially interesting: (a) They have
recently come under intense experimental scrutiny. (b) They are
amenable to shell model calculations, sometimes even exact no-core
ones (c) They exhibit very interesting behavior, such as halos and
sudden onset of deformation. (d) They achieve the highest $N/Z$ ratios
attained. (e) When all (or most of) the valence particles are
neutrons, the spherical shell model closures, dictated by the
isovector channel of the nuclear interaction alone, may differ from
those at the stability valley. These regions propose some exacting
tests for the theoretical descriptions, that the conventional shell
model calculations have passed satisfactorily.  For nearly unbound
nuclei, the SM description has to be supplemented by some refined
extensions such as the shell model in the continuum, which falls
outside the scope of this review.  References to the subject can be
found in \citet{Bennaceur.Nowacki.ea:1999,Bennaceur.Nowacki.ea:2000},
and in \citet{IdBetan.Liotta.ea:2002,Michel.Nazarevicz.ea:2002,
  Michel.Nazarevicz.ea:2003} that deal with the Gamow shell model.

Section VIII. There is a characteristic of the Shell Model we have not
yet stressed: it is the approach to nuclear structure that can give
more precise quantitative information. SM wave functions are, in
particular, of great use in other disciplines. For example: Weak decay
rates are crucial for the understanding of several astrophysical
processes, and neutrinoless $\beta\beta$ decay is one of the main
sources of information about the neutrino masses. In both cases, shell
model calculations play a central role. The last section deal with
these subjects, and some others...

Appendix B  contains a full derivation of the general form of
the monopole field.

Two recent reviews by \citet{Brown:2001} and
\citet{Otsuka.Honma.ea:2001} have made it possible to simplify our
task and avoid redundancies. However we did not feel dispensed from
quoting and commenting in some detail important work that bears
directly on the subjects we treat.

Sections II, IV and the Appendices are based on unpublished notes by APZ.

\section{The interaction}
\label{sec:interaction}

The following remarks from~\citet*{Abzouzi.Caurier.Zuker:1991} still
provide a good introduction to the subject: 

``The use of realistic potentials ({\it i.e.}, consistent with $NN$
scattering data) in shell-model calculation was pioneered by
\citet{Kuo.Brown:1966}.
Of the enormous body of work that followed we would like to extract
two observations.  The first is that whatever the forces (hard or soft
core, ancient or new) and the method of regularization (Brueckner $G$
matrix~\citep{Kuo.Brown:1966,Kahana.Lee.Scott:1969a}, Sussex direct
extraction~\citep{Elliott.Jackson.ea:1968} or Jastrow
correlations~\citep{Fiase.Hamoudi.ea:1988}) the effective matrix elements
are {\em extraordinarily
  similar}~\citep{Pasquini.Zuker:1978,Rutsgi.Kung.ea:1971}. The most
recent results~\citep{Jiang.Machleidt.Stout:1989} amount to a vindication of
the work of Kuo and Brown. We take this similarity to be the great
strength of the realistic interactions, since it confers on them a
model-independent status as direct links to the phase shifts.

The second observation is that when used in shell-model calculations
and compared with data these matrix elements give results that
deteriorate rapidly as the number of particle
increases~\citep{Halbert.Mcgrory.ea:1971}
and~\citep{Brown.Wildenthal:1988}. It was
found~\citep{Pasquini.Zuker:1978} that in the $pf$ shell a
phenomenological cure, confirmed by exact diagonalizations up to
A=48~\citep{Caurier.Zuker.ea:1994}, amounts to very simple
modifications of some average matrix elements ($centroids$) of the KB
interaction~\citep{Kuo.Brown:1968}.''

In~\citet{Abzouzi.Caurier.Zuker:1991} very good spectroscopy in the
$p$ and $sd$ shells could be obtained only through more radical
changes in the centroids, involving substantial three-body (3b) terms.
In 1991 it was hard to interpret them as effective and there were no
sufficient grounds to claim that they were real.

Nowadays, the need of true 3b forces has become irrefutable: In the
1990 decade several two-body (2b) potentials were
developed---Nijmegen I and II~\citep{Stoks.ea:1993},
AV18~\citep{Wiringa.ea:1995}, CD-Bonn-\citep{Machleidt.ea:1996}---that
fit the $\approx 4300$ entries in the Nijmegen data
base~\citep{Stoks.ea:1994} with $\chi^2\approx\, 1$, and none of them
seemed capable to predict perfectly the nucleon vector analyzing power
in elastic $(N,d)$ scattering (the $A_y$ puzzle). Two recent additions
to the family of high-precision 2b potentials---the charge-dependent
``CD-Bonn''~\citep{Machleidt:2001} and the chiral Idaho-A and
B~\citep{Entem.Machleidt:2002a}---have dispelled any hopes of solving
the $A_y$ puzzle with 2b-only
interactions~\citep{Entem.Machleidt:2002b}.

Furthermore, quasi-exact 2b Green Function Monte Carlo (GFMC)
results~\citep{Pudliner.Pandharipande.ea:1997,Wiringa.Pieper.ea:2000}
which provided acceptable spectra for $A\le 8$ (though they had
problems with binding energies and spin-orbit splittings), now
encounter serious trouble in the spectrum of $^{10}$B, as found
through the No Core Shell Model (NCSM) calculations of
~\citet{Navratil.Ormand:2002} and~\citet{Caurier.Navratil.ea:2002},
confirmed by~\citet*{Pieper.Varga.Wiringa:2002}, who also show that
the problems can be remedied to a large extent by introducing the new
Illinois 3b potentials developed
by~\citet{Pieper.Pandharipande.ea:2001}.
 
It is seen that the trouble detected with a 2b-only description---with
binding energies, spin-orbit splittings and spectra---is always
related to {\em centroids}, which, once associated to operators that
depend only on the number of particles in subshells, determine a
``monopole'' Hamiltonian ${\cal H}_m$ that is basically in charge of Hartree
Fock selfconsistency. As we shall show, the full ${\cal H}$ can be separated
rigorously as ${\cal H}={\cal H}_m+{\cal H}_M$. The multipole part ${\cal H}_M$ includes pairing,
quadrupole and other forces responsible for collective behavior,
and---as checked by many calculations---{\em is well given by the}
2b {\em potentials}.
 
The preceding paragraph amounts to rephrasing the two observations
quoted at the beginning of this section with the proviso that the
blame for discrepancies with experiment cannot be due to the---now
nearly perfect---2b potentials. Hence the necessary ``corrections'' to
${\cal H}_m$ must have a 3b origin. Given that we have no complaint with
${\cal H}_M$, the primary problem is the monopole contribution to the
3b  potentials. This is welcome because a full 3b treatment would
render most shell model calculations impossible, while the
phenomenological study of monopole behavior is quite simple.
Furthermore it is quite justified because there is little {\em ab
  initio} knowledge of the 3b potentials\footnote{Experimentally
  there will never be enough data to determine them. The hope for a
  few parameter description comes from chiral perturbation
  theory~\cite{Entem.Machleidt:2002a}.}. Therefore, whatever
information that comes from nuclear data beyond A=3 is welcome.

In Shell Model calculations, the interaction appears as matrix
elements, that soon become far more numerous than the number of
parameters defining the realistic potentials.  Our task will consist
in analyzing the Hamiltonian in the oscillator representation (or Fock
space) so as to understand its workings and simplify its form. In
Section~\ref{sec:effective} we sketch the theory of effective
interactions. In Section~\ref{sec:monopole-hamiltonian} it is
explained how to {\em construct from data} a minimal ${\cal H}_m$, while
Section~\ref{sec:mult-hamilt} will be devoted to {\em extract from
  realistic forces} the most important contributions to ${\cal H}_M$.  The
basic tools are symmetry and scaling arguments, the clean separation
of bulk and shell effects and the reduction of ${\cal H}$ to sums of
factorable terms~\citep{Dufour.Zuker:1996}.

\subsection{Effective interactions}\label{sec:effective}
The Hamiltonian is written in an oscillator basis as
\begin{eqnarray}
  \label{eq:h}
 {\cal H}&=&{\cal K}+\sum_{r\leq s, \;  t\leq u,\; \Gamma} 
{{\cal V}}_{rstu}^{\Gamma} Z_{rs\Gamma}^{+}\cdot  Z_{tu\Gamma}\nonumber \\
&&+\sum_{r\leq s \leq t, \; u\leq
  v\leq w,\; \Gamma}
{\cal V}_{rstuvw}^{\Gamma} Z_{rst\Gamma}^{+}\cdot  Z_{uvw\Gamma},
\end{eqnarray}
where ${\cal K}$ is the kinetic energy, ${\cal V}_{\bm{rr'}}^{\Gamma}$ the
interaction matrix elements, $Z^{+}_{{\bm r}\Gamma}$ ( $Z_{{\bm
    r}\Gamma}$) create (annihilate) pairs (${\bm r}\equiv rs$) or
triples (${\bm r}\equiv rst$) of particles in orbits ${\bm r}$,
coupled to $\Gamma=JT$. Dots stand for scalar products. 
The basis and the matrix elements are large but never infinite\footnote{
  Nowadays, the non-relativistic potentials must be thought of as
  derived from an effective field theory which has a cutoff of about 1
  GeV~\cite{Entem.Machleidt:2002a}. Therefore, truly hard cores are
  ruled out, and $\mathcal{H}$ should be understood to act on a
  sufficiently large vector space, \emph{not over the whole Hilbert
    space}.}.

The aim of an effective interaction theory is to reduce the secular
problem in the large space to a smaller model space by treating
perturbatively the coupling between them, thereby transforming the
full potential, and its repulsive short distance behavior, into a
smooth pseudopotential.

In what follows, if we have to distinguish between large ($N\hbar\omega$)
and model spaces we use ${\cal H},\, {\cal K}\,{\rm and }\, {\cal
  V}$ for the pseudo-potential in the former and $H,\, K\, {\rm and }\,
V$ for the effective interaction in the latter.

\subsubsection{Theory and calculations}\label{sec:theory-calculations}

The general procedure to describe an exact eigenstate in a restricted
space was obtained independently by~\citet{Suzuki.Lee:1980}
and~\citet{Poves.Zuker:1981b}\footnote{The notations in both papers
  are very different but the perturbative expansions are probably
  identical because the Hermitean formulation of~\citet{Suzuki:1982}
  is identical to the one given in~\citet{Poves.Zuker:1981b}. This
  paper also deals extensively with the coupled cluster formalism.}. It 
consists in dividing the full space into model ($\vert i\rangle$) and
external ($\vert \alpha\rangle$) determinants, and introducing a
transformation that respects strict orthogonality between the spaces
and decouples them exactly:
 \begin{gather}
  \vert \overline{\imath}\rangle=\vert i\rangle + \sum_{\alpha }
  A_{i\alpha } \vert \alpha \rangle \; \; \vert \overline{\alpha
  }\rangle=\vert \alpha \rangle - \sum_{i} A_{i\alpha } \vert
  i\rangle \label{eq:decij} \\
  \langle\overline{\imath}|\overline{\alpha }\rangle=0, \; \; A_{i\alpha
  }\; \text{is defined through}\; \;\langle\overline{\imath}\vert {\cal H}
  \vert \overline{\alpha }\rangle=0.
\label{eq:dec}
\end{gather}
The idea is that the model space can produce one or several starting
wavefunctions that can evolve to exact eigenstates through
perturbative or coupled cluster evaluation of the amplitudes
$A_{i\alpha }$, which can be viewed as matrix elements of a many body
operator $A$. In coupled cluster theory (CCT or $\exp{S}$) 
(\citet{Coester.Kummel:1960},
~\citet{Kummel.Luhrmann.Zabolitzky:1978}) one sets $A=\exp{S}$, where
$S=S_1+S_2+\cdots +S_k$ is a sum of $k$-body (kb) operators. The
decoupling condition $\langle\overline{\imath}\vert {\cal H} \vert
\overline{\alpha }\rangle=0$ then leads to a set of coupled integral
equations for the $S_i$ amplitudes. When the model space reduces to a
single determinant, setting $S_1=0$ leads to Hartree Fock theory if
all other amplitudes are neglected. The $S_2$ approximation contains
both low order Brueckner theory (LOBT) and the random phase
approximation (RPA) . In the presence of hard-core potentials, the
priority is to screen them through LOBT, and the matrix elements
contributing to the RPA are discarded. An important implementation of
the theory was due to Zabolitzky, whose calculations for $^{4}$He,
$^{16}$O and $^{40}$Ca, included $S_3$ (Bethe Faddeev) and $S_4$ (Day
Yacoubovsky) amplitudes~\citep*{Kummel.Luhrmann.Zabolitzky:1978}.
This ``Bochum truncation scheme'' that retraces the history of nuclear
matter theory has the drawback that at each level terms that one would
like to keep are neglected.

The way out of this problem ~\cite{Heisenberg.Mihaila:1999,
  Mihaila.Heisenberg:2000} consists on a new truncation scheme in
which some approximations are made, but no terms are neglected,
relying on the fact that the matrix elements are finite. The
calculations of these authors for $^{16}$O (up to $S_3$) can be ranked
with the quasi-exact GFMC and NCSM ones for lighter nuclei. 

In the quasi-degenerate regime (many model states), the coupled
cluster equations determine an effective interaction in the model
space. The theory is much simplified if we enforce the decoupling
condition for a {\em single} state whose exact wavefunction is written as
\begin{multline}
  \label{eq:lct}
  |\overline{ref}\rangle =(1+A_1+A_2+\dots)(1+B_1+B_2+\dots)|ref\rangle\\
=(1+C_1+C_2+\dots)|ref\rangle, \qquad C=\exp{S},
\end{multline}
where $|ref\rangle$ is a model determinant. The internal amplitudes
associated with the $B$ operators are those of an eigenstate obtained
by diagonalizing $H_{eff}={\cal H}(1+A)$ in the model space. As it is
always possible to eliminate the $A_1$ amplitude, at the $S_2$ level
there is no coupling {\it i.e.}, the effective interaction is a state
independent G-matrix~\cite{Zuker:1984}, which has the advantage of
providing an initialization for $H_{eff}$. Going to the $S_3$ level
would be very hard, and we examine what has become standard
practice.

The power of CCT is that it provides a unified framework for the two
things we expect from decoupling: to smooth the repulsion and to
incorporate long range correlations. The former demands jumps of say,
50$\hbar\omega$, the latter much less. Therefore it is convenient to
treat them separately. Standard practice {\em assumes} that
G-matrix elements can provide a smooth pseudopotential in some
sufficiently large space, and then accounts for long range correlations
through perturbation theory. The equation to be solved is 
\begin{equation}
  \label{eq:G}
  G_{ijkl}={\cal V}_{ijkl}-\sum_{\alpha\beta}\frac{{\cal
  V}_{ij\alpha\beta}G_{\alpha\beta
  kl}}{\epsilon_{\alpha}+\epsilon_{\beta}-\epsilon_i-\epsilon_j+\Delta},  
\end{equation}
where $ij$ and $kl$ now stand for orbits in the model space while in
the pair $\alpha \beta$ at least one orbit is out of it, $\epsilon_x$
is an unperturbed (usually kinetic) energy, and $\Delta$ a free parameter
called the ``starting'' energy. \citet*{Hjorth-Jensen.Kuo.Osnes:1995}
describe in detail a sophisticated partition that amounts to having
two model spaces, one large and one small. 

In the NCSM calculations an $N\hbar\omega$ model space is chosen with
$N\approx 6$-10.  When initiated by~\citet{Zheng.Barrett.ea:1993} the
pseudo-potential was a G-matrix with starting energy. Then the
$\Delta$ dependence was eliminated either by arcane perturbative
maneuvers, or by a truly interesting proposal: direct decoupling of 2b
elements from a very large space~\cite{Navratil.Barrett:1996}, further
implemented by~\citet*{Navratil.Vary:2000}, and extended to 3b
effective forces~\cite{Navratil.Kamuntavicious.ea:2000},
\cite{Navratil.Ormand:2002}. It would be of interest to compare the
resulting effective interactions to the Brueckner and Bethe Faddeev
amplitudes obtained in a full $\exp{S}$ approach (which are also free
of arbitrary starting energies).

A most valuable contribution of NCSM is the {\em proof} that
it is possible to work with a pseudo-potential in $N\hbar\omega$
spaces. The method relies on exact diagonalizations. As they soon
become prohibitive, in the future, CCT may become the standard
approach: going to $S_3$ for $N=50$ (as Heisenberg and Mihaila did) is
hard. For $N=10$ it should be much easier.

Another important NCSM indication is that the excitation spectra
converge well before the full energy, which validates formally the
$0\hbar\omega$ diagonalizations with rudimentary
potentials\footnote{Even old potentials fit well the low energy $NN$
  phase shifts, the only that matter at $0\hbar\omega$.} and second
order corrections. 

The $0\hbar\omega$ results are very good for the spectra. However, to
have a good pseudopotential to describe energies is not enough. The
transition operators also need dressing. For some of them, notably
$E2$, the dressing mechanism (coupling to 2\hw quadrupole excitations)
has been well understood for years~\citep[see][for a detailed
analysis]{Dufour.Zuker:1996}, and yields the, abundantly tested and
confirmed, recipe of using effective charges of $\approx$ 1.5e for the
protons and $\approx$ 0.5e for neutrons. For the Gamow Teller (GT)
transitions, mediated by the spin-isospin $\sigma\tau_{\pm}$ operator
the renormalization mechanism involves an overall ``quenching'' factor
of 0.7--0.8 whose origin is far subtler. It will be examined in
Sections~\ref{meaning-val-space}~and~\ref{sec:quenching:-no-core}.
The interested reader may wish to consult them right away.

\subsection{The monopole Hamiltonian}\label{sec:monopole-hamiltonian}
A many-body theory usually starts by separating the Hamiltonian into
an ``unperturbed'' and a ``residual'' part, ${\cal H}={\cal H}_0+{\cal
  H}_r$. The traditional approach consists in choosing for ${\cal
  H}_0$ a one body (1b) single-particle field. Since ${\cal H}$
contains two and three-body (2b and 3b) components, the separation is
not mathematically clean.  Therefore, we propose the following
\begin{equation}
  \label{eq:hmM}
  {\cal H}={\cal H}_m+{\cal H}_M,
\end{equation}
where ${\cal H}_m$, the monopole Hamiltonian, contains ${\cal K}$ and
all quadratic and cubic (2b and 3b) forms in the scalar products of
fermion operators $a_{r_x}^{\dagger}\cdot a_{s_y}$\footnote{$r$ and
  $s$ are subshells of the same parity and angular momentum, $x$ and
  $y$ stand for neutrons or protons}, while the multipole ${\cal H}_M$
contains all the rest.

Our plan is to concentrate on the 2b part, and introduce 3b elements
as the need arises.

${\cal H}_m$ has a {\em diagonal} part, ${\cal H}_m^d$, written in
terms of number and isospin operators ($a_{r_x}^{\dagger}\cdot a_{r_y}$). It
reproduces the average energies of configurations at fixed number of
particles and isospin in each orbit ($jt$ representation) or,
alternatively, at fixed number of particles in each orbit and each
fluid (neutron-proton, $np$, or $j$ representation).
In $jt$ representation the
centroids 
\begin{subequations}
\label{eq:vab}
\begin{eqnarray}
  \label{eq:vjt}
  {\cal V}_{st}^T&=&\frac{\sum_J
  {\cal V}_{stst}^{JT}(2J+1)[1-(-)^{J+T}\, \delta_{st}]}{\sum_J
  (2J+1)[1-(-)^{J+T} 
 \,  \delta_{st}]},\\ 
\label{eq:ab}
a_{st}&=&\frac{1}{4}(3{\cal V}_{st}^1+{\cal V}_{st}^0), \hspace{10pt}
b_{st}={\cal V}_{st}^1-{\cal V}_{st}^0,
\end{eqnarray}
\end{subequations}
are associated to the 2b quadratics in number ($m_s$) and isospin
operators ($T_s$), 
\begin{subequations}
\label{eq:hmjt}
\begin{eqnarray}
  \label{eq:nrs}
  m_{st}&=&\frac{1}{1+\delta_{st}}m_s(m_t-\delta_{st}), \\
\label{eq:trs}
  T_{st}&=&\frac{1}{1+\delta_{st}}(T_s\cdot
  T_t-\frac{3}{4}m_{st}\, \delta_{st}), 
\end{eqnarray}
\end{subequations}
 to define
the diagonal 2b part of the monopole
Hamiltonian~(\citet{Bansal.French:1964},~\citet{French:1969})
\begin{equation} 
\label{eq:hmjtd}
 {\cal H}_{mjt}^d={\cal K}^d+\sum_{s\le t} (a_{st}\,
 m_{st}+b_{st}\, T_{st})+ {\cal V}^{d_3}_m,
\end{equation}
The 2b part is a standard result, easily extended to include the 3b term
\begin{equation}
\label{hm3}
{\cal V}^{d_3}_m=\sum_{s\, t\, u} (a_{stu}\, m_{stu}+b_{stu}\, T_{stu}),
\end{equation} 
where $m_{stu}\equiv m_{st}m_u$, or $m_s(m_s-1)(m_s-2)/6$ and
$T_{stu}\equiv m_sT_{tu}$ or $(m_s-2)T_{su}$.  The extraction of the
full ${\cal H}_m$ is more complicated and it will be given in detail
in Appendix~\ref{sec:hm}. Here we only need to note that ${\cal H}_m$
is closed under unitary transformations of the underlying fermion
operators, and hence, {\em under spherical Hartree-Fock} (HF)
{\em  variation}. This property explains the interest of the ${\cal
  H}_m+{\cal H}_M$ separation.

\subsubsection{Bulk properties. Factorable forms}
\label{sec:bulk-prop-fact-1}
${\cal H}_m$ must contain all the information necessary to produce the
parameters of the Bethe Weisz\"acker mass formula, and we start by
extracting the bulk energy. The key step involves the reduction to a
sum of factorable forms valid for any
interaction~\citep{Dufour.Zuker:1996}. Its enormous power derives from
the strong dominance of a single term in all the cases considered so
far.

For clarity we restrict attention to the full isoscalar centroids
defined in Eqs.~\eqref{(II.22)} and~\eqref{(II.24)}\footnote{They
ensure the right average energies for $T=0$ closed shells, which is
not the case in Eq.~\eqref{eq:hmjtd} if one simply drops the $b_{st}$
terms.}
\begin{gather}
  \label{eq:vm}
  {\cal V}_{st}=a_{st}-\frac{3\delta_{st}b_{st}}{4(4j_s+1)}.
\end{gather}

Diagonalize $ {\cal V}_{st}$
\begin{gather}
U^{-1}{\cal V}^d_m U={\cal E}  \Longrightarrow {\cal V}_{st} =\sum_k
U_{sk} U_{tk}{\cal  E}_k,\qquad \bm {\therefore} \label{eq:u1}
\\
{\cal V}^d_m=\sum_{k}{\cal E}_k\sum_sU_{sk}
m_s\sum_t U_{tk}m_t-\sum_s {\cal V}_{ss}m_s.
\label{eq:u2}
\end{gather}
For the 3b interaction, the corresponding centroids ${\cal V}_{stu}$
are treated as explained in \citep[][Appendix B1b]{Dufour.Zuker:1996}:
The $st$ pairs are replaced by a single index $x$. Let $L$ and $M$ be
the dimensions of the $x$ and $s$ arrays respectively. Construct and
diagonalize an $(L+M)\times (L+M)$ matrix whose non-zero elements are
the rectangular matrices ${\cal V}_{xs}$ and ${\cal V}_{sx}$.
Disregarding the contractions, the strictly 3b part ${\cal V}^{(3)}$
can then be written as a sum of factors
\begin{equation}
{\cal V}^{d_3}_m=\sum_{k}{\cal E}_k^{(3)}\sum_sU_{sk}^{(1)}
m_s\sum_{x\equiv tu}U_{tu,k}^{(2)}m_tm_u.\label{eq:u3}
\end{equation}
Full factorization follows by applying Eqs.~(\ref{eq:u1},\ref{eq:u2})
to the $U_{tu,k}^{(2)}$ matrices for each $k$.

\begin{figure}[h]
\begin{center}
  \includegraphics[width=0.9\linewidth]{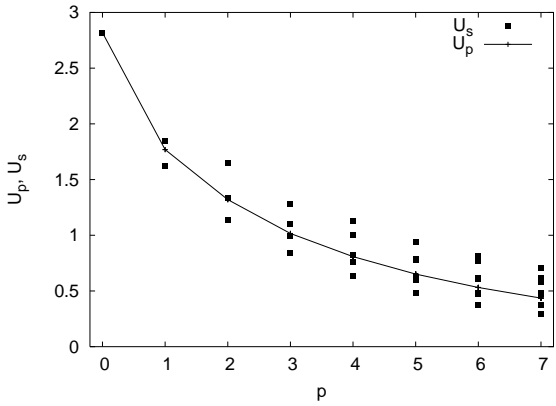}
  \caption {The $U_s$ terms in Eq.~(\ref{eq:master}). Arbitrary scale.}
\label{fig:dzn}
\end{center}
\end{figure}
In Eq.~(\ref{eq:u2}) a single term strongly dominates all others. For
the KLS interaction~\cite{Kahana.Lee.Scott:1969b} and including the
first eight major shells, the result in Fig.~\ref{fig:dzn} is roughly
approximated by
\begin{gather}
  \label{eq:master}
U_s\approx \frac{4.-0.5\, (l-\langle
  l\rangle)+(j-l)}{\sqrt{D_p}}, 
\end{gather}
where $p$ is the principal quantum number of an oscillator shell with
degeneracy $D_p=\sum_s (2j_s+1)=(p+1)(p+2)$, $\langle l\rangle=\sum_l
l(2l+1)/\sum_l (2l+1)$. The (unitary) $U$ matrices have been
affected by an arbitrary factor 6 to have numbers of order unity.
Operators of the form ($D_s=2j_s+1$)
\begin{gather}
  \hat\Omega=\sum_s m_s\Omega_s \qquad {\rm with}\qquad \sum_s
  D_s\Omega_s=0 \label{eq:Omega}
\end{gather}
vanish at closed shells and are responsible for shell effects.
 As $l-\langle l\rangle)$ and $j-l $ are of
this type, only the the $U_p$ part contributes to the bulk energy. 

To proceed it is necessary to know how interactions depend on the
oscillator frequency $\omega$ of the basis, related to the observed
square radius (and hence to the density) through the
estimate~\citep[Eq.~(2-157)]{Bohr.Mottelson:1969}:
\begin{equation}
  \label{eq:hw}
\frac{\hbar\omega}{(A)^{1/3}}=
\frac{35.59}{\langle r^2\rangle}\Longrightarrow \hbar\omega\approx
\frac{40}{A^{1/3}} {\rm MeV}
\end{equation}
A $\delta$ force scales as $(\hbar\omega)^{3/2}$. A 2b potential of
short range is essentially linear in $\hbar\omega$; for a 3b one we
shall tentatively assume an $(\hbar\omega)^2$ dependence while
the Coulomb force goes exactly as $(\hbar\omega)^{1/2}$.

To calculate the bulk energy of nuclear matter we average out subshell
effects through uniform filling $m_s\Rightarrow m_p\, D_s/D_p$. Though the
$\Omega$-type operators vanish, we have kept them for reference in
Eqs.~(\ref{eq:W},\ref{eq:W3}) below.  The latter is an educated guess
for the 3b contribution. The eigenvalue ${\cal E}_0$ for the dominant
term in Eq.~\ref{eq:u2} is replaced by $\hbar\omega{\cal V}_0$ defined
so as to have $U_p=1$. The subindex $m$ is dropped throughout.
Then, using Boole's factorial powers, e.g., $p^{(3)}=p(p-1)(p-2)$, we
obtain the following asymptotic estimates for the leading terms ({\it i.e.},
disregarding contractions)
\begin{eqnarray}
&&{\cal K}^d=\frac{\hbar \omega}{2}\sum_p m_p(p+3/2)\nonumber\\
&&\Longrightarrow
  \frac{\hbar 
  \omega}{4}(p_f+3)^{(3)}(p_f+2)  \label{eq:K} \\
&&{\cal V}^d\approx {\hbar\omega}{\cal V}_0\left(\sum_p
  \frac{m_p}{\sqrt{D_p}}+\hat\Omega\right)^2\nonumber\\
&&\Longrightarrow{\hbar\omega}{\cal
  V}_0[p_f(p_f+4)]^2,   \label{eq:W}\\ 
&&{\cal V}^{d_3}\approx {(\hbar\omega})^2\beta{\cal
  V}_0\left(\sum_p 
  \frac{m_p}{D_p}+\hat\Omega_1\right)\left(\sum_p
  \frac{m_p}{\sqrt{D_p}}+\hat\Omega_2\right)^2\nonumber\\
&&\Longrightarrow {(\hbar\omega})^2\beta{\cal V}_0p_f^3(p_f+4)^2.
  \label{eq:W3}\\
&&\; {\rm Finally, relate}\; p_f\; {\rm to}\; A: \nonumber\\
&&\sum_p m_p=\sum_{p=0}^{p_f}2(p+1)(p+2)\Longrightarrow
A=\frac{2(p_f)^{(3)}}{3}.   \label{eq:A}   
\end{eqnarray}
Note that in Eq.~\eqref{eq:W3} we can replace $(\hbar\omega)^2$ by
$(\hbar\omega)^{1+\kappa}$ and change the powers of $D_p$ in the
denominators accordingly. 

Assuming that non-diagonal ${\cal K}^d$ and ${\cal V}^d$ terms cancel
we can vary with respect to \hw\ to obtain the saturation energy
 
\begin{gather}
E_s={\cal K}^d+(1-\beta\omega^\kappa p_f){\cal V}^d,\nonumber\\
 \frac{\partial E}{\partial\omega}=0\; \Rightarrow\; 
\beta\omega^\kappa_e p_f=\frac{{\cal K}^d+ {\cal V}^d}{(\kappa+1)\, 
  {\cal V}^d}\nonumber\\
{\bm \therefore}\quad E_s=\frac{\kappa}{\kappa+1}({\cal K}^d+ {\cal V}^d)
\nonumber\\  
=\frac{\hbar\omega_e}{4}\frac{\kappa}{\kappa+1}
\left(1-4 {\cal V}_0\right)
{\left(\frac{3}{2}A\right)}^{4/3}.
\label{eq:sat}
\end{gather}

The correct saturation properties are obtained by fixing
${\cal V}_0$ so that $E_s/A\approx 15.5$ MeV.  The 
$\hbar\omega_e\approx 40A^{-1/3}$ choice ensures the correct density.
It is worth noting that the same approach leads to  $V_C\approx
3e^2Z(Z-1)/5R^c$~\cite{Duflo.Zuker:2002}.

It should be obvious that nuclear matter properties---derived from
finite nuclei---could be calculated with techniques designed to treat
finite nuclei: A successful theoretical mass table must necessarily
extrapolate to the Bethe Weisz\"acker
formula~\citep{Duflo.Zuker:1995}. It may be surprising though, that
such calculations could be conducted so easily in the oscillator
basis. The merit goes to the separation and factorization properties
of the forces.

\begin{figure}[htb]
  \includegraphics[width=0.9\linewidth]{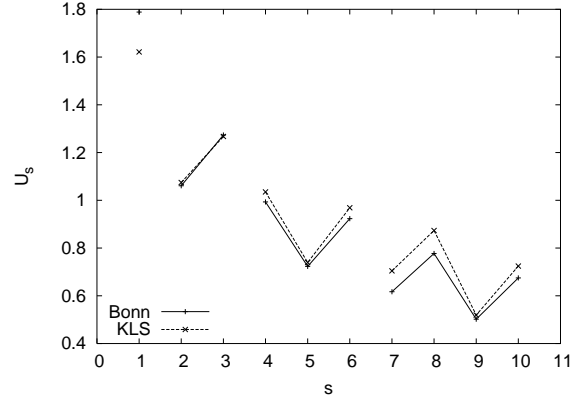}
  \caption{The KLS~\cite{Kahana.Lee.Scott:1969b} shell effects in
    Fig.~\ref{fig:dzn} for the first 10 orbits compared to those
    produced by a Bonn potential~\cite{Hjorth-Jensen:1996} (to within
    an overall factor). Points in the same major shell are connected
    by lines and ordered by increasing $j$. \label{fig:shelleff}}
\end{figure}

Clearly, Eq.~\eqref{eq:sat} has no (or trivial) solution for
$\kappa=0$, {\it i.e.}, without a 3b term. Though 2b forces do saturate,
they do it at the wrong place and at a heavy price because their
short-distance repulsion prevents direct HF variation. The crucial
question is now: Can we use realistic (2+3)b potentials soft enough to
do HF? Probably we can. The reason is that the nucleus is quite
dilute, and nucleons only ``see'' the low energy part of the 2b
potential involving basically $s$ wave scattering, which has been
traditionally well fitted by realistic potentials. Which in turn
explains why primitive versions of such potentials give results close
to the modern interactions for G-matrix elements calculated at
reasonable \hw\ values. This is a recurrent theme in this review (see
especially Section~\ref{sec:univ-real-inter}) and
Fig.~\ref{fig:shelleff} provides an example of particular relevance
for the study of shell effects in next
Section~\ref{sec:shell-formation}. Before we move on, we insist on
what we have learned so far: It may be possible to describe nuclear
structure with soft (2+3)b potentials consistent with the low energy
data coming from the $A=2$ and 3 systems.

\subsubsection{Shell formation}\label{sec:shell-formation}
Fig.~\ref{fig:shelleff} provides a direct reading of the expected
single particle sequences produced by realistic interactions above
$T=0$ closed shells. Unfortunately, they do not square at all with
what is seen experimentally. In particular, the splitting between
spin-orbit partners appears to be nearly constant, instead of being
proportional to $l$. As it could no longer be claimed that the 2b
realistic forces are ``wrong'', the solution must come from the 3b
operators. The ``educated guess'' in Eq.~\eqref{eq:W3} may prove of
help in this respect but it has not been implemented yet. What we
propose instead is to examine an existing, strict (1+2)b, model of
${\cal H}_m^d$ that goes a long way in explaining shell formation.
Though we know that 3b ingredients are necessary, they may be often
mocked by lower rank ones, in analogy with the kinetic energy, usually
represented as 1b, when in reality it is a 2b operator, as explained
in Appendix~\ref{sec:center-mass-problem}. Something similar seems to
happen with the spin-orbit force which must have a 3b origin, while a
1b picture is phenomenologically quite satisfactory, as our study will
confirm. We shall also find out that some mechanisms have an
irreducible 3b character.

${\cal H}_m^d$ manifests itself directly in nuclear spectra through
the doubly magic closures and the particle and hole states built on
them. The members of this set, which we call $cs\pm 1$ are well
represented by single determinants.  Configuration mixing may alter
somewhat the energies but, most often, enough experimental evidence
exists to correct for the mixing.  Therefore, a minimal
characterization, $\tilde{\cal H}_m^d$, can be achieved by demanding
that it reproduce all the observed $cs\pm 1$ states. The task was
undertaken by~\citet{Duflo.Zuker:1999}. With a half dozen parameters a
fit to the 90 available data achieved an rmsd of some 220 keV. The
differences in binding energies (gaps) $2BE(cs)-BE(cs+1)-BE(cs-1)$,
not included in the fit, also came out quite well and provided a test
of the reliability of the results.

The model concentrates on single particle splittings which are of
order $A^{-1/3}$. The valence spaces involve a number of particles of
the order of their degeneracy, {\it i.e.}, $\approx (p_f+3/2)^2\approx
(3A/2)^{2/3}$ according to Eq.~(\ref{eq:A}). As a consequence, at
midshell, the addition of single particle splittings generates shell
effects of order $A^{1/3}$.  To avoid bulk contributions all
expressions are written in terms of $\hat\Omega$-type operators
[Eq.~(\ref{eq:Omega})] and the model is defined by
\begin{gather}
  \label{eq:hmdz}
 \tilde{\cal H}_m^d\equiv (W-4K)+  \hat{\Omega}(ls) + \hat{\Omega}(ll)
  \nonumber \\  + \hat{\Omega}(ffi) +
 \hat{\Omega}(zni) + \hat{\Omega}(ffc) + \hat{\Omega}(znc),
\end{gather} 

Where:

\begin{equation}
  \label{eq:w-4k}
[{\rm I}]\;
W-4K=\left(\sum_p\frac{m_p}{\sqrt{D_p}}\right)^2-2\sum_p m_p(p+3/2). 
\end{equation}
comes from ${\cal K}^d$ and ${\cal V}^d$.  Both terms lead to the same
gaps, and the chosen combination has the
advantage of canceling exactly to orders $A$ and $A^{2/3}$.
\begin{figure}[t]
\begin{center}
  \includegraphics[width=0.9\linewidth]{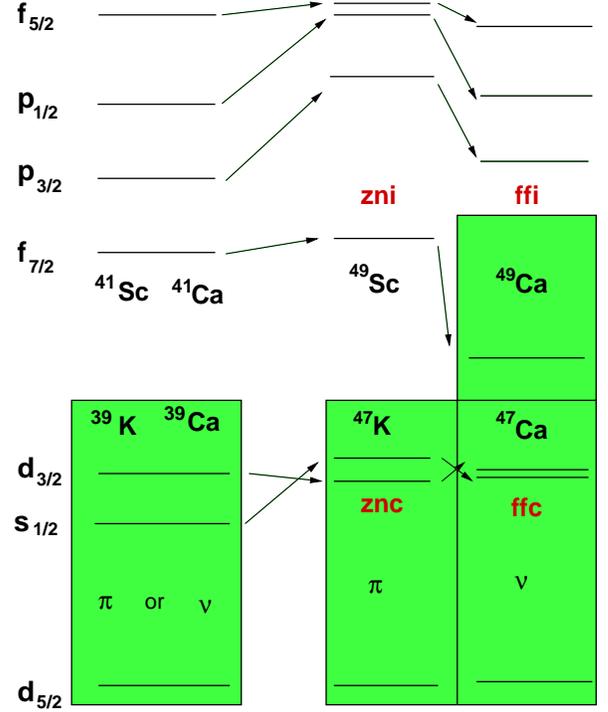}
  \caption {Evolution of $cs\pm 1$ states from $^{40}$Ca to $^{48}$Ca}
  \label{fig:pm1}
\end{center}
\end{figure} 
\begin{figure}[b]
\begin{center}
  \includegraphics[width=0.9\linewidth]{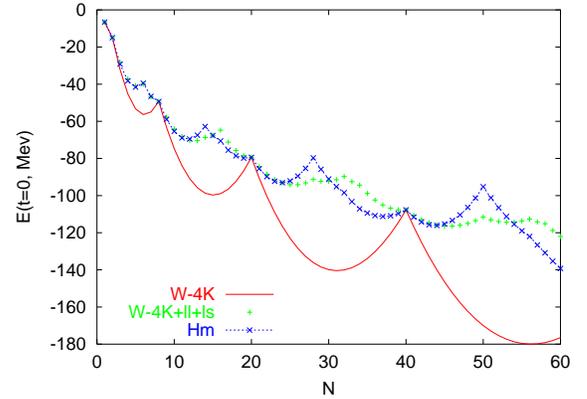}
  \caption {The different contributions to $\tilde{\cal H}^d_m$ for
    N=Z nuclei \label{fig:Hm0}}
\end{center}
\end{figure} 

[II] $\hat{\Omega}(ls)\approx -22\, l\cdot s/A^{2/3}$ (very much the value
in~\cite[][Eq.~(2-132)]{Bohr.Mottelson:1969}  and $\hat{\Omega}(ll)\approx
-22[l(l+1)-p(p+3)/2]/A^{2/3}$ (in MeV), are  $\hat\Omega$-type
operators that reproduce the single-particle spectra above HO closed
shells.
\begin{figure}[t]
\begin{center}
  \includegraphics[width=0.9\linewidth]{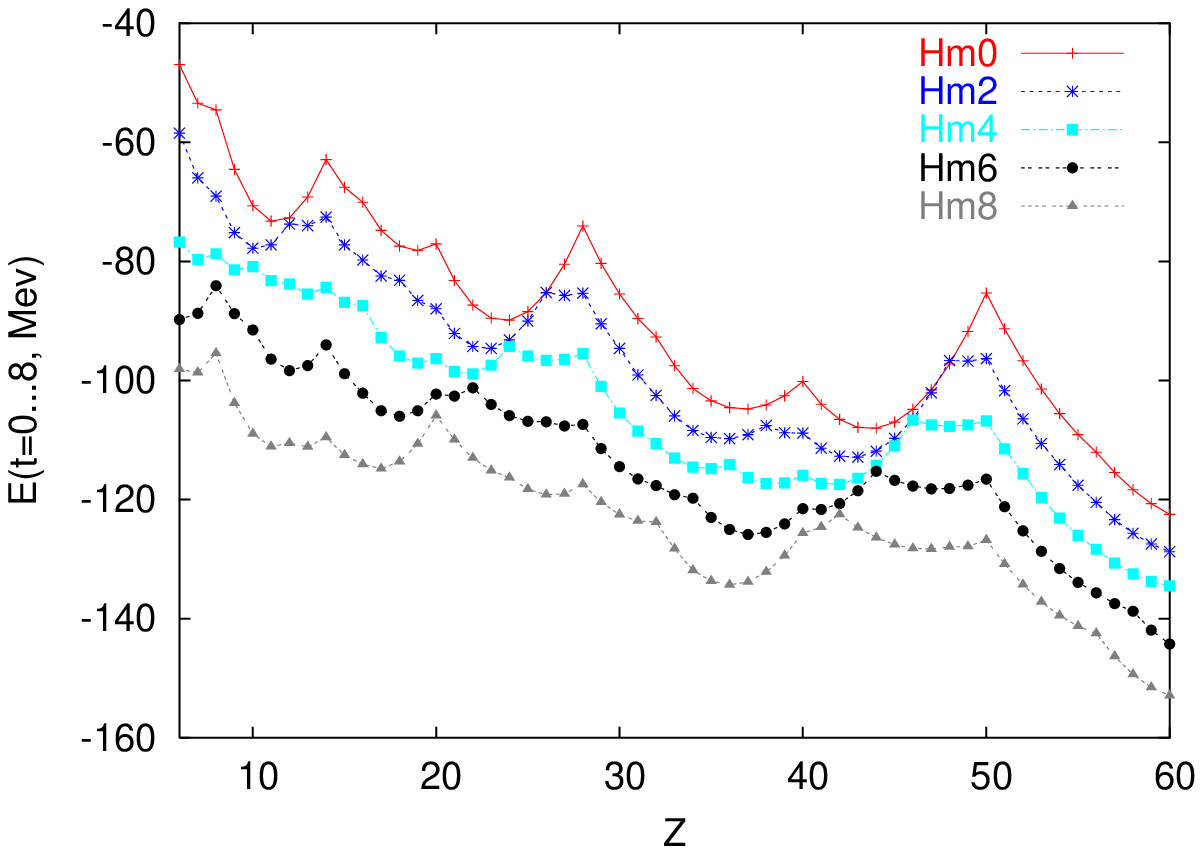}
  \includegraphics[width=0.9\linewidth]{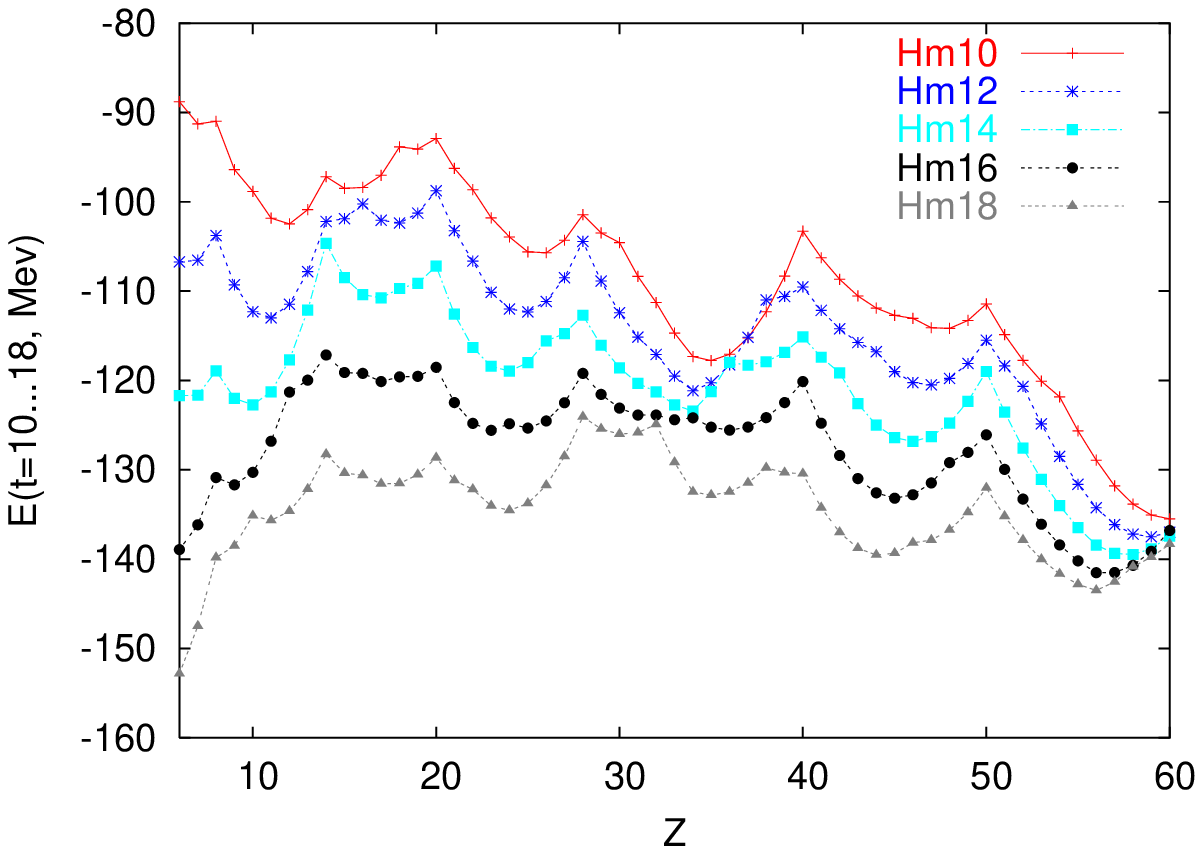}
  \caption {$\tilde{\cal H}^d_m$ for even $t=N-Z=10\ldots 18$. 
    \label{fig:Hmz}}
\end{center} 
\end{figure}

[III] $\hat{\Omega}(ffi)$, $\hat{\Omega}(zni)$, $\hat{\Omega}(ffc)$ 
 and $\hat{\Omega}(znc)$\footnote{ $ff$ stands for ``same
  fluid'', $zn$ for ``proton-neutron'', $i$ for ``intra-shell'' and
  $c$ for ``cross-shell''} are $\hat\Omega$-type 2b operators whose
effect is illustrated in Fig.~\ref{fig:pm1}: In going from $^{40}$Ca
to $^{48}$Ca the filling of the $f_{7/2}$ ($f$ for short) neutron
orbit modifies the spectra in $^{49}$Ca, $^{49}$Sc, $^{47}$Ca and
$^{47}$K.

As mentioned, realistic 2b forces will miss the particle spectrum in
$^{41}$Ca. Once it is phenomenologically corrected by $ls+ll$ such
forces account reasonably well for the $zni$ and $eei$ mechanism,
within one major exception: they fail to depress the $f_{7/2}$ orbit
with respect to the others. As a consequence, and quite generally,
they fail to produce the EI closures. Though $zni$ and $eei$ can be
made to cure this serious defect, it is not clear how well they do it,
since the correct mechanism necessarily involves 3b terms, as will be
demonstrated in Section~\ref{sec:monop-probl-three}. Another problem
that demands a three body mechanism is associated with the $znc$ and
$ffc$ operators which always depress orbits of larger $l$ [as in the
hole spectra of $^{47}$Ca and $^{47}$K in Fig.~\ref{fig:pm1}], which
is not the case for light nuclei as seen from the spectra of $^{13}$C
and $^{29}$Si.  Only 3b operators can account for this inversion.

In spite of these caveats, $\tilde{\cal H}_m^d$ should provide a
plausible view of shell formation, illustrated in Figure~\ref{fig:Hm0}
for $t=N-Z=0$ nuclei. $\tilde{\cal H}_m^d$ is calculated by filling
the oscillator orbits in the order dictated by $ls+ll$. The $W-4K$ term
produces enormous HO closures (showing as spikes). They are
practically erased by the 1b contributions, and it takes the 2b ones
to generate the EI\footnote{Remember: EI valence spaces consist in
  the HO orbits shell $p$, except the largest (extruder), plus the
  largest (intruder) orbit from shell $p+1$.}  spikes. Note that the
HO magicity is strongly attenuated but persists.  The drift toward
smaller binding energies that goes as $A^{1/3}$ is an artifact of the
$W-4K$ choice and should be ignored.

In Figs.~\ref{fig:Hmz} we show the situation for even $t=N-Z=0$ to 18.
Plotting along lines of constant $t$ has the advantage of detecting
magicity for both fluids. In other words, the spikes appear when
either fluid is closed, and are reinforced when both are closed.  The
spikes are invariably associated either to the HO magic numbers (8,
20, 40) or to the EI ones (14, 28, 50), but the latter always show
magicity while the former only do it at and {\em above} the double
closures ($Z,N$)=(8,14),(20,28) and (40,50). For example, at $Z=20$
$^{40}$Ca shows as a weak closure. $^{42,\, 44,\, 46}$Ca are not
closed (which agrees with experiment), $^{48}$Ca is definitely magic,
and spikes persist for the heavier isotopes. At $Z=40$, $^{80}$Zr
shows a nice spike, a bad prediction for a strongly rotational
nucleus. However, there are no (or weak) spikes for the heavier isotopes
except at $^{90}$Zr and $^{96}$Zr, both definitely doubly magic. There
are many other interesting cases, and the general trend is the
following: No known closure fails to be detected. Conversely, not all
predicted closures are real. They may be erased by deformation---as in
the case of $^{80}$Zr---but this seldom happens, thus suggesting
fairly reliable predictive power for $\tilde{\cal H}^d_m$.

\subsection{The multipole Hamiltonian}\label{sec:mult-hamilt}
The multipole Hamiltonian is defined as \HM=\H-\Hm. As we are no
longer interested in the full \H, but its restriction to a finite
space, ${\cal H}_M$ will be more modestly called $H_M$, with
monopole-free matrix elements given by
\begin{equation}
  \label{VW}
  W_{rstu}^{JT}=V_{rstu}^{JT}-\delta_{rs}\delta_{tu}V_{rs}^T.
\end{equation}
We shall describe succinctly the main results
of~\citet{Dufour.Zuker:1996}, insisting on points that were not
stressed sufficiently in that paper
(Section~\ref{sec:collapse-avoided}), and adding some new information
(Section~\ref{sec:univ-real-inter}).

There are two standard ways of writing $H_M$:
 \begin{equation}
 H_M=\sum\limits_{r\leq s,t\leq u,\Gamma}
W_{rstu}^\Gamma Z_{rs\Gamma}^{\dagger}
\cdot Z_{tu\Gamma},\quad {\rm or}
\label{1}
\end{equation}
\begin{equation}
 H_M=\sum_{rstu \Gamma}[\gamma]^{1/2}f_{rtsu}^\gamma (S_{rt}^\gamma
S_{su}^\gamma)^0,
\label{2}
\end{equation}
where
$f_{rtsu}^{\gamma}=\omega_{rtsu}^{\gamma}
\sqrt{(1+\delta_{rs})(1+\delta_{tu})}/4$.
The matrix elements $\omega_{rtsu}^{\gamma}$ and $W_{rstu}^{\Gamma}$
are related through Eqs.~(\ref{3A}) and~(\ref{3B}).

Replacing pairs by single indices $rs\equiv x$, $tu\equiv y$ in
eq.~(\ref{1}) and $rt\equiv a$, $su\equiv b$ in eq.~(\ref{2}), we
proceed as in Eqs.~(\ref{eq:u1}) and~(\ref{eq:u2})  to bring the
matrices $W_{xy}^\Gamma\equiv W_{rstu}^\Gamma$ and $f_{ab}^\gamma\equiv
f_{rtsu}^\gamma$, to diagonal form through unitary
transformations $U_{xk}^\Gamma,u_{ak}^\gamma$ to obtain factorable
expressions 
\begin{equation}
H_M=\sum_{k,\Gamma}E_k^\Gamma\sum_xU_{xk}^\Gamma
Z_{x\Gamma}^{\dagger}\cdot\sum\limits_yU_{yk}^\Gamma Z_{y\Gamma},\label{4a}
\end{equation}
\begin{equation}
H_M=\sum_{k,\gamma}e_k^\gamma\left(\sum_au_{ak}^\gamma
S_a^\gamma\sum_bu_{bk}^\gamma S_b^\gamma\right)^0
[\gamma]^{1/2},\label{4b}
\end{equation}
which we call the $E$ (or normal or particle-particle or pp) and $e$
(or multipole or particle-hole or ph) representations.  Since ${\cal
  H}_m$ contains all the $\gamma=00$ and 01 terms, for ${\cal H}_M$,
$\omega_{rstu}^{00}=\omega_{rstu}^{01}=0$ (see Eq.~(\ref{(II.15)}).
There are no one body contractions in the $e$ representation because
they are all proportional to $\omega_{rstu}^{0\tau}$.

\begin{figure}[htbp]
  \includegraphics[angle=270,width=0.9\linewidth]{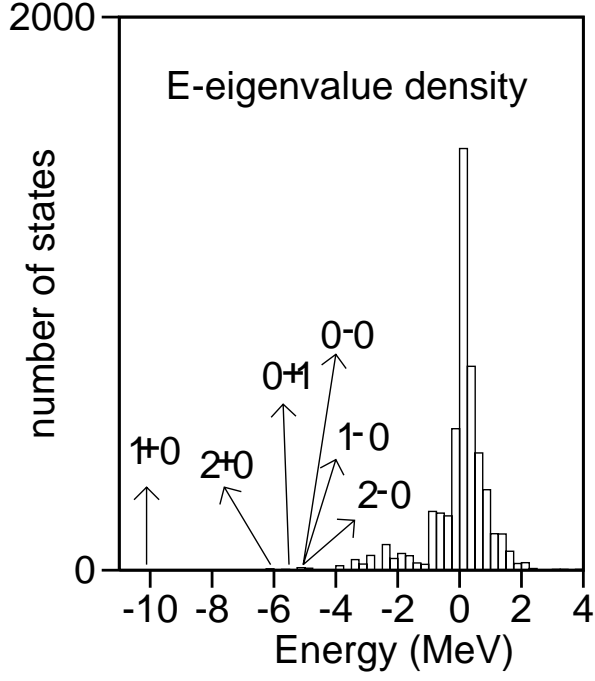}
  \caption{$E$-eigenvalue density for the KLS interaction in the pf+sdg
    major shells $\hbar\omega=9$. Each eigenvalue has multiplicity
    $[\Gamma]$.  The largest ones are shown by arrows.\label{Edef}}
\end{figure}

The eigensolutions in eqs.~(\ref{4a}) and (\ref{4b}) using the KLS
interaction~\cite{Kahana.Lee.Scott:1969b,Lee:1969}, yield the
density of eigenvalues (their number in a given interval) in the $E$
representation that is shown in Fig.~\ref{Edef} for a typical two-shell
case.

It is skewed, with a tail at negative energies which is what we expect
from an attractive interaction.

\begin{figure}[htb]
  \includegraphics[angle=270,width=0.9\linewidth]{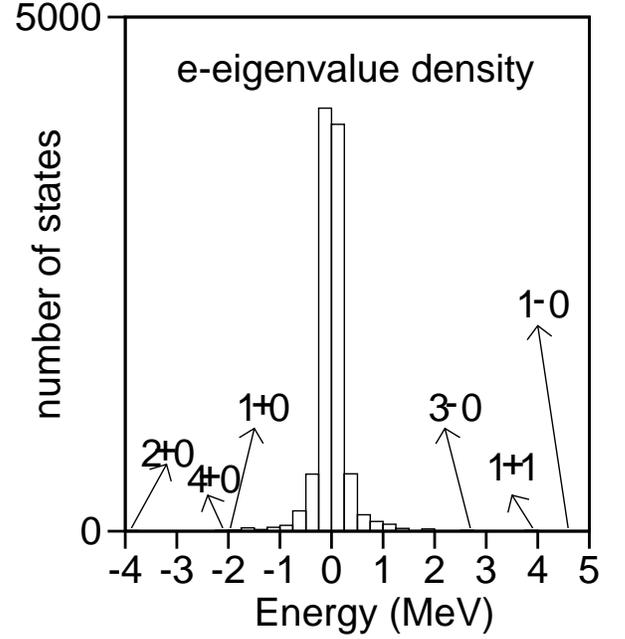}
  \caption{$e$-eigenvalue density for the KLS interaction in the pf+sdg
    major shells. Each eigenvalue has multiplicity $[\gamma]$.
    The largest ones are shown by arrows.\label{edef}}
\end{figure}

The $e$ eigenvalues are plotted in fig.~\ref{edef}.  They are very
symmetrically distributed around a narrow central group, but a few of
them are neatly detached. The strongest have
$\gamma^\pi=1^-0,\;1^+1,~2^+0,~3^-0, ~4^+0$. ``If the corresponding
  eigenvectors are eliminated from H in'' eq.~(\ref{4b}) and the
  associated H in eq.~(\ref{4a})  is recalculated\footnote{Inverted commas are
  meant to call attention on an erratum in~\cite{Dufour.Zuker:1996}.}, the E
  distribution becomes quite symmetric, as expected for a random interaction.

If the diagonalizations are restricted to one major shell, negative
parity peaks are absent, but for the positive parity ones the results
are practically identical to those of Figs.~\ref{Edef} and \ref{edef},
except that the energies are halved. This point is crucial: 

{\em If $u_{p_1}$ and $u_{p_2}$ are the eigenvectors obtained in shells
  $p_1$ and $p_2$, their eigenvalues are approximately equal
  $e_{p_1}\approx e_{p_2}=e$. 
When diagonalizing in $p_1+p_2$, the {\em unnormalized}
  eigenvector turns out to be $u_{p_1}+u_{p_2}$ with eigenvalue $e$}. 

In the figures the eigenvalues for the two shell case are doubled,
because they are associated with normalized eigenvectors.
To make a long story short: The contribution to $H_M$ associated to
the large $\Gamma=01$, and $\gamma=20$ terms,
\begin{eqnarray}
H_{\bar P}&=&-\frac{\hbar\omega}{\hbar\omega_0}{|E^{01}|}
(\overline P\,^{\dagger}_p+\overline P\,^{\dagger}_{p+1})
\cdot(\overline P_p+\overline P_{p+1})\label{Hp}\\
H_{\bar q}&=&-\frac{\hbar\omega}{\hbar\omega_0}{|e^{20}|}
(\bar q_p+\bar q_{p+1})
\cdot(\bar q_p+\bar q_{p+1}),\label{Hq}
\end{eqnarray}
turns out to be 
the usual pairing plus quadrupole Hamiltonians, except that
the operators for each major shell of principal quantum number $p$ are
affected by a normalization. $E^{01}$ and $e^{20}$ are the  one
shell values called generically $e$ in the discussion above. To be
precise 
\begin{gather}
\bar P^{\dagger}_p=
\sum\limits_{r\in p}Z_{rr01}^{\dagger}\Omega_r^{1/2}/\Omega_p^{1/2},\label{P}\\
\bar q_p=
\sum\limits_{rs\in p}S_{rs}^{20}q_{rs}/{\cal N}_p,\label{q}\\
{\rm where}\quad
\Omega_r=j_r+1,\quad q_{rs}=\sqrt\frac{1}{5}\langle r\| r^2Y^2\|
s\rangle\label{normpq1}\\
{\rm and}\quad
\Omega_p=\frac{1}{2}D_p, \quad
{\cal N}_p^2=\Sigma q_{rs}^2\cong \frac{5}{ 32\pi}(p+3/2)^4
\label{normpq2}
\end{gather}

\subsubsection{Collapse avoided}\label{sec:collapse-avoided}

The pairing plus quadrupole ($P+Q$) model has a long and glorious
history~\cite{Baranger.Kumar:1968,Bes.Sorensen:1969}, and one big
problem: as more shells are added to a space, the energy grows,
eventually leading to collapse.  The only solution was to stay within
limited spaces, but then the coupling constants had to be readjusted
on a case by case basis. The normalized versions of the operators
presented above are affected by universal coupling constants that do
not change with the number of shells.  Knowing that $\hbar \omega_0=9$
MeV, they are $|E^{01}|/{\hbar \omega_0}=g'=0.32$ and $|e^{20}|/{\hbar
  \omega_0}=\kappa '=0.216$ in Eqs.~(\ref{Hp}) and (\ref{Hq}). 
 
Introducing $A_{mf}\approx\frac{2}{ 3}(p_f+3/2)^3$, the total number
of particles at the middle of the Fermi shell $p_f$, the relationship
between $g'$, $\kappa '$, and their conventional
counterparts~\cite{Baranger.Kumar:1968} is, for one
shell\footnote{According to~\citet{Dufour.Zuker:1996} the bare
  coupling constants, should be renormalized to 0.32(1+0.48) and
  0.216(1+0.3)}
\begin{eqnarray}
\frac{0.32\hbar\omega}{ \Omega_p}&\cong&
\frac{19.51}{ A^{1/3}A_{mf}^{2/3}}= G \equiv {G_0}A^{-1},
\nonumber\\
\frac{0.216\hbar\omega}{{\cal N}_p^2}&\cong&\frac {1}{ 2}\,
\frac{216}{A^{1/3}A_{mf}^{4/3}}=\frac{\chi'}{ 2}\equiv 
\frac{\chi_0'}{2}A^{-5/3}.
\label{1s}
\end{eqnarray}
To see how collapse occurs assume $m=O(D_f)=O(A^{2/3})$ in the Fermi
shell, and promote them to a higher shell of degeneracy $D$. The
corresponding 
pairing and quadrupole energies can be estimated as
\begin{gather}
E_P=-{|G|}{ 4}m(D-m+2)=-|G|O(mD),\\  {\rm and}\quad E_q\approx
-|\chi'|Q_0^2=-|\chi'|O(m^2D), 
\end{gather}
respectively, which become for the two possible scalings
\begin{gather}
E_P({\rm old})=O(\frac{mD}{ A})\Longrightarrow E_P({\rm
  new})=O(\frac{m}{ A^{1/3}}) \\
 E_q({\rm old})=O(\frac{m^2D}{ A^{5/3}})\Longrightarrow  E_q({\rm
  new})=O(\frac{m^2}{ A^{1/3}D}).
\end{gather}
If the $m$ particles stay at the Fermi shell, all energies go as
$A^{1/3}$ as they should. If $D$ grows both energies grow in the old
version.  For sufficiently large $D$ the gain will become larger than
the monopole loss $O(mD^{1/2}\hbar\omega)=O(D^{1/2}A^{1/3})$. Therefore
the traditional forces lead the system to collapse. In the new form
there is no collapse: $E_P$ stays constant, $E_q$ decreases and the
monopole term provides the restoring force that guarantees that
particles will remain predominantly in the Fermi shell.

As a model for \HM, $P+Q$ is likely to be a reasonable first
approximation for many studies, provided it is supplemented by a
reasonable \Hm, to produce the $P+Q+m$ model.

\subsection{Universality of the realistic
  interactions}\label{sec:univ-real-inter}
To compare sets of matrix elements we define the overlaps
\begin{gather}
  \label{eq:overlap}
  O_{AB}=V_A\cdot
  V_B=\sum_{rstu\Gamma}V^{\Gamma}_{rstuA}V^{\Gamma}_{rstuB}\\ 
\overline{O}_{AB}=\frac{O_{AB}}{\sqrt{O_{AA}O_{BB}}}.
\label{eq:normovlap}
\end{gather}
Similarly, $O_{AB}^T$ is the overlap for matrix elements with the same
$T$. $O_{AA}$ is a measure of the strength of an interaction. If the
interaction is referred to its centroid (and {\em a fortiori} if it is
monopole-free), $O_{AA}\equiv \sigma^2_A$.
\begin{table}[htb]
  \caption{The overlaps between the KLS($A$) and Bonn($B$)
    interactions for the first ten oscillator shells (2308 matrix
    elements), followed by those for the $pf$ shell (195 matrix
    elements) for the BonnC ($C$) and KB ($K$) interactions.}
\begin{ruledtabular}
\begin{tabular}{ccccc}
$ O_{AA}^0= 17.56$&&  $ O_{AA}^1=2.61$&&   $O_{AA}=6.49$\\
$ O_{BB}^0= 11.84$&&  $ O_{BB}^1=2.31$&&   $O_{BB}=4.78$\\
$\overline O_{AB}^0=\ 0.98$&&  $\overline O_{AB}^1=0.99$&&
$\overline O_{AB}=0.98$\\ 
\\
$ O_{CC}^0=\ 3.71$&&  $ O_{CC}^1=0.56$&&   $O_{CC}=1.41$\\
$ O_{KK}^0=\ 3.02$&&  $ O_{KK}^1=0.46$&&   $O_{KK}=1.15$\\
$\overline O_{CK}^0=\ 0.99$&&  $\overline O_{CK}^1=0.97$&&
$\overline O_{CK}=0.99$\\ 
\end{tabular}
\end{ruledtabular}
\label{tab:ovklsbonn}
\end{table}
The upper set of numbers in Table~\ref{tab:ovklsbonn} contains what is
probably the most important single result concerning the interactions:
a 1969 realistic potential~(\citet{Kahana.Lee.Scott:1969b,Lee:1969})
and a modern Bonn one~\cite{Hjorth-Jensen:1996} differ in total
strength(s) but the normalized $\overline O(AB)$ are very close to
unity.

Now: all the modern realistic potentials agree closely with one
another in their predictions {\em except} for the binding energies.
For nuclear matter the differences are
substantial~\cite{Pieper.Pandharipande.ea:2001}, and for the BonnA,B,C
potentials studied by~\citet{Hjorth-Jensen.Kuo.Osnes:1995}, they
become enormous. However, the matrix elements given in this reference
for the $pf$ shell have normalized cross overlaps of {\em better than
  0.998}. At the moment, the overall strengths must be viewed as free
parameters.

At a fundamental level the discrepancies in total strength stem for
the degree of non-locality in the potentials and the treatment of the
tensor force. In the old interactions, uncertainties were also due to
the starting energies and the renormalization processes, which, again,
affect mainly the total strength(s), as can be gathered from
the bottom part of Table~\ref{tab:ovklsbonn}.
 
The dominant terms of $H_M$ are central and table~\ref{tab:duzu}
collects their strengths (in MeV) for effective interactions in the
$pf$-shell:~\citet{Kuo.Brown:1968} (KB), the potential fit
of~\citet{Richter.Brown:1991} (FPD6), the Gogny force---successfully
used in countless mean field studies---~\cite{Decharge.Gogny:1980} and
BonnC~\cite{Hjorth-Jensen.Kuo.Osnes:1995}. 

There is not much to choose between the different forces which is a
nice indication that overall nuclear data (used in the FPD6 and Gogny
fits) are consistent with the $NN$ data used in the realistic KB and
BonnC G-matrices. The splendid performance of Gogny deserves
further study. The only qualm is with the weak quadrupole strength of
KB, which can be understood by looking again at the bottom part of
Table~\ref{tab:ovklsbonn}. In assessing the $JT=01$ and $JT=10$  pairing
terms, it should be borne in mind that their renormalization remains
an open question~\cite{Dufour.Zuker:1996}. 

Conclusion: Some fine tuning may be in order and 3b forces may (one
day) bring some multipole news, but as of now the problem is \Hm, not
\HM.

\begin{table}[htb]
   \caption{Leading terms of the multipole Hamiltonian}
    \label{tab:duzu}
\begin{ruledtabular}
    \begin{tabular}{cccccc}\noalign{\smallskip}
  Interaction & \multicolumn{2}{c}{particle-particle} &
                \multicolumn{3}{c}{particle-hole}\\ \noalign{\smallskip}
                \hline\noalign{\smallskip}
                & JT=01  & JT=10 & $\lambda \tau$=20 &
                                 $\lambda \tau$=40 &
                                 $\lambda \tau$=11   \\ \hline\noalign{\smallskip}
 KB & -4.75 & -4.46 & -2.79 & -1.39 & +2.46 \\
 FPD6 & -5.06 & -5.08 & -3.11 & -1.67 & +3.17 \\
 GOGNY & -4.07 & -5.74 & -3.23 & -1.77 & +2.46 \\
 BonnC&-4.20 & -5.60 &  -3.33 & -1.29 & +2.70\\
    \end{tabular}
\end{ruledtabular}
 \end{table}

\section{The solution of the secular problem in a finite space}
\label{sec:solut-secul-probl}

Once the interaction and the valence space ready, it is time to 
 construct and 
diagonalize   the many body secular matrix of the Hamiltonian
describing the (effective) interaction between valence particles in the
valence space. Two questions need now consideration:
which basis to take to
calculate the non-zero many-body matrix elements (NZME) and which method to use
for the diagonalization of the matrix.

\subsection{The Lanczos method}
\label{sec:lanczos-method}

\begin{figure}[htb]
  \includegraphics[angle=270,width=0.9\linewidth]{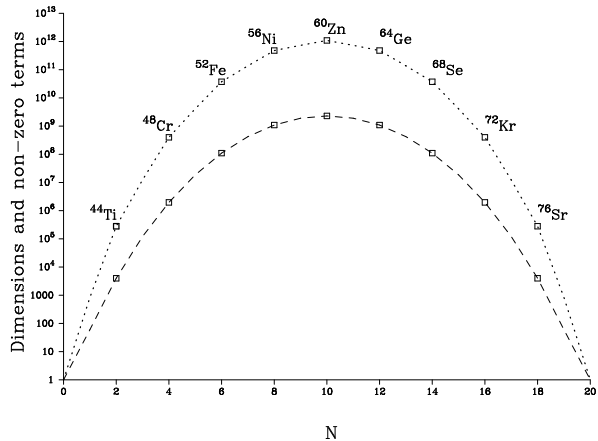}
  \caption{$m$-scheme dimensions and total number of non-zero
    matrix elements in the $pf$-shell for nuclei with
    $M=T_z=0$\label{fig:dim}}
\end{figure}

In the standard diagonalization methods~\cite{Wilkinson:1965} the CPU
time increases as $N^3$, $N$ being the dimension of the matrix.
Therefore, they cannot be used in large scale shell model (SM)
calculations.  Nuclear SM calculations have two specific features. The
first one is that, in the vast majority of the cases, only a few (and
very often only one) eigenstates of a given angular momentum ($J$) and
isospin ($T$) are needed.  Secondly, the matrices are very sparse.  As
can be seen in figure~\ref{fig:dim}, the number of NZME varies
linearly instead of quadratically with the size of the matrices.  For
these reasons, iterative methods are of general use, in particular the
Lanczos method~\cite{Lanczos:1950}.  As an alternative, the Davidson
method~\cite{Davidson:1975} has the advantage of avoiding the storage
of a large number of vectors, however, the large increase in storage
capacity of modern computers has somehow minimized these advantages.
 
The Lanczos method consists in the construction of an orthogonal basis
in which the Hamiltonian matrix ($H$) is tridiagonal. 

A normalized starting vector (the pivot) $|1\rangle$ is chosen. The vector
$|a_1\rangle = H|1\rangle$ has necessarily the form
\begin{equation}
  \label{eq:lanc0}
|a_1\rangle=H_{11}|1\rangle +|2'\rangle,\quad  
{\rm with\ } \langle 1|2'\rangle=0.  
\end{equation}
Calculate $H_{11}=\langle 1|H|1\rangle$ through $\langle
1|a_1\rangle=H_{11}$. Normalize $|2\rangle=|2'\rangle/\langle
2'|2'\rangle ^{1/2}$ to find $H_{12}=\langle 1|H|2\rangle=\langle
2'|2'\rangle ^{1/2}$.  Iterate until state $|k\rangle$ has been found.
The vector $|a_k\rangle = H |k\rangle$ has necessarily the form
\begin{equation}
  \label{lanc}
  |a_k\rangle= H_{k\, k-1} |k-1\rangle+ H_{k\, k} |k\rangle   
                        + |(k+1)'\rangle. 
\end{equation}
Calculate $\langle k|a_k\rangle=H_{k\, k}$. Now $|(k+1)'\rangle$ is
known. Normalize it to find $H_{k\, k+1} = \langle
(k+1)'|(k+1)'\rangle ^{1/2}$.

The (real symmetric) matrix is diagonalized at each iteration and the
iterative process continues until all the required eigenvalues are
converged according to some criteria. The number of
iterations depends little on the dimension of the matrix.
Besides, the computing time is directly proportional to the number of
NZME and for this reason it is nearly linear (instead of cubic as in
the standard methods) in the dimension of the matrix.  It depends on
the number of iterations, which in turn depends on the number of
converged states needed, and also on the choice of the starting
vector.
 
The Lanczos method can be used also as a projection method.  For
example in the $m$-scheme basis, a Lanczos calculation with the
operator $J^2$ will generate states with well defined $J$. Taking
these states as starting vectors, the Lanczos procedure (with
$H$) will remain inside a fixed $J$ subspace and therefore will
improve the convergence properties. When only one converged state is
required, it is convenient to use as pivot state the solution obtained
in a previous truncated calculation. For instance, starting with a
random pivot the calculation of the ground state of $^{50}$Cr in the
full $pf$ space (dimension in $m$-scheme 14,625,540 Slater
determinants) needs twice more iterations than if we start with the
pivot obtained as solution in a model space in which only 4 particles
are allowed outside the $1f_{7/2}$ shell (dimension 1,856,720). The
overlap between these two $0^+$ states is 0.985. When more eigenstates
are needed ($n_c > 1$), the best choice for the pivot is a linear
combination of the $n_c$ lower states in the truncated space.  Even if
the Lanczos method is very efficient in the SM framework, there may be
numerical problems. Mathematically the Lanczos vectors are orthogonal,
however numerically this is not strictly true due to the limited
floating point machine precision. Hence, small numerical precision
errors can, after many iterations, produce catastrophes, in
particular, the states of lowest energy may reappear many times,
rendering the method inefficient. To solve this problem it is
necessary to orthogonalize each new Lanczos vector to all the
preceding ones.  The same precision defects can produce the appearance
of non expected states.  For example in a $m$-scheme calculation with
a $J=4$, $M=0$ pivot, when many iteration are performed, it may
happen that one $J=0$ and even one $J=2$ state, lower in energy than the
$J=4$ states, show up abruptly.  This specific problem can be solved by
projecting on $J$ each new Lanczos vector.

\subsection{The choice of the basis}
\label{sec:basis}
 
Given a valence space, the optimal choice of the basis is related to
the physics of the particular problem to be solved. As we  
discuss later, depending on what states or properties we want to
describe (ground state, yrast band, strength function,\ldots) and
depending on the type of nucleus (deformed, spherical,\ldots)
different choices of the basis are favored.  There are essentially
three possibilities depending on the
underlying symmetries: $m$-scheme, $J$ coupled scheme, and $JT$
coupled scheme.

\begin{table}[htb]
\caption{Some dimensions in the $pf$ shell \label{tab:dim}}
\begin{tabular}{c|ccccc}
  \hline\hline
  A        &   4   &   8      &  12         &   16        &   20    \\
  \hline
  $M=T_z=0$  & 4000  & $2\times10^6$ & $1.10\times10^8$ &
  $1.09\times10^9$ &$2.29\times10^9$ \\ 
  $J=T_z=0$    & 156   & 41355    & $1.78\times10^6$ &
  $1.54\times10^7$ &$3.13\times10^7$\\ 
$J=T=0$      & 66    & 9741     & $3.32\times10^5$ &
$2.58\times10^6$ & $5.05\times10^6$\\ 
\hline\hline
\end{tabular}
\end{table}

As the $m$-scheme basis consists of Slater Determinants (SD) the
calculation of the NZME is trivial, since they are equal to the
decoupled two body matrix element (TBME) up to a phase. This means
that, independently of the size of the matrix, the number of possible
values of NZME is relatively limited.  However, the counterpart of the
simplicity of the $m$-scheme is that only $J_{z}$ and $T_{z}$ are good
quantum numbers, therefore all the possible $(J,T)$ states are
contained in the basis and as a consequence the dimensions of the
matrices are maximal.  For a given number of valence neutrons, $n_v$,
and protons, $z_v$ the number of different Slater determinants that
can be built in the valence space is :
\begin{equation}
d= \left( \begin{array}{c} D_n \\ n_v \end{array} \right) \cdot
\left( \begin{array}{c} D_p \\ z_v \end{array} \right)
\end{equation} 
where $D_n$ and $D_p$ are the degeneracies of the neutron and proton
valence spaces.  Working at fixed M and T$_z$ the bases are
smaller ($d = \sum_{M, T_z} d\;(M, T_z)$). 
The $J$ or $JT$ coupled bases, split the full
$m$-scheme matrix in boxes whose dimensions are much smaller.  This is
especially spectacular for the $J=T=0$ states (see table~\ref{tab:dim}).

It is often convenient to truncate the space. In the particular case
of the $pf$ shell, calling $f$ the largest subshell ($f_{7/2}$), and
$r$, generically, any or all of the other subshells of the $p=3$
shell, the possible $t$-truncations involve the spaces
\begin{equation}
\label{eq:t}
f^{m-m_0} r^{m_0}+f^{m-m_0-1} r^{m_0+1}+\cdots+
f^{m-m_0-t} r^{m_0+t},
\end{equation}
where $m_0\neq 0$ if more than 8 neutrons (or protons) are present.
For $t=m-m_0$ we have the full space $(pf)^m$ for $A=40+m$.

In the late 60's, the Rochester group developed the algorithms needed
for an efficient work in the $(J,T)$ coupled basis and implemented them
in the Oak-Ridge Rochester Multi-Shell
code~\cite{French.Halbert.ea:1969}. The structure of the calculation
is as follows: In the first place, the states of $n_i$ particles in a
given $j_i$ shell are defined: $|\gamma_i \rangle = |(j_i)^{n_i} v_i
J_i x_i \rangle $ ($v_i$ is the seniority and $x_i$ any extra quantum
number).  Next, the states of $N$ particles distributed in several
shells are obtained by successive angular momentum couplings of the
one-shell basic states:
\begin{equation}
\left[\left[\left[| \gamma_1 \rangle | \gamma_2 \rangle\right]^{\Gamma_2}
| \gamma_3 \rangle\right]^{\Gamma_3} \ldots
| \gamma_k \rangle\right]^{\Gamma_k}.
\end{equation}
Compared to the simplicity of the $m$-scheme, the calculation of the
NZME is much more complicated. It involves products of  9j symbols and
coefficients of fractional parentage (cfp), {\it i.e.}, the single-shell
reduced matrix elements of operators of the form
\begin{equation}
  (a^{\dagger}_{j_1}a^{\dagger}_{j_2})^{\lambda},\;
  (a^{\dagger}_{j_1}a_{j_2})^{\lambda},\;
  a^{\dagger}_{j_1},\;
  ((a^{\dagger}_{j_1}a^{\dagger}_{j_2})^{\lambda}a_{j_3})^{j_4}.
\label{eq:ops}
\end{equation}

This complexity explains why in the OXBASH
code~\cite{Brown.Etchegoyen.ea:1985}, the JT-coupled basis states are
written in the $m$-scheme basis to calculate the NZME. The
Oxford-Buenos Aires shell model code, widely distributed and used has
proven to be an invaluable tool in many calculations. Another recent
code that works in JT-coupled formalism is the Drexel University
DUPSM~\cite{Novoselsky.Vallieres:1997} has had a much lesser use.

In the case of only $J$ (without $T$) coupling, a strong
simplification in the calculation of the NZME can be achieved using
the Quasi-Spin formalism~\cite{Lawson.Macfarlane:1965,%
  Kawarada.Arima:1964,Ichimura.Arima:1966}, as in the code NATHAN
described below. The advantages of the coupled scheme decrease also
when $J$ and $T$ increase.  As an example, in $^{56}$Ni the ratio
dim($M=J$)/dim($J$) is 70 for $J=0$ but only 5.7 for $J=6$.  The
coupled scheme has another disadvantage compared to the $m$-scheme. It
concerns the percentage of NZME.  Let us give the example of
the state $4^+$ in $^{50}$Ti (full $pf$ space); the percentages of
NZME are respectively 14\% in the $JT$-basis
\citep[see][]{Novoselsky.Vallieres.ea:1997}, 5\% in the J-basis and
only 0.05\% in $m$-scheme. For all these reasons and with the present
computing facilities, we can conclude that the $m$-scheme is the most
efficient choice for usual SM calculations, albeit with some notable
exceptions that we shall mention below.

\subsection{The Glasgow $m$-scheme code}
\label{sec:m-scheme-formalism}
The steady and rapid increase of computer power in recent years has
resulted in a dramatic increase in the dimensionality of the SM
calculations. A crucial point nowadays is to know what are the limits
of a given computer code, their origin and their evolution.  As far as
the NZME can be calculated and stored, the diagonalization with the
Lanczos method is trivial.  It means that the fundamental limitation
of standard shell model calculations is the capacity to store the
NZME. This is the origin of the term ``giant'' matrices, that we apply
to those for which it is necessary to recalculate the NZME during the
diagonalization process.  The first breakthrough in the treatment of
giant matrices was the SM code developed by the Glasgow
group~\cite{Whitehead.Watt.ea:1977}.  Let us recall its basic ideas:
It works in the $m$-scheme and each SD is represented in the computer
by an integer word.  Each bit of the word is associated to a given
individual state $|nljm\tau \rangle$. Each bit has the value 1 or 0
depending on whether the state is occupied or empty.  A two-body
operator $a^{\dagger}_{i}a^{\dagger}_{j}a_{k}a_{l}$ will select the
words having the bits $i,j,k,l$ in the configuration 0011, say, and
change them to 1100, generating a new word which has to be located in
the list of all the words using the bi-section method.
  
\subsection{The $m$-scheme code ANTOINE}
\label{sec:antoine}

\begin{figure}[htb]
  \includegraphics[width=0.9\linewidth]{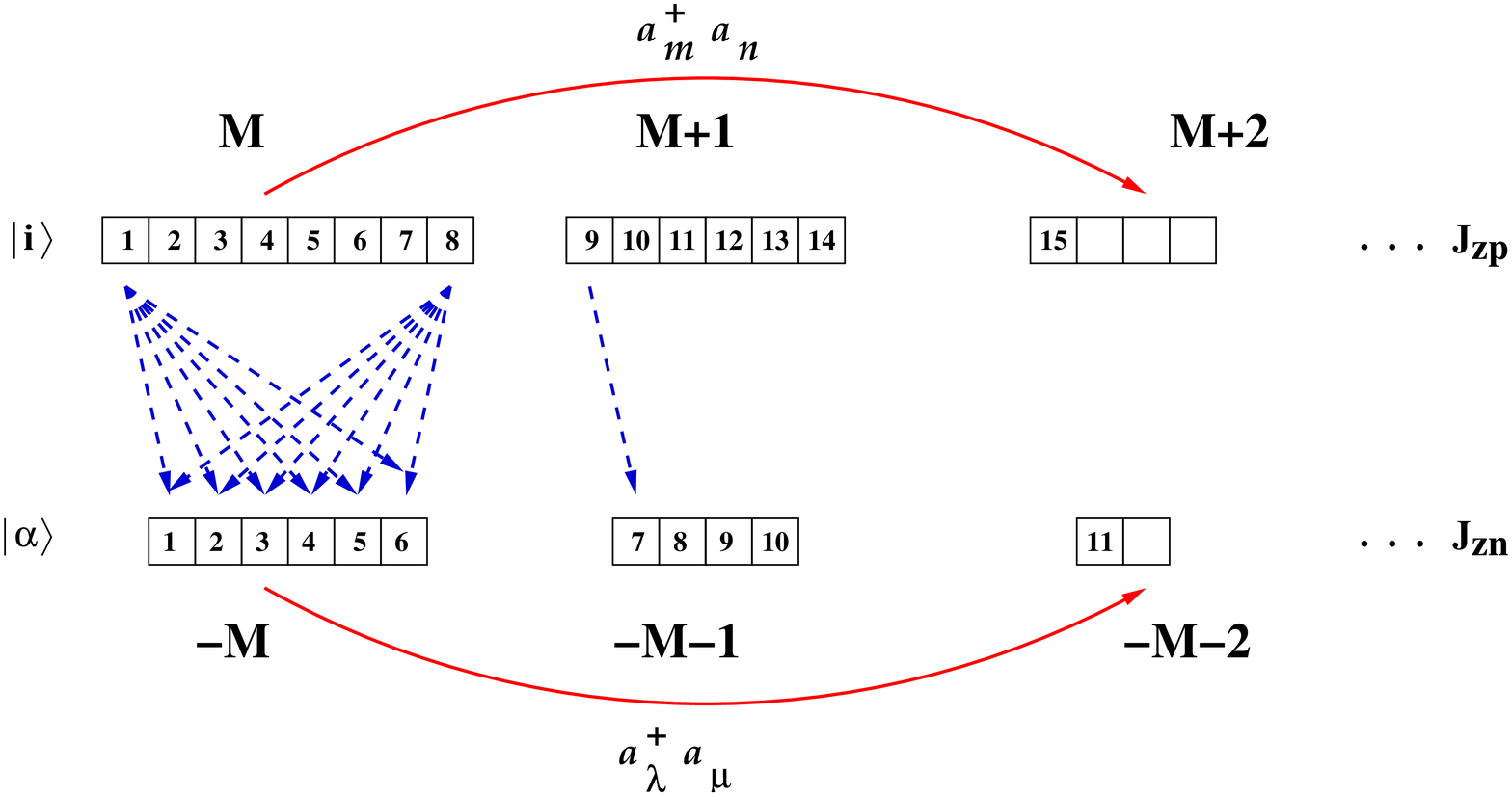}
  \caption{Schematic representation of the basis\label{fig:anto}}
\end{figure}

The shell model code ANTOINE\footnote{A version of the code can be
  downloaded from the URL
  \url{http://sbgat194.in2p3.fr/~theory/antoine/main.html}}
\cite{Caurier.Nowacki:1999} has retained many of these ideas,
while improving upon the Glasgow code in the several aspects. To start
  with, it takes
advantage of the fact that the dimension of the proton and neutron
spaces is small compared with the full space dimension, with the
obvious exception of semi-magic nuclei. For example, the 1,963,461 SD
with $M=0$ in $^{48}$Cr are generated with only the 4,865 SD
  (corresponding to all the
possible $M$ values) in $^{44}$Ca.  The states of the basis
are written as the product of two SD, one for protons and one for
neutrons: $ |I\rangle = |i,\alpha\rangle $. We use, $I, J,$ capital
letters for states in the full space, $i, j,$ small case Latin letters
for states of the first subspace (protons or neutrons), $\alpha$,
$\beta$, small case Greek letters for states of the second subspace
(neutrons or protons). The Slater determinants $i$ and $\alpha$ can be
classified by their $M$ values, $M_1$ and $M_2$. The total $M$ being
fixed, the SD's of the 2 subspaces will be  associated only if $M_1 +
M_2 = M$. A pictorial example is given in fig.~\ref{fig:anto}.

It is clear that for each $|i\rangle$ state the allowed
$|\alpha\rangle$ states run, without discontinuity, between a minimum
and a maximum value. Building the basis in the total space by means of
an $i$-loop and a nested $\alpha$-loop, it is possible to construct
numerically an array $R(i)$ that points to the $|I\rangle$
state\footnote{We use Dirac's notation for the quantum mechanical state
  $|i\rangle$, the position of this state in the basis
  is denoted by $i$.}:
 \begin{equation}
   \label{eq:anto}
   I=R(i)+\alpha
\end{equation}
For example, according to figure~\ref{fig:anto}, the numerical values
of $R$ are: $R(1)=0$, $R(2)=6$, $R(3)=12$,\ldots  The
relation~\eqref{eq:anto} holds even in the case of truncated spaces,
provided we define sub-blocks labeled with $M$ and $t$ (the truncation
index defined in \eqref{eq:t}). Before the diagonalization, all the
calculations that involve only the proton or the neutron spaces
separately are carried out and the results stored. For the
proton-proton and neutron-neutron NZME the numerical values of
$R(i),R(j), W_{ij}$ and $\alpha,\beta,W_{\alpha\beta}$, where $\langle
i|H|j\rangle=W_{ij}$ and $\langle
\alpha|H|\beta\rangle=W_{\alpha\beta}$, are pre-calculated and
stored.  Therefore, in the Lanczos procedure a simple loop on $\alpha$
and $i$ generates all the proton-proton and neutron-neutron NZME,
$W_{I,J}=\langle I|H|J\rangle$. For the proton-neutron matrix
elements the situation is slightly more complicated. Let's assume that
the $|i\rangle$ and $|j\rangle$ SD are connected by the one-body
operator $a^{\dagger}_{q}a_{r}$ (that in the list of all possible
one-body operators appears at position $p$), with $q=nljm$ and
$r=n'l'j'm'$ and $m'-m=\Delta m$. Equivalently, the $|\alpha\rangle$
and $|\beta\rangle$ SD are connected by a one-body operator whose
position is denoted by $\mu$.  We pre-calculate the numerical values
of $R(i),R(j),p$ and $\alpha,\beta,\mu$.  Conservation of the total
$M$ implies that the proton operators with $\Delta m$ must be
associated to the neutron operators with $-\Delta m$. Thus we could
draw the equivalent to figure~\ref{fig:anto} for the proton and
neutron one-body operators. In the same way as we did before for
$I=R(i)+\alpha$, we can now define an index $K=Q(p)+\mu$ that labels
the different two-body matrix elements (TBME).  Then, we denote $V(K)$
the numerical value of the proton-neutron TBME that connects the
states $|i,\alpha\rangle$ and $|j,\beta\rangle$.  Once
$R(i),R(j),Q(p)$ and $\alpha,\beta,\mu$ are known, the non-zero
elements of the matrix in the full space are generated with three
integer additions:
\begin{equation}
   I=R(i)+\alpha,  \quad   J=R(j)+\beta, \quad K=Q(p)+\mu.
\end{equation}
The NZME of the Hamiltonian between states $|I\rangle$ and $|J\rangle$
is then:
\begin{equation}
  \langle I|H|J\rangle=\langle J|H|I\rangle=V(K).
\end{equation} 

The performance of the code is optimal when the two subspaces have
comparable dimensions. It becomes less efficient for asymmetric nuclei
(for semi-magic nuclei all the NZME must be stored) and for large
truncated spaces. No-core calculations are typical in this respect.
If we consider a $N=Z$ nucleus like $^{6}$Li, in a valence space that
comprises up to 14$\hbar\omega$ configurations, there are 50000 Slater
determinants for protons and neutrons with $M=1/2$ and $t=14$. Their
respective counterparts can only have $M=-1/2$ and $t=0$ and have
dimensionality 1. This situation is the same as in semi-magic nuclei.
 
As far as two Lanczos vectors can be stored in the RAM memory of the
computer, the calculations are straightforward. Until recently this
was the fundamental limitation of the SM code ANTOINE. It is now
possible to overcome it dividing the Lanczos vectors in segments:
\begin{equation}
\Psi_f= \sum_{k}\Psi^{(k)}_f.
\end{equation}
The Hamiltonian matrix is also divided in blocks so that the action of
the Hamiltonian during a Lanczos iteration can be expressed as:
\begin{equation}
\Psi_{f}^{(k)}= \sum_{q} H^{(q,k)}\Psi^{(q)}_i
\end{equation}
The $k$ segments correspond to specific values of $M$ (and
occasionally $t$) of the first subspace. The price to pay for the
increase in size is a strong reduction of performance of the
code. Now, $\langle I|H|J\rangle$ and $\langle J|H|I\rangle$
are not generated simultaneously when $|I\rangle$ and $|J\rangle$ do
not belong at the same $k$ segment of the vector and the amount of 
disk use increases. As a counterpart, it gives a
natural way to parallelize the code, each processor calculating some
specific $\Psi^{(k)}_f$. This technique allows the diagonalization of
matrices with dimensions in the billion range.  All the nuclei of the
$pf$-shell can now be calculated without truncation.
Other $m$-scheme codes have recently joined ANTOINE in the run,
incorporating some of its algorithmic findings as
MSHELL~\cite{Mizusaki:2000} or REDSTICK~\cite{Ormand.Johnson:2002}.
 
\subsection{The coupled code NATHAN.}
\label{sec:nathan}
 
As the mass of the nucleus increases, the possibilities of performing
SM calculations become more and more limited. To give an order of
magnitude the dimension of the matrix of $^{52}$Fe with 6 protons and
6 neutrons in the $pf$ shell is $10^8$. If now we consider $^{132}$Ba
with also 6 protons and 6 neutron-holes in a valence space of 5
shells, four of them equivalent to the $pf$ shell plus the $1h_{11/2}$
orbit, the dimension reaches $2\times 10^{10}$. Furthermore, when many
protons and neutrons are active, nuclei tend to get deformed and their
description requires even larger valence spaces.  For instance, to
treat the deformed nuclei in the vicinity of $^{80}$Zr, the "normal"
valence space, $2p_{3/2}$, $1f_{5/2}$, $2p_{1/2}$, $1g_{9/2}$, must be
complemented with, at least, the orbit $1d_{5/2}$.  This means that
our span will be limited to nuclei with few particles outside closed
shells ($^{130}$Xe remains an easy calculation) or to almost
semi-magic, as for instance the Tellurium isotopes. These nuclei are
spherical, therefore the seniority can provide a good truncation
scheme. This explains the interest of a shell model code in a coupled
basis and quasi-spin formalism.
 
In the shell model code NATHAN \cite{Caurier.Nowacki:1999} the fundamental idea
of the code ANTOINE is kept, i.\ e.\ splitting the valence space in
two parts and writing the full space basis as the product of states
belonging to these two parts.  Now, $|i\rangle$ and $|\alpha\rangle$
are states with good angular momentum. They are built with the usual
techniques of the Oak-Ridge/Rochester group
\cite{French.Halbert.ea:1969}.  Each subspace is now partitioned with
the labels $J_1$ and $J_2$. The only difference with the $m$-scheme is
that instead of having a one-to-one association ($M_1+M_2=M$), for a
given $J_1$ we now have all the possible $J_2$, $J_{min} \le J_2 \le
J_{max}$, with $J_{min}=|J_0-J_1|$ and $J_{max}=J_0+J_1$.  The
continuity between the first state with $J_{min}$ and the last with
$J_{max}$ is maintained and consequently the fundamental relation
$I=R(i)+\alpha$ still holds.  The generation of the proton-proton and
neutron-neutron NZME proceeds exactly as in $m$-scheme.  For the
proton-neutron NZME the one-body operators in each space can be
written as $O_{p}^{\lambda}=(a^{\dagger}_{j_1}a_{j_2})^{\lambda}$.
There exists a strict analogy between $\Delta m$ in $m$-scheme and
$\lambda$ in the coupled scheme. Hence, we can still establish a
relation $K=Q(p)+\omega$. The NZME read now:
\begin{equation} 
   \langle I|H|J\rangle=\langle J|H|I\rangle=h_{ij} \cdot
   h_{\alpha\beta} \cdot W(K), 
\end{equation}  
with $h_{ij}=\langle i | O_p^{\lambda} | j \rangle$,
$h_{\alpha\beta}=\langle \alpha | O_{\omega}^{\lambda} | \beta
\rangle$, and
\begin{equation}  
 W(K) \propto V(K) \cdot \left\{\begin{array}{ccc}
                     i & \alpha  & J \\
                     j & \beta   & J  \\
               \lambda & \lambda & 0
   \end{array}\right\},
\end{equation}  
where $V(K)$ is a TBME.  We need to perform --as in the $m$-scheme
code-- the three integer additions which generate I, J, and K, but, in
addition, there are two floating point multiplications to be done,
since $h_{ij}$ and $h_{\alpha\beta}$, which in $m$-scheme were just
phases, are now a product of cfp's and 9j symbols (see formula 3.10 in
\citet{French.Halbert.ea:1969}).  Within this formalism we can
introduce seniority truncations, but the problem of the semi-magic
nuclei and the asymmetry between the protons and neutrons spaces
remains. To overcome this difficulty in equalizing the dimension of
the two subspaces we have generalized the code as to allow that one of
the two subspaces contains a mixing of proton and neutron orbits.  For
example, for heavy Scandium or Titanium isotopes, we can include the
neutron $1f_{7/2}$ orbit in the proton space.  For the $N=126$
isotones we have only proton shells. We then take the $1i_{13/2}$ and
$1h_{9/2}$ orbits as the first subspace, while the $2f_{7/2}$,
$2f_{5/2}$, $3p_{3/2}$ and $3p_{1/2}$ form the second one.  The states
previously defined with $J_p$ and $J_n$ are now respectively labelled
$A_{1},J_{1}$ and $A_{2},J_{2}$, $A_1$ and $A_2$ being the number of
particles in each subspace. In $h_{ij}$ and $h_{\alpha\beta}$ now
appear mean values of all the operators in Eq.(\ref{eq:ops})

The fact that in the coupled scheme the dimensions are smaller and
that angular momentum is explicitly enforced makes it possible to
perform a very large number of Lanczos iterations without storage or
precision problems. This can be essential for the calculation of
strength functions or when many converged states are needed, for
example to describe non-yrast deformed bands.  The coupled code has
another important advantage. It can be perfectly parallelized without
restriction on the number of processors. The typical working dimension
for large matrices in NATHAN is 10$^7$ (10$^9$ with ANTOINE). It means
that there is no problem to define as many final vectors as processors
are available. The calculation of the NZME is shared between the
different processors (each processor taking a piece of the
Hamiltonian, $H=\sum_k H^{(k)}$) leading to different vectors that are
added to obtain the full one:
\begin{equation} 
\Psi^{(k)}_f= H^{(k)}\Psi_i, \qquad
\Psi_f= \sum_{k} \Psi^{(k)}_i.
\end{equation}

\subsection{No core shell model} 
\label{sec:no-core}

The \emph{ab initio} no-core shell model (NCSM)
\cite{Navratil.Vary:2000,Navratil.Vary.Barrett:2000} is a method to
solve the nuclear many body problem for light nuclei using realistic
inter-nucleon forces. The calculations are performed using a large but
finite harmonic-oscillator (HO) basis. Due to the basis truncation, it
is necessary to derive an effective interaction from the underlying
inter-nucleon interaction that is appropriate for the basis size
employed. The effective interaction contains, in general, up to
$A$-body components even if the underlying interaction had, e.g. only
two-body terms.  In practice, the effective interaction is derived in
a sub-cluster approximation retaining just two- or three-body terms. A
crucial feature of the method is that it converges to the exact
solution when the basis size increases and/or the effective
interaction clustering increases.

In a first phase, their applications were limited to the use of
realistic two-nucleon interactions, either G-matrix-based two-body
interactions~\cite{Zheng.Vary.ea:1994}, or derived by the Lee-Suzuki
procedure~\cite{Suzuki.Lee:1980} for the
NCSM~\cite{Navratil.Barrett:1996}. This resulted in the elimination of
the purely phenomenological parameter $\Delta$ used to define G-matrix
starting energy. A truly \emph{ab initio} formulation of the formalism
was presented by~\citet{Navratil.Barrett:1998}, where convergence to
the exact solutions was demonstrated for the $A=3$ system. The same
was later accomplished for the $A=4$
system~\cite{Navratil.Barrett:1999}, where it was also shown that a
three-body effective interaction can be introduced to improve the
convergence of the method. The capability to derive a three-body
effective interaction and apply it in either relative-coordinate
\cite{Navratil.Kamuntavicious.ea:2000} or Cartesian-coordinate formalism
\cite{Navratil.Ormand:2002} together with the ability to
solve a three-nucleon system with a genuine three-nucleon force in the
NCSM approach \cite{Marsden.Navratil.ea:2002} now opens the
possibility to include a realistic three-nucleon force in the NCSM
Hamiltonian and perform calculations using two- and three-nucleon
forces for the $p$-shell nuclei.  Among successes of the NCSM approach
was the first published result of the binding energy of $^4$He with
the CD-Bonn NN potential \cite{Navratil.Barrett:1999}, the
near-converged results for $A=6$ using a non-local Hamiltonian
\cite{Navratil.Ormand.ea:2001}, the first observation of the incorrect
ground-state spin in $^{10}$B predicted by the realistic two-body
nucleon-nucleon interactions \cite{Caurier.Navratil.ea:2002}. This
last result, together with the already known problems of
under-binding, confirms the need of realistic three-nucleon forces, a
conclusion that also stems from the Green Function Monte Carlo
calculations of \citet{Wiringa.Pieper:2002} \citep[see also][for a
review of the results of the Urbana-Argonne collaboration for nuclei
A$\le$10]{Pieper.Wiringa:2001}.

Recently, a new version of the shell model code ANTOINE has been
developed for the NCSM applications \cite{Caurier.Navratil.ea:2001}.
This new code allows to perform calculations in significantly larger
basis spaces and makes it possible to reach full convergence for the
$A=6$ nuclei. Besides, it opens the possibility to investigate slowly
converging intruder states and states with unnatural parity.  The
largest bases  reached with this code so far are, according to
the number of HO excitations, the 14$\hbar\omega$ space for $^6$Li
and, according the matrix dimension, the 10$\hbar\omega$ calculations
for $^{10}$C.  In the latter case, the $m$-scheme matrix dimension
exceeds 800 millions.

\subsection{Present possibilities}
 
The combination of advances in computer technology and in algorithms
has enlarged the scope of possible SM studies. The rotational band of
$^{238}$U seems to be still very far from the reach of SM
calculations, but predictions for $^{218}$U are already available. Let
us list some of the present opportunities.
 
\begin{itemize}
\item[i)] No-core calculations.  One of the major problems of the
  no-core SM  is the convergence of the results
  with the size of the valence space; For $^6$Li we can handle
  excitations until 14$\hbar \omega$ and at least 8$\hbar \omega$ for
  all the $p$-shell nuclei.
  
\item[ii)] The $pf$-shell. This where the shell model has been most
  successful and exact diagonalizations are now possible throughout
  the region Beyond $^{56}$Ni, as the $1f_{7/2}$ orbit becomes more
  and more bound, truncated calculations are close to exact, for
  instance, in $^{60}$Zn \cite{Mazzocchi.Janas.ea:2001} the
  wave-functions are fully converged when 6p-6h excitations are
  included.
  
\item[iii)] The $r_3 \;g$ valence space.  We use the notation $r_p$
  for the set of the $nlj$ orbits with $2(n-1)+l=p$ excluding the orbit
  with maximum total angular momentum $j=p+1/2$. This space describes
  nuclei in the region $ 28 < N,Z < 50 $; $^{56}$Ni is taken as inert
  core.  Most of them are nearly spherical, and can be treated without
  truncations.  The $\beta\beta$ decay (with and without neutrinos)
  of $^{76}$Ge and $^{82}$Se are a prime example.  The deformed nuclei
  ($N \sim Z \sim 40$) are more difficult because they demand the
  inclusion of the $2d_{5/2}$ orbit to describe prolate states and
  oblate-prolate shape coexistence.
  
\item[iv)] The $pfg$ space.  To enlarge the $pf$ valence space to
  include the $1g_{9/2}$ orbit is hopeless nowadays. Furthermore,
  serious center of mass spuriousness is expected in the
  $1f_{7/2}^{-k} 1g_{9/2}^k$ configurations (see
  Appendix~\ref{sec:center-mass-problem}). For neutron rich nuclei in
  the Nickel region, a tractable valence space that avoids the center
  of mass problem can be defined: A $^{48}$Ca core on top of which $pf$
  protons and $r_3 \; g$ neutrons are active.

\item[v)] Heavy nuclei.  All the semi-magic nuclei, for instance the
  $N=126$ isotones, can be easily studied and the addition of few
  particles or holes remains tractable.  Some long chains of Tellurium
  and Bismuth isotopes have been recently studied
  \cite{Caurier.Grawe.Rejmund:2003}.
\end{itemize}

\section{The Lanczos basis}\label{sec:lancz-diag}
There exists a strong connection between the Lanczos algorithm, the
partition function, $Z(\beta)=\sum_i \langle i|\exp(-\beta
{\cal H})|i\rangle$,
 and the evolution operator $\exp(i {\cal H} t)$. In
the three cases, powers of the Hamiltonian determine the properties of
the systems. The partition function can be written as $Z(\beta)=\sum_E
\rho(E)\exp(-\beta E)$ {\it i.e.}, the Laplace transform of the
density of states, a quantity readily accessible once the Hamiltonian
has been fully diagonalized. The evolution operator addresses more
specifically the problem of evolving from a starting vector into the
exact ground state. We shall discuss both questions in turn.

\subsection{Level densities}\label{sec:level-densities} 
\begin{figure}[htbp]
  \includegraphics[width=0.9\linewidth]{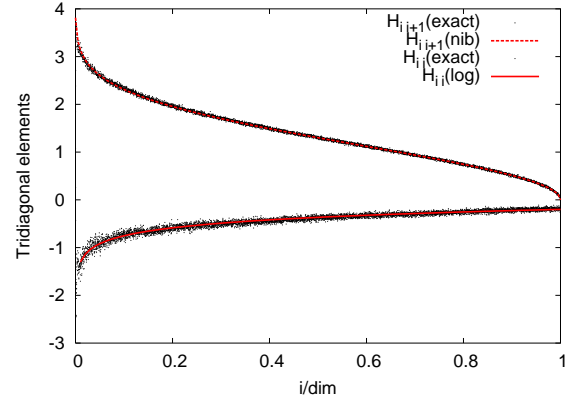}  
  \caption{Tridiagonal matrix elements for a 6579 dimensional matrix,
    and the logarithmic and inverse binomial approximations
    (nib)~\citep{Zuker.Waha.ea:2001}.\label{fig:fpgd6}}
\end{figure}

For many, shell model is still synonymous with diagonalizations, in
turn synonymous with black box. One still hears questions such as: Who
is interested in diagonalizing a large matrix? As an answer we propose
to examine Fig.~\ref{fig:fpgd6} showing the two point-functions that
define the diagonal, $H_{ii}$, and off-diagonal elements, $H_{i\,
  i+1}$ in a typical Lanczos construction.
The continuous lines are calculated knowing the first four moments of
the matrix (\emph{i.e.} of the 1+2-body Hamiltonian $\mathcal{H}$ it
represents). These results hold at fixed quantum numbers, \emph{i.e.},
when the matrix admits no block decomposition.

When the matrix is diagonalized the level density is well reproduced
by a continuous binomial 
\begin{equation}
  \label{rbin}
\rho_b(x,N,p,S)=p^{xN}q^{\bar
  xN}d\frac{\Gamma(N+1)}{\Gamma(xN+1)\Gamma(\bar xN+1)}\frac{N}{S},  
\end{equation}
where $x$ is an a-dimensional energy, $\bar x=1-x$, $N$ the  number of valence
particles, $p$ an asymmetry parameter,  $p+q=1$, and $S$ the span of the
spectrum (distance between lowest and highest eigenstates).
Introducing the energy scale $\varepsilon$, $S$, the centroid $E_c$, the
variance $\sigma^2$ and $x$ are given by
\begin{equation}
  \label{npe}
  S=N\varepsilon,\; E_c=Np\varepsilon ,\;
  \sigma^2=Np(1-p)\varepsilon^2,\; x=\frac{E}{S}.
\end{equation}
Note that $\rho_b(x,N,p,S)$ reduces to a discrete binomial, ${N\choose
  n}$ if $x=n/N=n\varepsilon/S$, with integer $n$.

\begin{figure}[htbp]
  \includegraphics[width=0.9\linewidth]{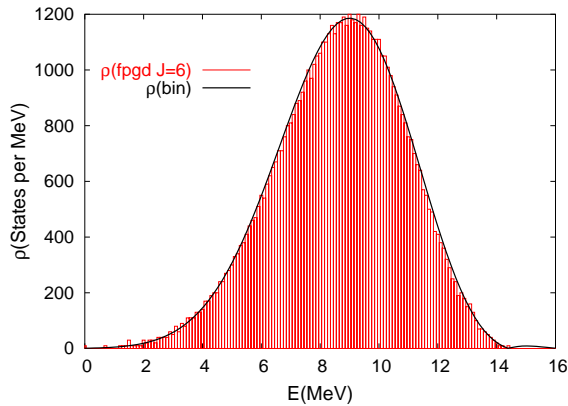}
  \caption{Exact and binomial level densities for the matrix in
    Fig.~\ref{fig:fpgd6}.\label{fig:fpgd6_r}} 
\end{figure}

$N$, $p$ and $S$ are calculated using the moments of the Hamiltonian
${\cal H}$, {\em i.e.}, averages given by the traces of ${\cal H}^K$, to be
equated to the corresponding moments of $\rho_b(x,N,p,S)$, which for
low $K$ are the same as those of a discrete binomial. The necessary
definitions and equalities follow.
\begin{eqnarray}
  \label{moms}
&&  d^{-1}\text{tr}({\cal H}^K)=\langle{\cal H}^K\rangle, \;
  E_c=\langle{\cal H}^1\rangle,\; {\cal M}_K=\langle({\cal
  H}-E_c)^K\rangle\nonumber \\
&&\sigma^2={\cal M}^2,\quad \overline{\cal M}_K=\frac{{\cal
  M}_K}{\sigma^K},\quad \gamma_1=\overline{\cal
  M}_3=\frac{q-p}{\sqrt{Npq}} \nonumber \\
&&\gamma_2=\overline{\cal M}_4-3=\frac{1-6pq}{Npq}; \qquad d=d_0(1+p/q)^N.
\end{eqnarray}
These quantities also define the logarithmic and inverse binomial
(nib) forms of $H_{ii}$ and $H_{i\, i+1}$ in Fig.~\ref{fig:fpgd6}.
Note that the corresponding lines are almost impossible to distinguish
from those of the exact matrix. The associated level densities are
found in Fig.~\ref{fig:fpgd6_r}.

The mathematical status of these results is somewhat mixed.
\citet{Mon.French:1975} proved that the total density of a Hamiltonian
system is to a first approximation a Gaussian. \citet{Zuker:2001}
extended the approximation to a binomial, but the result remains valid
only in some neighborhood of the centroid. Furthermore, one does not
expect it to hold generally because binomial thermodynamics is
trivial and precludes the existence of phase transitions. In the
example given above, we do not deal with the total density, which
involves all states, $\rho=\sum_{JT}(2J+1)(2T+1)\rho_{JT}$ but with a
partial $\rho_{JT}$ at fixed quantum numbers. In this
case~\citet{Zuker.Waha.ea:2001} conjectured that the tridiagonal
elements given by the logarithmic and nib forms are valid, and hence
describe the full spectrum. The conjecture breaks down if a dynamical
symmetry is so strong as to define new (approximately) conserved
quantum numbers. 

\begin{figure}[htbp]
  \includegraphics[width=0.9\linewidth]{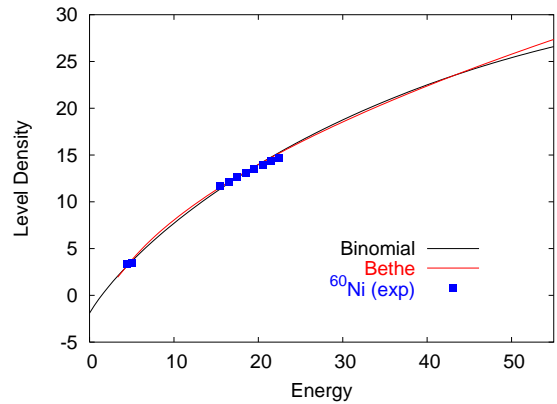}  
  \caption{Binomial, Bethe and experimental level densities for
    $^{60}$Ni.\label{binbethe60}} 
\end{figure}

Granted that a single binomial cannot cover all situations
we may nonetheless explore its validity in nuclear physics,
where the observed level densities are extremely
well approximated by the classical formula of~\citet{Bethe:36} with a
shift $\Delta$,
\begin{equation}
  \label{bet}
  \rho_B(E,a,\Delta)=
\frac{\sqrt{2\pi}}{12}\frac{e^{\sqrt{4a(E+\Delta)}}}
{(4a)^{1/4}(E+\Delta)^{5/4}}. 
\end{equation}
Obviously, if binomials are to be useful, they must reproduce---for
some range of energies---Eq.~(\ref{bet}). They do indeed, as shown in
Fig.~\ref{binbethe60}, where the experimental points for
$^{60}$Ni~\cite{Iljinov.ea:1992} are also given.
Though Bethe and binomial forms are seen to be equivalent, the latter
has the advantage that the necessary parameters are well defined and
can be calculated, while in the former, the precise meaning of the $a$
parameter is elusive. The shift $\Delta$ is necessary to adjust the
ground state position. The problem also exists for the
binomial\footnote{ The energies are referred to the mean (centroid)
  of the distribution, while they are measured with respect to the
  ground state.}  and the result in the figure~\cite{Zuker:2001}
solves it phenomenologically.  The Shell Model Monte Carlo method
provides the only parameter-free approach to level
densities~\cite{Dean.Koonin.ea:1995,Nakada.Alhassid:1997,Langanke:1998}
whose reliability is now established~\cite{Alhassid.Liu.ea:1999}. The
problem is that the calculations are hard.

As the shape of the level density is well reproduced by a binomial {\em
  except} in the neighborhood of the ground state, to reconcile
  simplicity with  full rigor we have to examine the tridiagonal
  matrix at the origin.   

\subsection{The ground state}\label{sec:ground-state}
\begin{figure}[htbp]
  \includegraphics[width=0.9\linewidth]{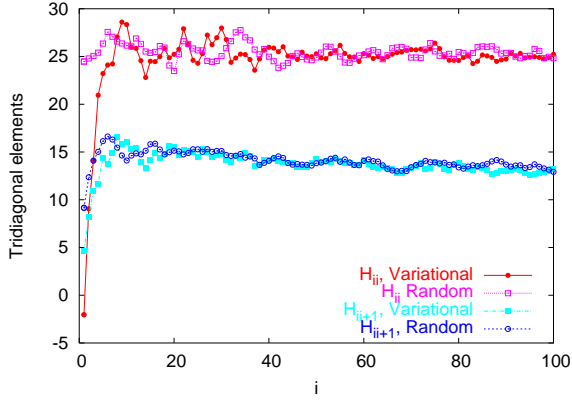}  
  \caption{Tridiagonal matrix elements after 100 iterations for a
    random and a variational pivot.\label{fig:sc48_tri}}
\end{figure}

Obviously, some dependence on the pivot should exist. We examine it
through the $J=1\, T=3$ $pf$ states in $^{48}$Sc\footnote{They are
  reached in the $^{48}$Ca$(p,n)^(48)$Sc reaction, which will be our
  standard example of Gamow Teller transitions.}.  The matrix is
8590-dimensional, and we calculate the ground state with two pivots,
one random (homogeneous sum of all basic states), the other
variational (lowest eigenstate in $f_{7/2}^8$ space).  A zoom on the
first matrix elements in Fig.~\ref{fig:sc48_tri} reveals that they are
very different for the first few iterations, but soon they merge into
the canonical patterns discussed in the preceding section.
The ground state wavefunction is unique of course, but it takes the
different aspects shown in Fig.~\ref{fig:sc48_wfs}. In both cases the
convergence is very fast, and it is not difficult to show in general
that it occurs for a number of iterations of order $N\log N$ for
dimensionality $d=2^N$. However, the variational pivot is clearly
better if we are interested in the ground state: If its overlap
with the exact solution exceeds 50\%, all other contributions are
bound to be smaller and in general they will decrease
uniformly\footnote{There are some very interesting counter-examples.
  One is found in Fig.~\ref{fig:spec_f5}, where the natural pivot is
  heavily fragmented.}.
  We shall see how to exploit this property of good pivots to
simplify the calculations.

\subsubsection {The exp(S) method }\label{sec:the-exps-method}
\begin{figure}[htbp]
  \includegraphics[width=0.9\linewidth]{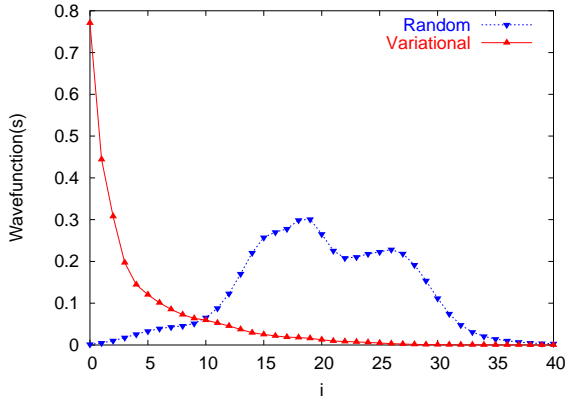}  
  \caption{The ground state wavefunction in the Lanczos basis for a
    random, and a variational pivot. The plotted values are the
    squared overlaps of the successive Lanczos vectors with the final
    wave-function \label{fig:sc48_wfs}} 
\end{figure}
In Section~\ref{sec:theory-calculations} we have sketched the
coupled-cluster (or ${\exp (S)}$) formalism. The formulation in the
Lanczos basis cannot do justice to the general theory, but it is a
good introduction to the underlying ideas.  Furthermore, it turns out
to be quite useful.

The construction is as in Eqs.~(\ref{eq:lanc0},\ref{lanc}), {\em but
  the succeeding vectors---except the pivot---are not normalized}.
Then, the full wavefunction takes the form
\begin{equation}
  \label{eqn:w0}
 |\bar 0\rangle =(1+c_1P_1+c_2P_2+\cdots +c_IP_I) |0\rangle, 
\end{equation}
where $P_m$ is a polynomial in $H$ that acting on the pivot produces
orthogonal unnormalized vectors in the Lanczos basis:
$P_m|0\rangle=|m\rangle$. Eq.~(\ref{lanc}) becomes 
\begin{equation}
  \label{lancu}
{\cal H}|m\rangle= V_m |m-1\rangle+ E_m|m\rangle + |m+1\rangle. 
\end{equation}
To relate with the normalized version ($m\Rightarrow \bar m$); Divide
by $\langle m|m\rangle^{1/2}$, then multiply and divide
$|m-1\rangle$ and $|m+1\rangle$ by their norms to recover
Eq.~(\ref{lanc}), and obtain $E_{m}=H_{\bar m\, \bar m}$, $V_m=\langle
m|m\rangle/\langle m-1|m-1\rangle= H_{\bar m\, \bar m-1}^2$.  The
secular equation $({\cal H}-E)|\bar 0\rangle=0$ leads to the recursion
\begin{equation}
  \label{eqn:rec1}
c_{m-1}+(E_m-E)\, c_m+V_{m+1}\, c_{m+1}=0,
\end{equation}
whose solution is equivalent to diagonalizing a matrix with $V_{m+1}$
in the upper diagonal and 1 in the lower one, which is of course
equivalent to the symmetric problem. However, here we shall solve the
recursion by transforming it into a set of non-linear coupled
equations for the $c_m$ amplitudes. The profound reason for using an
unnormalized basis is that first term in Eq.~(\ref{eqn:rec1}),
$E=E_0+V_1c_1$ implies that once $c_1$ is known the problem is solved.
Calling $\varepsilon_m=E_m-E_0$ and replacing $E=E_0+V_1c_1$ in
Eq~(\ref{eqn:rec1}) leads to
\begin{equation}
  \label{eqn:rec2}
c_{m-1}+(\varepsilon_m-V_1c_1)\, c_m+V_{m+1}\, c_{m+1}=0.
\end{equation}
Now introduce
\begin{equation}
  \label{eqn:expS1}
 \sum c_m P_m=\exp(\sum S_m P_m), 
\end{equation}
Expanding the exponential and equating terms
\begin{align}
  \label{eqn:coeffS}
c_1&=S_1,\quad c_2=S_2+\frac{1}{2}S_1^2, \\
c_3&=S_3+S_1S_2+\frac{1}{3!}S_1^3, \; \; {\rm etc}.   
\end{align}
Note that we have used the formal identification $P_m\, P_n=P_{n+m}$
as a heuristic way of suggesting Eqs.~\eqref{eqn:coeffS}.
Inserting in Eq.~(\ref{eqn:rec2}) and regrouping, a system of coupled
equations obtains. The first two are
 \begin{align}
   \label{eqn:expS3a}
0=&S_2V_2+S_1^2(\frac{1}{2}V_2-V_1)+S_1\varepsilon_1+1 \\
0=&S_3V_3+S_2\varepsilon_2+S_1S_2(V_3-V_1)+\nonumber\\   
&\frac{1}{2}S_1^2\varepsilon_2+
S_1^3(\frac{1}{3!}V_3-\frac{1}{2}V_1)+S_1.
\label{eqn:expS3b}
\end{align}

\begin{figure}[htb]
  \includegraphics[width=0.9\linewidth]{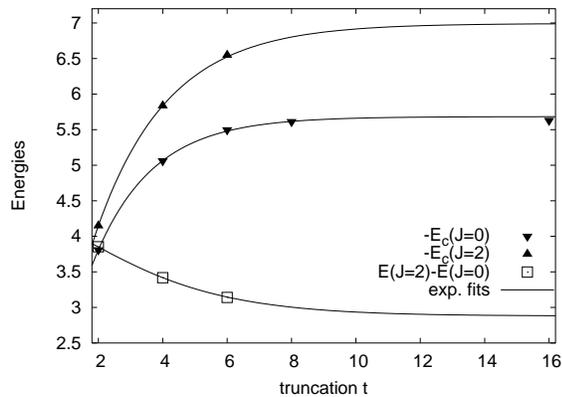}  
  \caption{Energy convergence for ground and first excited state in
    $^{56}$Ni as a function of the truncation level 
   $t$~\cite{Caurier.Martinez-Pinedo.ea:1999a}.\label{fig:ni56.con}} 
\end{figure}
Hence, instead of the usual $c_m$-truncations we can use
$S_m$-truncations, which have the advantage of providing a model for
the wavefunction over the full basis. At the first iteration
Eq.~(\ref{eqn:expS3a}) is solved neglecting $S_2$. The resulting value
for $S_1$ is inserted in Eq.~(\ref{eqn:expS3b}) which is solved by
neglecting $S_3$. The resulting value for $S_2$ is reinserted in
Eq.~(\ref{eqn:expS3a}) and the process is repeated. Once $S_1$ and
$S_2$ are known, the equation for $S_3$ (not shown) may be
incorporated, and so on.  Since the energy depends only on $S_1$,
convergence is reached when its value remains constant from step to
step. With a very good pivot, Eq.~(\ref{eqn:expS3a}) should give a fair
approximation, improved by incorporating Eq.~(\ref{eqn:expS3b}), and
checked by the equation for $S_3$. 

If all amplitudes except $S_1$ are neglected, the difference
scheme~(\ref{eqn:rec1}) is the same that would be obtained for an
harmonic ${\cal H}\equiv \varepsilon S^0+V(S^++S_-)$. Successive
approximations amount to introduce anharmonicities.  An equivalent
approach---numerically expedient---consists in diagonalizing the
matrices at each $c_m$ truncation level. Then, after some iteration,
the energies converge exponentially: Introduce
\begin{equation}
  \label{eqn:conv}
  \text{conv}(i,a,e_0,i_0)=e_0\frac{\exp(-ai)-1}{\exp(-ai_0)-1},
\end{equation}
which equals $e_0$ at point $i=i_0$. Choose $a$ so as to yield
conv$(i=i_0+1)=e(i_0+1)$ {\em i.e.}, the correct energy at the next
point . Check that $e(i_0+2)$ is well reproduced.
Fig.~\ref{fig:ni56.con} provides an example of the efficiency of the
method for the lowest two states in $^{56}$Ni. The index $i$ is
replaced by the truncation level $i\equiv t/2$ [Eq.~(\ref{eq:t}, with
$m=16,\, m_0=0$]: We set $i_0=1$, fix $a$ so as to reproduce the
energy at the second iteration and check that the curve gives indeed
the correct value at the third point. The results reproduce those of
the $S_2$ truncation, confirming that ``exponential convergence'' and
$\exp (S)$ are very much the same thing.  In this example, the closed
shell pivot is particularly good, and the exponential regime sets in
at the first iteration. In general, it will always set in for some
value of $i_0$ that may be quite large (Fig.~\ref{fig:sc48_wfs}
suggests $i_0\approx 25$ for the random pivot). Fortunately, it is
almost always possible to find good pivots, and the subject deserves a
comment.

In the case of $^{56}$Ni the good pivot is a closed shell. As a
consequence, the first iterations are associated to truncated spaces
of much smaller dimensionalities than the total one ($d_m\approx
10^9$). This also happens for lighter $pf$ shell nuclei, for which the
$f_{7/2}^m$ [or eventually $(f_{7/2}p_{3/2})^m$] subspaces provide a
good pivot. The same argument applies for other regions. For well
deformed nuclei Hartree Fock should provide good determinantal
pivots, and hence enormous gains in dimensionality, once projection
to good angular momentum can be tackled efficiently. 

As we shall see next, the Lanczos and $\exp S$ procedures provide
a convenient framework in which to analyze other approaches.

\subsubsection{Other numerical approximation methods}
\label{sec:approx}
The exponential convergence method introduced in the previous section
was first described by~\citet{Horoi.Volya.Zelevinsky:1999}, under a
different but equivalent guise. In later work, a hierarchy of
configurations determined by their average energy and width was
proposed. Successive diagonalizations make it possible to reach the
exact energy by exponential extrapolation. The method has been
successfully applied to the calculation of the binding energies of the
$pf$ shell nuclei \cite{Horoi.Brown.Zelevinsky:2002} and to
calculation of the excitation energies of the deformed 0$^+$ states of
$^{56}$Ni and $^{52}$Cr \cite{Horoi.Brown.Zelevinsky:2003}.

The work of~\citet{Mizusaki.Imada:2002,Mizusaki.Imada:2003}, is based
on the fact that the width of the total Hamiltonian in a truncated
space tends to zero as the solution approaches the exact one. They
have devised different extrapolation methods and applied  them to some
$pf$-shell nuclei. 

\citet*{Andreozzi.LoIudice.ea:2003} have recently proposed a
factorization method that allows for importance sampling to
approximate the exact eigenvalues and transition matrix elements.

A totally different approach which has proven its power in other
fields, the density matrix renormalization group, has been
proposed~\citep{Dukelsky.Pittel:2001,Dukelsky.Pittel.ea:2002}, and
\citet{Papenbrock.Dean:2003} have developed a method based in the
optimization of product wavefunctions of protons and neutrons that
seems very promising.

 Still, to our
knowledge, the only prediction for a truly large matrix that has
preceded the exact calculation remains that of $^{56}$Ni, $J=2$ in
Fig.~\ref{fig:ni56.con}. It was borne out when the diagonalization
became feasible two years later.

\subsubsection{Monte Carlo methods}\label{sec:monte-carlo-methods}
Monte Carlo methods rely on the possibility to deal with the
imaginary-time evolution operator acting on some trial wavefunction
$\exp(-\beta {\cal H})|0\rangle$ which tends to the exact ground state
$|\bar 0\rangle$ as $\beta\Rightarrow \infty$. This is very much what
the Lanczos algorithm does, but no basis is constructed. Instead, the
energy (or some observable $\hat \Omega$) is calculated through
\begin{equation}
  \label{eq:MC}
\frac{\langle 0|e^{-\beta\mathcal{H}/2} \; \hat\Omega \;
  e^{-\beta\mathcal{H}/2}|0\rangle}{\langle 0|e^{-\beta{\cal
  H}}|0\rangle}\stackrel{\beta\rightarrow\infty}{\Longrightarrow}
\frac{\langle\bar 0|\hat\Omega|\bar 0\rangle}{\langle \bar 0|\bar 0\rangle}    
\end{equation}
which is transformed into a quotient of multidimensional integrals
evaluated through Monte Carlo methods with importance sampling. A
``sign problem'' arises because the integrands are not definite
positive, leading to enormous precision problems. The Green Function
Monte Carlo studies mentioned at the beginning of
Section~\ref{sec:interaction} are conducted in coordinate space. The
Shell Model Monte Carlo (SMMC) variant is formulated in Fock space,
and hence directly amenable to comparisons with standard SM
results~\citep*[see][for a review]{Koonin.Dean.Langanke:1997}. The
approach relies on the Hubbard-Stratonovich transformation, and the
sign problem is circumvented either by an extrapolation method or by
choosing Hamiltonians that are free of it while remaining quite
realistic ({\em e.g.} pairing plus quadrupole). At present it remains
the only approach that can deal with much larger valence spaces than
the standard shell model. It should be understood that SMMC does not
lead to detailed spectroscopy, as it only produces ground state
averages, but it is very well suited for finite temperature
calculations.
  The introduction of Monte Carlo techniques in the Lanczos
construction is certainly a tempting project\ldots.

The quantum Monte Carlo diagonalization method (QMCD)
of~\citet*{Otsuka.Honma.ea:1998} consists in exploring the mean field
structure of the valence space by means of Hartree-Fock calculations
that break the symmetries of the Hamiltonian. Good quantum numbers are
enforced by projection techniques~\cite{Peierls.Yoccoz:1957}. Then the
authors borrow from SMMC to select an optimal set of basic states and
the full Hamiltonian is explicitly diagonalized in this basis. More
basic states are iteratively added until convergence is achieved. For
a very recent review of the applications of this method
see~\cite{Otsuka.Honma.ea:2001}.  A strong connection between mean
field and shell model techniques is also at the heart of the Vampir
approach~\cite{Petrovici.Schmid.ea:1999},~\cite{Schmid:2001} and of
the projected shell model (PSM) \cite{Hara.Sun:1995}.

\subsection{ Lanczos Strength Functions}
\label{sec:lancz-strength-funct}
The choice of pivot in the Lanczos tridiagonal construction is
arbitrary and it can be adapted to special problems. One of the most
interesting is the calculation of strength
functions~\cite{Whitehead:1980,Bloom:1984}: If $U_{i \, j}$
is the unitary matrix that achieves diagonal form, its first column
$U_{i \,0}$ gives the amplitudes of the ground state wavefunction in
the tridiagonal basis while the first row $U_{0\, j}$ determines the
amplitude of the pivot in the $j$-th eigenstate. $U_{0\, j}^2$ plotted
against the eigenenergies $E_j$ is called the ``strength function''
for that pivot. 

In practice, given a transition operator $\mathcal{T}$, act with it on
a target state $|t\rangle$ to define a pivot $|0'\rangle={\cal
  T}|t\rangle$ that exhausts the sum rule $\langle 0'|0'\rangle$ for
${\cal T}$. Normalize it. 
\begin{figure}[htb]
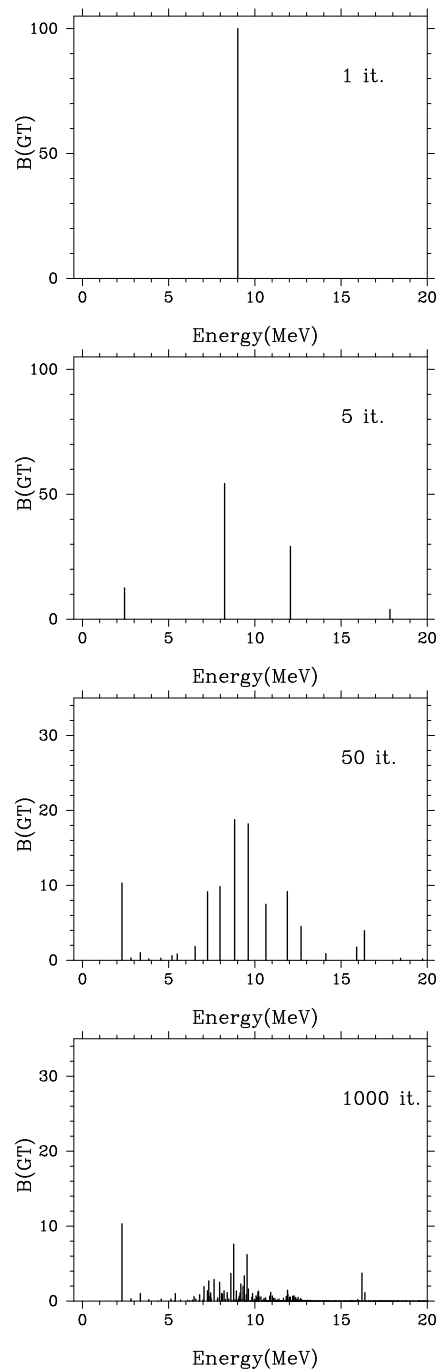

  \includegraphics[angle=270,width=0.65\linewidth]{rmpgif/poves_fig16a.ps}\\
  \includegraphics[angle=270,width=0.65\linewidth]{rmpgif/poves_fig16b.ps}\\
  \includegraphics[angle=270,width=0.65\linewidth]{rmpgif/poves_fig16c.ps}\\
  \includegraphics[angle=270,width=0.65\linewidth]{rmpgif/poves_fig16d.ps}
  \caption{Evolution of the Gamow-Teller strength function of
    $^{48}$Ca as the number of Lanczos iterations on the doorway state
    increases. \label{fig:sfevol}}
\end{figure}
Then, by definition 
\begin{gather}
  \label{lsf1}
  |0\rangle=\frac{{\cal T}|t\rangle}{\sqrt{\langle 0'|0'\rangle}}
   =\sum_j U_{0j}|j\rangle,\; {\rm whose\ 
    moments,}\\\label{lsf2} \langle 0|{\cal H}^k|0\rangle=\sum_j
  U_{0j}^2 E_j^k, \; {\rm are\ those\ of\ the}\\\label{lsf3} {\rm
    strength\ function\ }\; S(E)=\sum_j\delta(E-E_j)U_{0j}^2,
\end{gather}
\begin{figure}[htb]
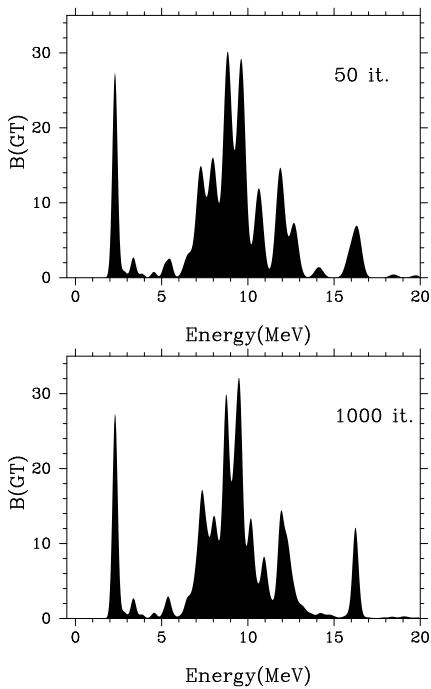

  \includegraphics[angle=270,width=0.65\linewidth]{rmpgif/poves_fig17a.ps}\\
  \includegraphics[angle=270,width=0.65\linewidth]{rmpgif/poves_fig17b.ps}
  \caption{Strength functions convoluted with gaussians: Upper
    panel; 50 iterations, Bottom panel; 1000 iterations.
  \label{fig:smooth}} 
\end{figure}
As the Lanczos vector $|I\rangle$ is obtained by orthogonalizing
${\cal H}^I|0\rangle$ to all previous Lanczos vectors $|i\rangle,\; i<I$, the
tridiagonal matrix elements are linear combinations of the moments of
the strength distribution. Therefore the eigensolutions of the
$I\times I$ matrix define an approximate strength function
$S_I(E)=\sum_{i=1,I}\, \delta(E-E_i)\, \langle i|{\cal T}|0\rangle^2$,
whose first $2I-1$ moments are the exact ones. The eigenstates act as
``doorways'' whose strength will be split until they become exact
solutions for $I$ large enough, as illustrated in
Fig.~\ref{fig:sfevol} that retraces the fragmentation process of the
sum rule pivot; in this case  a $^{48}$Sc ``doorway'' obtained by applying
the Gamow-Teller operator to the $^{48}$Ca ground state. The term
doorway applies to vectors that have a physical meaning but are not
eigenstates.  After full convergence is achieved for all states in the
resonance region, the strength function has the aspect shown at the
bottom of the figure. In practice, all the spikes are affected by an
experimental width, and assuming a perfect calculation, the observed
profile would have the aspect shown at the bottom of
Fig.~\ref{fig:smooth}, after convoluting with gaussians of 150 keV
width. The upper panel shows the result for 50 iterations, and 250 keV
widths for the non converged states. The profiles become almost
identical.

\section{The $\bm{0\hbar\omega }$ calculations}
\label{sec:bm0hb--calc} 
In this section we first revisit the $p$, $sd$, and $pf$ shells to
explain how a 3b monopole mechanism solves problems hitherto
intractable. Then we propose a sample of $pf$ shell results that will
not be discussed elsewhere. Finally Gamow Teller transitions and
strength functions are examined in some detail.

\subsection{The monopole problem and the three-body
  interaction}\label{sec:monop-probl-three}

The determinant influence of the monopole interaction was first
established through the mechanism of~\citet{Bansal.French:1964}. Its
efficiency in cross-shell calculations was further confirmed
by~\citet{Zamick:1965}, and the success of the ZBM
model~\citep*{Zuker.Buck.McGrory:1968} is implicitly due to a monopole
correction to a realistic force. \citet{Zuker:1969} identified the
main shortcoming of the model as due to what must be now accepted as a
3b effect\footnote{The ZBM model describes the region around $^{16}$O
  through a $p_{1/2} s_{1/2} d_{5/2}$ space. The French Bansal
  parameters $b_{pd}$ and $b_{ps}$ (see Eq.~(\ref{eq:ab})) must change
  when going from $^{14}$N to $^{16}$O; which demands a 3b mechanism}.
Trouble showed in 0\hw calculations somewhat later simply because it
takes larger matrices to detect it in the $sd$ shell than in the ZBM
space (dimensionalities of 600 against 100 for six particles). Up to 5
particles the results of~\citet{Halbert.Mcgrory.ea:1971} with a
realistic interaction were quite good, but at $^{22}$Na they were so
bad that they became the standard example of the unreliability of the
realistic forces~\cite{Brown.Wildenthal:1988} and lead to the
titanic\footnote{It took two years on a Vax. Nowadays it would take
  an afternoon on a laptop.} Universal SD (USD) fit of the 63 matrix
elements in the shell by~\citet{Wildenthal:84}.

Though the $pf$ shell demands much larger dimensionalities, it has the
advantage of containing two doubly magic nuclei $^{48}$Ca and
$^{56}$Ni for which truncated calculations proved sufficient to
identify very early the core of the monopole problem: The failure to
produce EI closures.

\begin{figure}[htb]
  \includegraphics[width=0.9\linewidth]{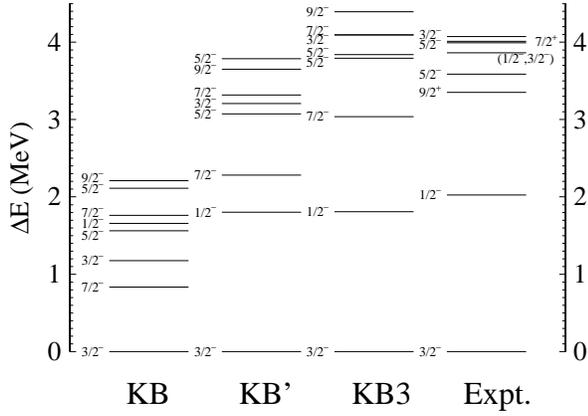}
\caption{The spectrum of $^{49}$Ca for the KB, KB' and KB3
interactions compared to experiment\label{fig:ca49el}}
\end{figure}

Fig.~\ref{fig:ca49el} gives an idea of what happens in $^{49}$Ca with the
KB interaction~\cite{Kuo.Brown:1968}: there are six states below 3
MeV, where only one
exists. In~\citet{Pasquini:1976} and \citet{Pasquini.Zuker:1978} the
following modifications were proposed      
($f\equiv f_{7/2},\, r\equiv f_{5/2},\,
p_{3/2},\, p_{1/2}$),
\begin{equation}
  \begin{array}{l}
V_{fr}^T(\text{KB1})=V_{fr}^T(\text{KB})-(-)^T\,300 \text{ keV},\\
V_{ff}^0(\text{KB1}) = V_{ff}^0(\text{KB})-350 \text{ keV},\\
V_{ff}^1(\text{KB1})= V_{ff}^1(\text{KB})-110 \text{ keV}.
  \end{array}
  \label{eq:kb1}
\end{equation}
The first line defines KB' in~Fig.~\ref{fig:ca49el}.  The variants KB2
and KB3 in~\citet{Poves.Zuker:1981a} keep the KB1 centroids and
introduce very minor multipole modifications. KB3 was adopted as
standard\footnote{The multipole changes---that were beneficial in the
  perturbative treatment of~\citet{Poves.Zuker:1981a}---had much less
  influence in the exact diagonalizations. } in successful
calculations in $A=47$-50 that will be described in the next
sections~\citep{Caurier.Zuker.ea:1994,Martinez-Pinedo.Poves.ea:1996a,
  Martinez-Pinedo.Zuker.ea:1997}.

\begin{figure}[htb]
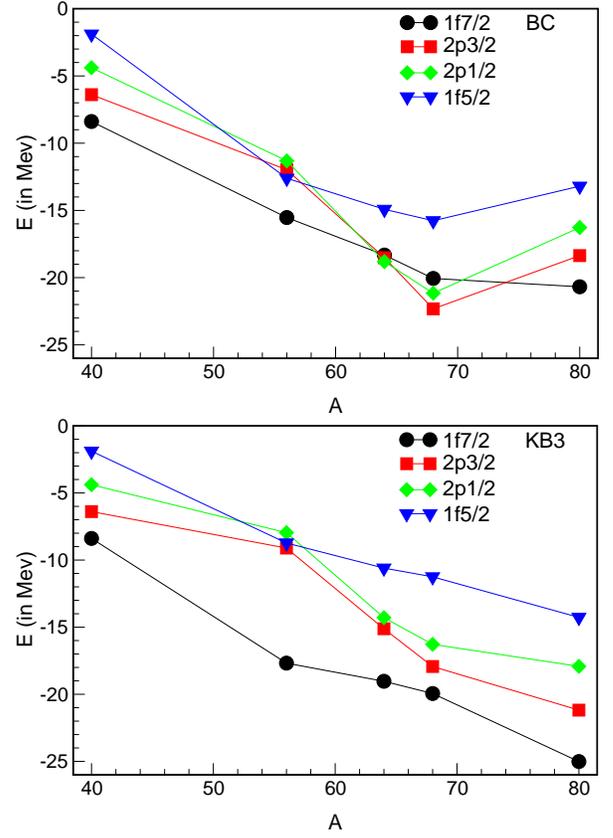

  \includegraphics[width=0.9\linewidth]{rmpgif/poves_fig19a.eps}\\
  \includegraphics[width=0.9\linewidth]{rmpgif/poves_fig19b.eps}
\caption{Effective single particle energies in the $pf$-shell along
  the $N=Z$ line, computed with the BC and KB3
  interactions.\label{fig:espe.nz}} 
\end{figure}

For higher masses, there are some problems, but nothing comparable to
the serious ones encountered in the $sd$ shell, where modifications
such as in Eq.~(\ref{eq:kb1}) are beneficial but (apparently)
insufficient.  The effective single particle energies in
Fig.~\ref{fig:espe.nz} are just the monopole values of the particle
and hole states  at the subshell closures.
 They give an idea of what happens in a 2b
description. At the origin, in $^{41}$Ca, the spectrum is the
experimental one. At $^{57}$Ni there is a bunching of the upper orbits
that the realistic BonnC (BC) and KB describe reasonably well, but they fail
to produce a substantial gap. Hence the need of KB1-type corrections.
At the end of the shell BC and KB reproduce an expanded version of the
$^{41}$Ca spectrum, which is most certainly incorrect. The indication
from $\tilde{\cal H}_m^d$ in Eq.~(\ref{eq:hmdz}) is that the bunching
of the upper orbits should persist. By incorporating this hint---which
involves the $V_{rr'}$ centroids---and fine-tuning the $V_{fr}$
ones~\citet{Poves.Sanchez-Solano.ea:2001} defined a KB3G interaction
that brings interesting improvements over KB3 in $A=50$-52, but around
$^{56}$Ni there are still some problems, though not as severe as the
ones encountered in the $sd$ and $p$ shells. Compared to KB3, the
interaction FPD6 has a better gap in $^{56}$Ni, however, the orbit
1f$_{5/2}$ is definitely too low in $^{57}$Ni. This produces problems
with the description of Gamow-Teller processes \citep[see][for a
recent experimental check in the beta decay of $^{56}$Cu]{Borcea.Aeystoe.ea:2001}.

The classic fits of~\citet{Cohen.Kurath:1965} (CK) defined state of
the art in the $p$ shell for a long time. As they preceded the
realistic G-matrices, they also contributed to hide the fact that the
latter produce in $^{10}$B a catastrophe parallel to the one in
$^{22}$Na. The work of ~\citet{Navratil.Ormand:2002},
and~\citet*{Pieper.Varga.Wiringa:2002} acted as a powerful reminder
that brought to the fore the 3b nature of the discrepancies.

Once this is understood, the solution follows: Eq.~(\ref{eq:kb1}) is
assumed to be basically sound but the corrections are taken to be
linear in the total number of valence particles $m$ (the simplest form
that a 3b term can take). 

\begin{figure}[htb]
  \includegraphics[width=0.9\linewidth]{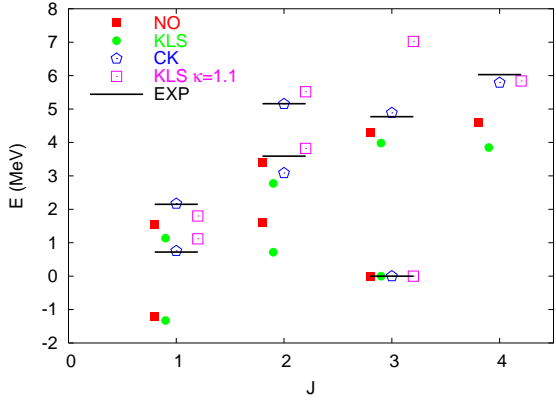}
  \caption{The excitation spectrum of $^{10}$B  for different
    interactions. \label{fig:b10}}
\end{figure}

Using $f\equiv (p_{3/2},\, d_{5/2},
\, f_{7/2}$) generically in the ($p,\, sd,\, pf$) shells respectively,
and $r= p_{1/2}$ and $r\equiv d_{3/2},s_{1/2}$ for the $p$ and $sd$
shells,~\citet{zuker:2003} proposes ($\kappa=\kappa_0+(m-m_0)\,
\kappa_1$)

\begin{equation}
  \begin{array}{l}
V_{fr}^T(\text{R})\Longrightarrow V_{fr}^T(\text{R})-(-)^T\,\kappa\\
V_{ff}^T(\text{R})\Longrightarrow V_{ff}^T(\text{R})-1.5\,\kappa \, \delta_{T_0}
  \end{array}
  \label{eq:R1}
\end{equation}
where R stands for any realistic 2b potential. The  results for
$^{10}$B are in Fig.~\ref{fig:b10}. The black squares (NO) are from
\cite[][Fig 4, 6$\hbar\Omega$]{Navratil.Ormand:2002}.  

\begin{figure}[htb]
  \includegraphics[width=0.9\linewidth]{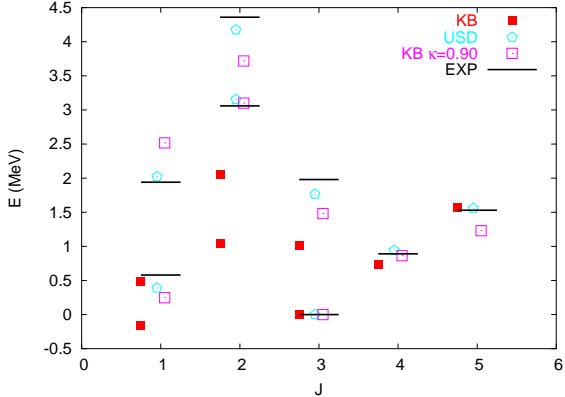}
  \caption{The excitation spectrum of $^{22}$Na for different
    interactions. \label{fig:na22}}
\end{figure}

The black circles correspond to the bare KLS G-matrix (\hw=17 MeV),
the white squares to the same with $\kappa=1.1$, and the pentagons to
the CK fit. NO and KLS give quite similar spectra, as expected from
the discussion in Sections~\ref{sec:effective} and
\ref{sec:univ-real-inter}. The $\kappa$ correction eliminates the
severe discrepancies with experiment and give values close to CK.

\begin{figure}[htb]
   \includegraphics[width=0.9\linewidth]{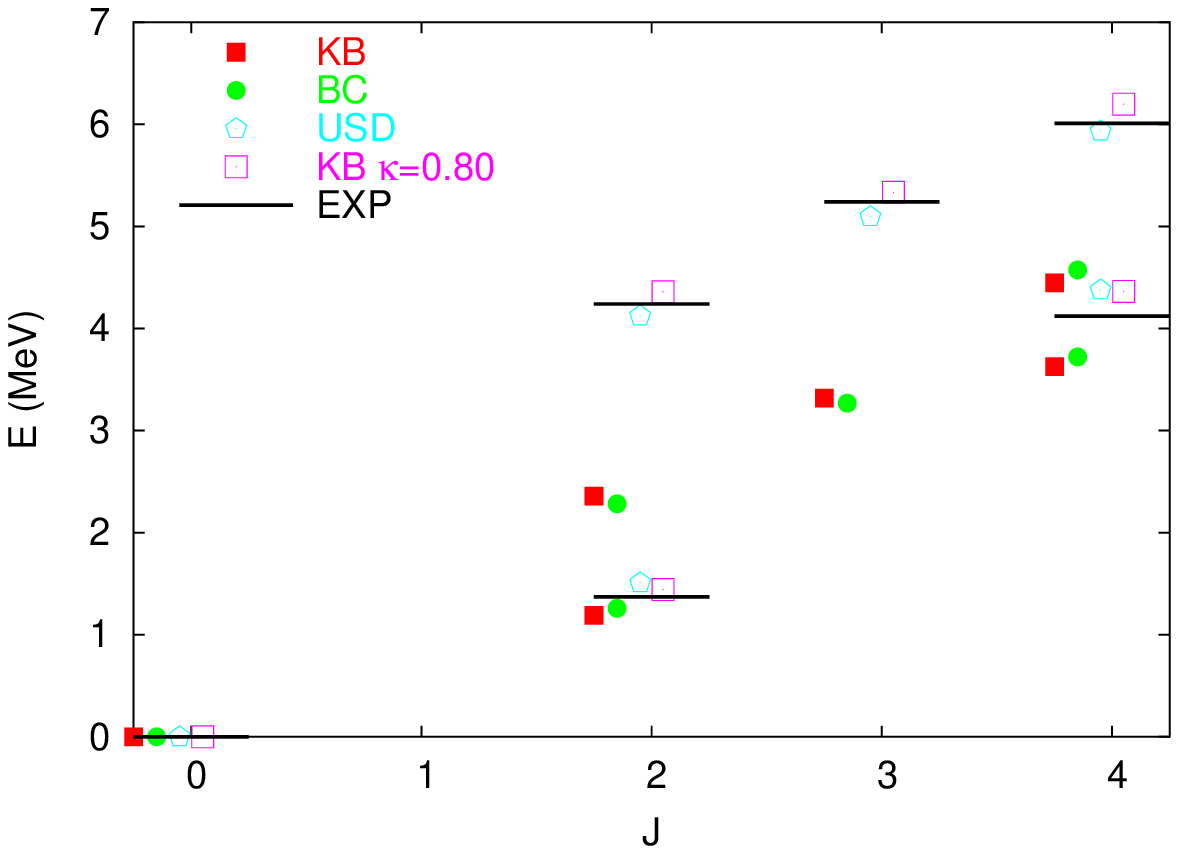}
   \includegraphics[width=0.9\linewidth]{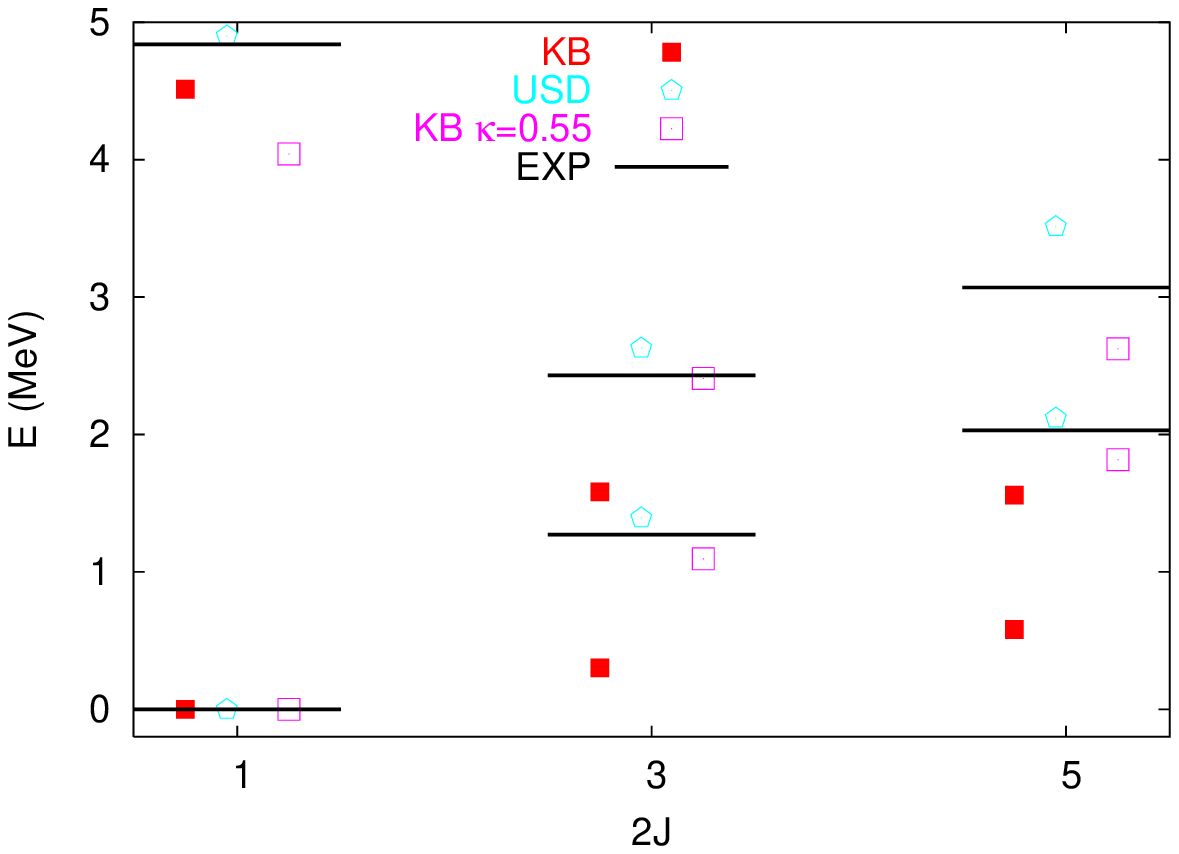}
  \caption{The excitation spectra of $^{24}$Mg and $^{29}$Si for different
  interactions. \label{fig:mg24si29}}
\end{figure}

In $^{22}$Na the story repeats itself: BC and KB are very close to one
another, the ground state spin is again $J=1$ instead of $J=3$ and the
whole spectrum is awful. The $\kappa$ correction restores the levels
to nearly correct positions, though USD still gives a better fit. This
simple cure was not discovered earlier because $\kappa$ {\em is not a
  constant}. In $^{24}$Mg we could still do with the value for
$^{22}$Na but around $^{28}$Si it must be substantially smaller. The
spectra in Fig.~\ref{fig:mg24si29} are obtained with
$\kappa(m)=0.9-0.05(m-6)$, and now the spectra  are
as good as those given by USD.

To determine a genuine 3b effect ({\it i.e.}, the linearity of $\kappa$) a
sufficiently large span of $A$ values is necessary. In the $p$ shell,
corrections to $V_{rr'}$ become indispensable very soon and must be
fitted simultaneously, as done successfully
by~\citet{Abzouzi.Caurier.Zuker:1991}, so we do not dwell on the
subject. As we have seen, in the $pf$ shell KB3, which is nearly
perfect in $A=47$-50, must be turned into KB3G for $A=50$-52, which
also does well at the lower masses. To find a real problem with a 2b
$R$-interaction ({\it i.e.}, compatible with $NN$ data)\footnote{It may be
  possible to fit the data with a purely 2b set of matrix elements:
  USD is the prime example, but it is
  $R$-incompatible~\cite{Dufour.Zuker:1996}.}  we have to move to
$^{56,58}$Ni. In particular the first $B(E2)(2\longrightarrow 0)$
transition in $^{58}$Ni falls short of the observed value (140
e$^2$fm$^4$) by a factor $\approx 0.4$ with {\em any monopole
  corrected KB interaction}. The problem can be traced to weak
quadrupole strength and it is not serious: as explained in
Section~\ref{sec:univ-real-inter} it is due to a normalization
uncertainty, and Table~\ref{tab:ovklsbonn} shows that with equal
normalizations BC and KB in the $pf$ shell are as close as in the
$sd$ shell.

In~\cite{zuker:2003} BC was adopted. Small modifications of the
$V_{fr}$ matrix were made to improve the (already reasonable)
$r$-spectrum in $^{57}$Ni (see Fig.~\ref{fig:espe.nz}). The particular
mixture in Eq.~(\ref{eq:R1}) was actually chosen to make possible, in
the simplest way, a good gap in $^{48}$Ca and a good single-particle
spectrum in $^{49}$Sc. It was found that for $A=48,\ 
\kappa(m=8)\approx 0.43$. For $A=56$, truncated calculations yielded
$\kappa(16)\lessapprox 0.28$. The B(E2)$(2\rightarrow 0)$ in
$^{58}$Ni indicated convergence to the right value. 

\begin{figure}[htb]
   \includegraphics[width=0.9\linewidth]{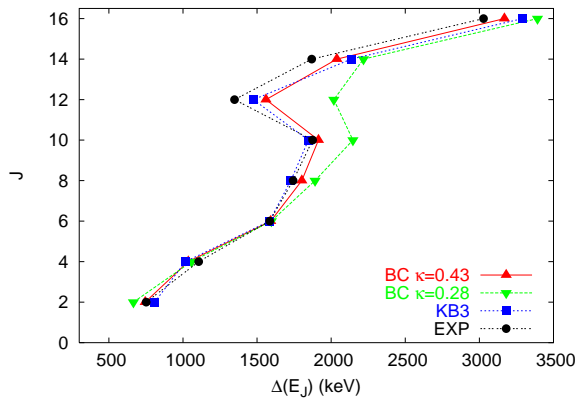}
    \caption{Backbending in $^{48}$Cr. See text. \label{fig:bb}}
\end{figure} 
With $\kappa(m=8)$, BC reproduces the yrast spectrum of $^{48}$Cr
almost as well as KB3, but not with $\kappa(m=16)$
(Fig.~\ref{fig:bb}), indicating that the three-body drift is
needed, though not as urgently as in the $sd$ shell. Another
indication comes from the T=0 spectrum of $^{46}$V, the counterpart of
$^{22}$Na and $^{10}$B in the $pf$-shell. The realistic interactions
place again the J=1 and  J=3 states in the wrong order, the correct
one is re-established by the three-body
monopole correction.

\subsection{The $pf$ shell}
\label{sec:miscelanea}

Systematic calculations of the A=47-52 isobars can be found in
\cite[$A=48$, KB3]{Caurier.Zuker.ea:1994}, \cite[$A=47$ and 49,
KB3]{Martinez-Pinedo.Zuker.ea:1997} and \cite[$A=50$-52,
KB3G]{Poves.Sanchez-Solano.ea:2001}.  Among the other full 0\hw
calculations let's highlight the following: The SMMC studies using
either the FPD6 interaction~\cite{Alhassid.Dean.ea:1994} or the KB3
interaction~\cite{Langanke.Dean.ea:1995}. A comparison of the exact
results with the SMMC can be found in
\cite{Caurier.Martinez-Pinedo.ea:1999b}. For a recollection of the
SMMC results in the $pf$ shell see also \cite{Koonin.Dean.ea:1997}.
The recent applications of the exponential extrapolation method
\cite{Horoi.Brown.Zelevinsky:2002,Horoi.Brown.Zelevinsky:2003}.  The
calculations of \citet{Novoselsky.Vallieres.ea:1997} for $^{51}$Sc and
$^{51}$Ti using the DUPSM code \citep[see also the erratum
in][]{Novoselsky.Vallieres.ea:1998} and for $^{52}$Sc and $^{52}$Ti
\cite{Novoselsky.Vallieres:1998}.  The extensive QMCD calculation of
the spectrum of $^{56}$Ni have been able to reproduce the exact result
for the ground state binding energy within 100$\sim$200~keV and to
give a fairly good description of the highly deformed excited band of
this doubly magic nucleus
\cite{Otsuka.Honma.ea:1998,Mizusaki.Otsuka.ea:1999,Mizusaki.Otsuka.ea:2002}.
Other applications to the $pf$ shell can be found in
\cite{Honma.Mizusaki.ea:1996}. The existence of excited collective
bands in the N=28 isotones is studied in
\cite{Mizusaki.Otsuka.ea:2001}.  Very recently, a new interaction for
the $pf$ shell (GXPF1) has been produced by a Tokyo-MSU collaboration
\cite{Honma.Otsuka.ea:2002}, following the fitting procedures that
lead to the USD interaction.  The fit starts with the G-matrix
obtained from the Bonn-C nucleon nucleon potential
\cite{Hjorth-Jensen.Kuo.Osnes:1995}. The fit privileges the upper part
of the $pf$ shell, as seen by the very large difference in single
particle energies between the 1f$_{7/2}$ and 12p$_{3/2}$ orbits (3~MeV
instead of the standard 2~MeV). In the calcium isotopes {\it i.e.},
when only the T=1 neutron neutron interaction is active) this new
interaction retains the tendency of the bare G-matrices (KB, Bonn-C,
KLS) to produce large gaps at N=32 and N=34, contrary to FPD6 or KB3G
that only predict a large gap at N=32. Some early spectroscopic
applications of GXPF1 to the heavy isotopes of Titanium, Vanadium and
Chromium can be found in~\cite{Janssens.Fornal.ea:2002} and
\cite{Mantica.Morton.ea:2003}, where the N=32-34 gaps are explored.

Each nucleus has its
interest, sometimes anecdotic, sometimes fundamental.  The agreement
with experiment for the energies, and for the quadrupole and magnetic
moments and transitions is consistently good, often excellent. These
results have been conclusive in establishing the soundness of the
minimally-monopole-modified realistic interaction(s).  There is no
point in reproducing them here, and we only present a few typical
examples concerning spectroscopic factors, isospin non-conserving
forces and  ``pure spectroscopy''.

\subsubsection{Spectroscopic factors}
\label{sec:spec}

\begin{figure}[htbp]
  \includegraphics[width=0.9\linewidth]{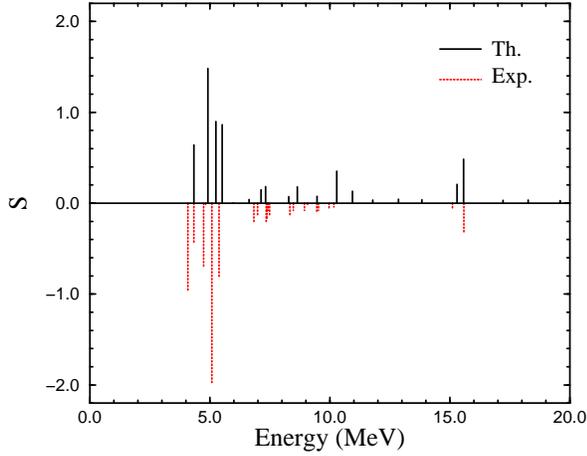}
    \caption{Spectroscopic factors, $(2j+1) S(j,t_z)$, corresponding
      to stripping of a particle in the orbit
      1$f_{5/2}$~\cite{Martinez-Pinedo.Zuker.ea:1997}.\label{fig:spec_f5}}
\end{figure}

The basic tenet of the independent particle model is that addition of
a particle to a closed shell $a^{\dagger}_r|cs\rangle$ produces an eigenstate
of the $|cs+1\rangle$ system. Nowadays we know better: it produces a
doorway that will be fragmented. If we choose $|cs\rangle=|^{48}{\rm
  Ca}\rangle$ and $|cs+1\rangle=|^{49}{\rm Sc}\rangle$, the four $pf$
orbits provide the doorways. The lowest, $f_{7/2}$, leads to an almost
pure eigenstate. The middle ones, $p_{3/2,\, 1/2}$, are more
fragmented, but the lowest level still has most of the strength and
the fragments are scattered at higher energies. Fig.~\ref{fig:spec_f5}
shows what happens to the $f_{5/2}$ strength: it remains concentrated
on the doorway but splits locally. The same evolution with energy of
the quasi-particles (Landau's term for single particle doorways) was
later shown to occur generally in finite
systems~\cite{Altshuler.Gefenand.ea:1997}.

\subsubsection{Isospin non conserving forces}\label{sec:isosp-non-cons}
\begin{figure}[htb]
    \includegraphics[width=0.9\linewidth]{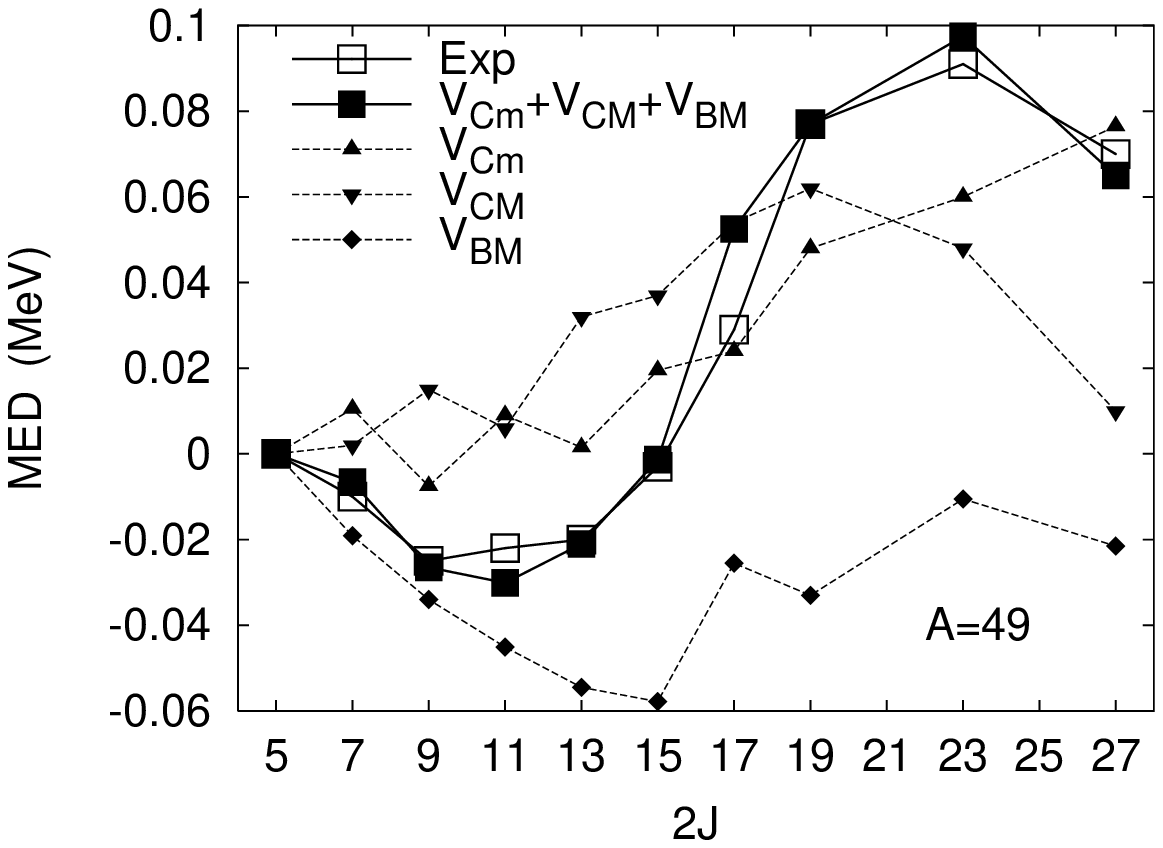}
    \caption {Experimental~\cite{OLeary.Bentley.ea:1997} and
      calculated~\cite{Zuker.Lenzi.ea:2002} MED for the pair
      $^{49}$Cr-$^{49}$Mn. \label{fig:tutti1}}
\end{figure}

Recent experiments have identified several yrast bands in mirror $pf$
nuclei~\cite{OLeary.Bentley.ea:1997,Bentley.OLeary.ea:1998,
Bentley.OLeary.ea:1999,Bentley.Williams.ea:2000,OLeary.Bentley.ea:2002,Lenzi.Marginean.ea:2001,
Brandolini.Sanchez-Solano.ea:2002}
for $A=47$, 49, 50 and 51. 
The naive view that the Coulomb energy should account for the MDE
(mirror energy differences) turns out to be
untenable. The four pairs were analyzed in~\cite{Zuker.Lenzi.ea:2002}
where it was shown that three
effects should be taken into account. A typical result is proposed in
Fig.~\ref{fig:tutti1}, where $V_{CM}$ and $V_{BM}$ stand for Coulomb
an nuclear isospin breaking multipole contributions, while $V_{Cm}$ is
a monopole Coulomb term generated by small differences of radii
between the members of the yrast band. The way these disparate
contributions add to
 reproduce the observed
pattern is striking.

Another calculation of the isospin non conserving effects,
and their influence in the location of the proton drip line,
 is due to \cite{Ormand:1997}. He has also analyzed the A=46 isospin
 triplet in \cite{Garrett.Ormand.ea:2001}.

\subsubsection{Pure spectroscopy}\label{sec:pure-spectroscopy}

The first nucleus we have chosen is also one that has been measured to
complete the mirror band~\cite{Bentley.Williams.ea:2000} in $A=51$,
whose MED are as well described as those of $A=49$ in
Fig.~\ref{fig:tutti1}. Such calculations need very good wavefunctions,
and the standard test they have to pass is the ``purely
spectroscopic'' one.

Work on the subject usually starts with a litany. Such as: Quadrupole
effective charges for neutrons q$_{\nu}$=0.5 and protons q$_{\pi}$=1.5 and bare $g$-factors
$g^s_{\pi}$=5.5857~$\mu_N$, $g^s_{\nu}$=-3.3826~$\mu_N$,
$g^l_{\pi}$=1.0~$\mu_N$ and $g^l_{\nu}$=0.0~$\mu_N$ in $M1$
transitions and moments are used. Except when the M1 transitions are
fully dominated by the spin term, the use of effective g-factors does
not modify the results very much due to the compensation between the
spin and orbital modifications. 

\begin{figure}[htb]
  \includegraphics[width=0.9\linewidth]{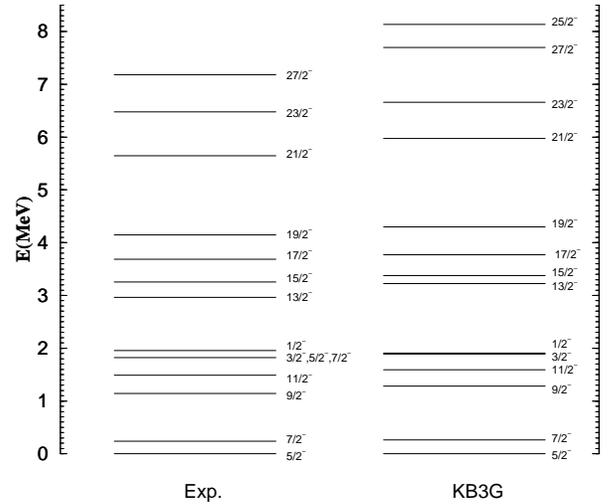}
  \caption{Yrast band of  $^{51}$Mn; experiment
    vs shell model calculation in the full $pf$-shell space.
    \label{fig:mn51}}
\end{figure}

Then some spectrum: The yrast band of $^{51}$Mn, calculated in the
full $pf$-shell space, is compared in Fig.~\ref{fig:mn51} with the
experimental data.  The first part of the test is passed. At most the
examiner will complain about slightly too high, high spin states.

Finally the transitions: The transitions in Table~\ref{tab:mn51} are
equally satisfactory.  Note the abrupt change in both $B(M1)$ and $B(E2)$
for $J=(17/2)^-$, beautifully reproduced by the calculation. The origin
of this isomerism is in the sudden alignment of two particles in the
1f$_{7/2}$ orbit, which provides an intuitive physical explanation for
the abrupt change in the MED~\cite{Bentley.Williams.ea:2000}.

\begin{table}[htb]
  \caption{Transitions in $^{51}$Mn.}
  \label{tab:mn51}
  \begin{tabular*}{\linewidth}{@{\extracolsep{\fill}}cccc}
    \hline\hline \noalign{\smallskip}
    & Exp. & \multicolumn{2}{c}{Th.}\\
    \hline\noalign{\smallskip}  
    $B(M1)$ & ($\mu_N^2$) & \multicolumn{2}{c}{($\mu_N^2$)} \\
$\frac{7}{2}^-\rightarrow \frac{5}{2}^-$ & 0.207(34)
&\multicolumn{2}{c}{0.177}\\
$\frac{9}{2}^-\rightarrow \frac{7}{2}^-$ & 0.16(5)
&\multicolumn{2}{c}{0.116}\\
$\frac{11}{2}^-\rightarrow \frac{9}{2}^-$ & 0.662(215)
&\multicolumn{2}{c}{0.421}\\
$\frac{17}{2}^-\rightarrow \frac{15}{2}^-$ & 0.00012(4) 
&\multicolumn{2}{c}{0.00003} \\
$\frac{19}{2}^-\rightarrow \frac{17}{2}^-$ & $>$0.572 
&\multicolumn{2}{c}{0.797} \\[0.8mm]
\hline\noalign{\smallskip}
$B(E2)$ & ($e^2\,fm^4$) & \multicolumn{2}{c}{($e^2\,fm^4$)} \\
$\frac{7}{2}^-\rightarrow \frac{5}{2}^-$ & 528(146) &
\multicolumn{2}{c}{305}
\\
$\frac{9}{2}^-\rightarrow \frac{5}{2}^-$ & 169(67) &
\multicolumn{2}{c}{84} \\ 
$\frac{9}{2}^-\rightarrow \frac{7}{2}^-$ & 303(112) &
\multicolumn{2}{c}{204} 
\\
$\frac{11}{2}^-\rightarrow \frac{7}{2}^-$ & 236(67) &
\multicolumn{2}{c}{154} 
\\
$\frac{11}{2}^-\rightarrow \frac{9}{2}^-$ & 232(75) &
\multicolumn{2}{c}{190}
 \\
$\frac{17}{2}^-\rightarrow \frac{13}{2}^-$ & 1.236(337) &
\multicolumn{2}{c}
{2.215}\\[2mm]
\hline\hline
\end{tabular*}
\end{table}

At the end one can add some extras: The electromagnetic moments of the
ground state are also known: Their values
$\mu_{exp}$=3.568(2)$\mu_{N}$ and
Q$_{exp}$=42(7)~efm$^2$~\cite{Firestone:1996} compare quite well with
the calculated $\mu_{th}$=3.397$\mu_{N}$ and Q$_{th}$=35~efm$^2$.

\begin{figure}[htb]
  \includegraphics[width=0.9\linewidth]{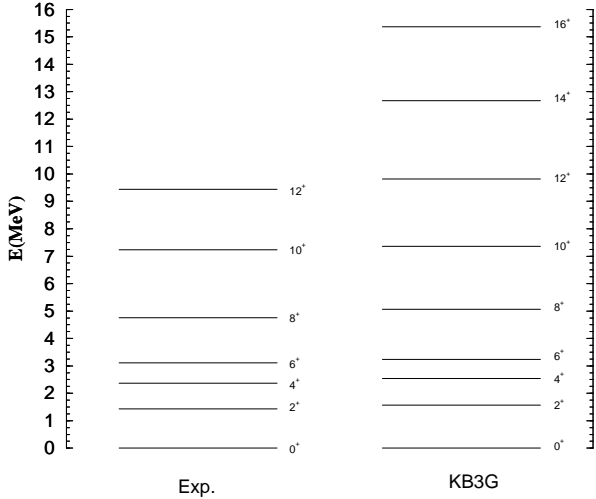}
    \caption{Yrast band of $^{52}$Cr; experiment
      $vs$ shell model calculation in the full $pf$-shell
      space. \label{fig:cr52}} 
\end{figure}

We complete this section with an even-even and an
odd-odd nucleus.  In figure~\ref{fig:cr52} we show the yrast states of
$^{52}$Cr, up to the band termination, calculated in the full
$pf$-shell space.  The agreement of the KB3G results with the
experiment is again excellent.

\begin{table}[htb]
\begin{center}
\caption{Transitions in $^{52}$Cr. $(a)$
  from~\protect\cite{Brown.Fossan.ea:1974}.}
\label{tab:cr52}
\vspace{0.3cm}
\begin{tabular*}{\linewidth}{@{\extracolsep{\fill}}cccc}
\hline\hline \noalign{\smallskip}
& Exp. & \multicolumn{2}{c}{Th.}\\
\hline\noalign{\smallskip}
$B(M1)$ & ($\mu_N^2$) & \multicolumn{2}{c}{($\mu_N^2$)} \\
$9^+\rightarrow 8^+$ & 0.057(38) &\multicolumn{2}{c}{0.040} \\[0.8mm]
\hline\noalign{\smallskip}
$B(E2)$ & ($e^2\,fm^4$) & \multicolumn{2}{c}{($e^2\,fm^4$)} \\
$2^+\rightarrow 0^+$ & 131(6) & \multicolumn{2}{c}{132} \\
$3^+\rightarrow 2^+$ & 7$^{+7}_{-5}$ & \multicolumn{2}{c}{5} \\
$4^+_1\rightarrow 2^+$ & 83(17)\,$^a$ & \multicolumn{2}{c}{107} \\
$4^+_2\rightarrow 2^+$ & 69(19) & \multicolumn{2}{c}{26} \\
%$4^+_2\rightarrow 4^+_1$ & 92$^{+37}_{-24}\,^a$ & \multicolumn{2}{c}{7} \\
$6^+\rightarrow 4^+$ & 59(2) & \multicolumn{2}{c}{68} \\
$8^+\rightarrow 6^+$ & 75(24) & \multicolumn{2}{c}{84} \\
$9^+\rightarrow 8^+$ & 0.5(20) & \multicolumn{2}{c}{0.6} \\[2mm]
\hline\hline
\end{tabular*}
\end{center}
\end{table}
The electromagnetic moments of the first $J=2^+$ state are known:
$\mu_{exp}$=2.41(13)~$\mu_{N}$ \cite{Speidel.Ernst.ea:2000} and
Q$_{exp}={\rm -8.2(16)~e~fm^2}$ \cite{Firestone:1996} and they are in
very good agreement with the theoretical predictions:
$\mu_{th}$=2.496~$\mu_{N}$ and Q$_{th}$=--9.4~e~fm$^2$ In
table~\ref{tab:cr52} the experimental values for the electromagnetic
transitions between the states of the yrast sequence are given.  The
calculated values are in very good correspondence with the experiment.

\begin{figure}[htb]
  \includegraphics[width=0.9\linewidth]{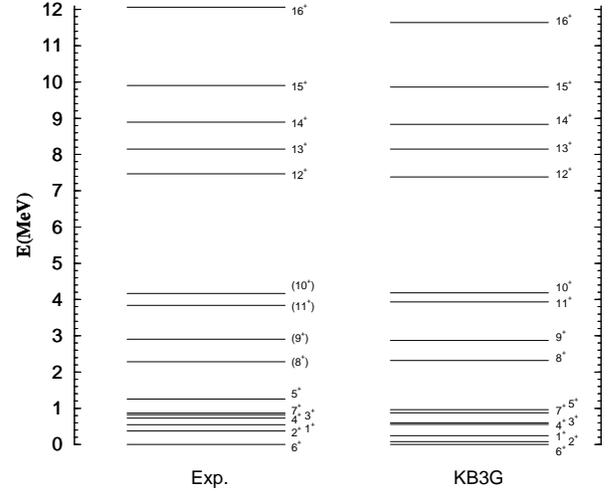}
    \caption{Yrast band of  $^{52}$Mn; experiment
    $vs$ shell model calculation in the full $pf$-shell space.
    \label{fig:mn52}} 
\end{figure}
The yrast states of the odd-odd nucleus $^{52}$Mn, calculated in the
full $pf$-shell, are displayed in fig.~\ref{fig:mn52}. Notice the perfect
correspondence between theory and experiment for the states belonging
to the multiplet below 1\,MeV. The experimental data for the spins
beyond the band termination ($11^+$ to $16^+$) come from a recent
experiment~\cite{Axiotis:2000}. The agreement between 
theory and experiment is spectacular.
There are only three experimental states with
spin assignment not drawn in the figure: a second $J=(5^+)$ at
1.42\,MeV, a third $J=(5^+)$ at 1.68\,MeV and a second $J=(6^+)$ at
1.96\,MeV. The calculation places the second $J=5^+$ at
1.37\,MeV, the third $J=5^+$ at 1.97\,MeV and the second $J=6^+$ at
1.91\,MeV.
 The experimental magnetic and quadrupole  moments for the  $6^+$ ground
state are also known; 
 $\mu_{exp}$=3.062(2)~$\mu_{N}$ 
 and Q$_{exp}$=+50(7)~e~fm$^2$ \cite{Firestone:1996}.
 The calculated values,
 $\mu_{th}$=2.952~$\mu_{N}$
 and Q$_{th}$=+50~e~fm$^2$  reproduce them nicely.
Some electromagnetic transitions along the
yrast sequence have been measured. The calculations are in very good
agreement with them (see table~\ref{tab:mn52}).
\begin{table}[htb]
\begin{center}
\caption{Transitions in $^{52}$Mn.}
\label{tab:mn52}
\vspace{0.3cm}
\begin{tabular*}{\linewidth}{@{\extracolsep{\fill}}cccc}
\hline\hline\noalign{\smallskip}
& Exp. & \multicolumn{2}{c}{Th.}\\
\hline\noalign{\smallskip}
$B(M1)$ & ($\mu_N^2$) & \multicolumn{2}{c}{($\mu_N^2$)}\\
$7^+\rightarrow 6^+$ & 0.501(251) & \multicolumn{2}{c}{0.667} \\
$8^+\rightarrow 7^+$ & $>$0.015 & \multicolumn{2}{c}{0.405} \\
$9^+\rightarrow 8^+$ & 1.074$^{+3.043}_{-0.537}$  
& \multicolumn{2}{c}{0.759} \\[0.8mm]
\hline\noalign{\smallskip}
$B(E2)$ & ($e^2\,fm^4$) & \multicolumn{2}{c}{($e^2\,fm^4$)} \\
$7^+\rightarrow 6^+$ & 92$^{+484}_{-81}$ & \multicolumn{2}{c}{126} \\
$8^+\rightarrow 6^+$ & $>$1.15 & \multicolumn{2}{c}{33} \\
$8^+\rightarrow 7^+$ & $>$4.15 & \multicolumn{2}{c}{126} \\
$9^+\rightarrow 7^+$ & 104$^{+300}_{-46}$ & \multicolumn{2}{c}{66} \\
$11^+\rightarrow 9^+$ & 54(6) & \multicolumn{2}{c}{53}\\[2mm]
\hline\hline
\end{tabular*}
\end{center}
\end{table}

\subsection{Gamow Teller and magnetic dipole strength.}
\label{sec:GT-M1}
Out of the approximately 2500 known nuclei that are bound with respect
to nucleon emission, only 253 are stable. The large majority of the
rest decay by $\beta$ emission or electron capture, mediated by the
weak interaction. When protons and neutrons occupy the same orbits, as
in our case, the dominant processes are allowed Fermi and Gamow-Teller
transitions. The information obtained from the weak decays has been
complemented by the $(p,n)$ and $(n,p)$ reactions in forward
kinematics, that make it possible to obtain total GT strengths and
strength functions that cannot be accessed by the decay data because
of the limitations due to the Q$_{\beta}$ windows.  From a theoretical
point of view, the comparison of calculated and observed strength
functions provide invaluable insight into the meaning of the valence
space and the nature of the deep correlations detected by the
``quenching'' effect.

The half-life for a transition between two nuclear states is given
by~\cite{Behrens.Buehring:1982,Schopper:1966}: 
\begin{equation}
 \label{eq:ft}
 (f_A+f^\epsilon)t=\frac{6144.4\pm1.6}{(f_V/f_A)B(F)+B(GT)}.
\end{equation}
The value $6144.4\pm1.6$ is obtained from the nine best-known
superallowed beta decays~\cite{Towner.Hardy:2002} \citep[see][for an
alternative study]{Wilkinson:2002a,Wilkinson:2002b}. $f_V$ and $f_A$
are the Fermi and Gamow-Teller phase-space factors,
respectively~\cite{Wilkinson.Macefield:1974,Chou.Warburton.Brown:1993}.
$f^\epsilon$ is the phase space for electron
capture~\cite{Bambynek.Behrens.ea:1977}, that is only present in
$\beta^+$ decays,
If $t_{1/2}$ is the total lifetime, the partial lifetime of a level
with branching ratio $b_r$ is $t=t_{1/2}/{b_r}$.

$B(F)$ and $B(GT)$ are defined as
\begin{subequations}
\begin{gather}
      \label{eq:fermidef}
     B(F)=\left[\frac{\langle f \parallel \sum_k
      \bm{t}^k_\pm \parallel i\rangle}{\sqrt{2J_i+1}}\right]^2\\ 
\label{eq:gamowteller}
  B(GT)=\left[\left(\frac{g_A}{g_V}\right)   
\frac{\langle f\parallel \sum_k \bm{\sigma}^k \bm{t}^k_\pm
  \parallel i\rangle}{\sqrt{2J_i+1}}\right]^2. 
\end{gather}
\end{subequations}
Matrix elements are reduced~(\ref{(I.26)}) with respect to spin only,
$\pm$ refers to $\beta^\pm$ decay, $\sigma=2S$ and $(g_A/g_V)=
-1.2720(18)$~\cite{Hagiwara.Hikasa.ea:2002} is the ratio of the weak
interaction axial-vector and vector coupling constants.

For states of good isospin the value of $B(F)$ is fixed. It can be
altered only by a small isospin-symmetry-breaking $\delta_C$
correction   
\cite{Towner.Hardy:2002}. Shell model estimates of this quantity can
be also found in \cite{Ormand.Brown:1995}.
\begin{equation}
  \label{eq:fermiiso}
  B(F)=\left[T(T+1)-T_{z_i} T_{z_f}\right] \delta_{if} (1-\delta_C),
\end{equation}
where $\delta_{if}$ allows only transitions between isobaric analog
states. Superallowed decays may shed light on the departures from
unitarity of the Cabibbo Kobayashi Maskawa matrix.

The total strengths $S_\pm$ are related by the sum rules
\begin{subequations}
  \label{eq:sumrules}
  \begin{gather}
    \label{eq:sfermi}
    S_-(F)-S_+(F)=N-Z,\\
    \label{eq:ikeda}
    S_-(GT)-S_+(GT)=3(N-Z),
  \end{gather}
\end{subequations}
 where $N$ and $Z$ refer to the initial state and $S_{\pm}(GT)$ does not
 contain the $g_A/g_V$ factor. The comparison of the 
 $(p,n)$ and $(n,p)$ data with the Gamow Teller sum rule
 \cite{Ikeda.Fujii.Fujita:1963} led to 
  the
 long standing ``quenching'' problem; only approximately one half of
 the sum rule value was found in the experiments.

The GT strength is not protected by a conservation principle and
depends critically on the wavefunctions used.  Full $0\hbar\omega$
calculations already show a large quenching with respect to the
independent particle limit as seen in Table~\ref{tab:gtstrength} where
the result of a full $pf$ calculation is compared with that obtained
with an uncorrelated Slater determinant having the same occupancies:
\begin{equation}
  \label{eq:uncorrelated}
  S=\sum_{i,k} \frac{n^p_i n^h_k}{(2j_i+1) (2j_k+1)} \langle i
  ||\sigma||k\rangle^2 
\end{equation}
the sum on $i$ runs over the proton (neutron) orbits in the valence
space and $k$ over the proton (neutron) orbits for $S_+$ ($S_-$).
$n^p$ and $n^h$ denote the number of particles and holes,
respectively. The determinantal state demands a quenching factor that
is almost twice as large as the standard quenching factor
Q$^2=(0.74)^2$) that brings the full calculation in line with
experiment.

\begin{table}[htbp]
  \caption{Comparison of GT$_+$ strengths. For $^{54}$Fe, $^{55}$Mn,
    $^{58}$Ni and $^{59}$Co the calculations are truncated to  $t=8$,
    $t=4$, $t=6$ and $t=4$ respectively. The data are
    from~\cite{Williams.Alford.ea:1995,El-Kateb.Jackson.ea:1994,%
    Alford.Brown.ea:1993,Vetterli.Haeusser.ea:1990}.
    \label{tab:gtstrength}}  
\vspace{0.2cm}
  \begin{tabular}{lcccc}
    \hline \hline 
    Nucleus & Uncorrelated & \multicolumn{2}{c}{Correlated} &
    Expt. \\ \cline{3-4}
    & & Unquenched & $Q=0.74$ & \\
    \hline \noalign{\smallskip}
 %   $^{48}$Ti & 4.16 & 1.21 & 0.66 & $1.19\pm20$\\
    $^{51}$V & 5.15 & 2.42 & 1.33 & $1.2\pm0.1$\\
    $^{54}$Fe & 10.19 & 5.98 & 3.27 & $3.3\pm 0.5$\\
    $^{55}$Mn & 7.96 & 3.64 & 1.99 & $1.7\pm 0.2$\\
    $^{56}$Fe & 9.44 & 4.38 & 2.40 & $2.8\pm 0.3$\\
    $^{58}$Ni & 11.9 & 7.24 & 3.97 & $3.8\pm0.4$\\
    $^{59}$Co & 8.52 & 3.98 & 2.18 & $1.9\pm0.1$\\
    $^{62}$Ni & 7.83 & 3.65 & 2.00 & $2.5\pm0.1$\\
    \hline \hline
  \end{tabular}
\end{table}

\subsubsection{The meaning of the valence space}\label{meaning-val-space}

\begin{figure}[htb]
  \includegraphics[width=0.9\linewidth]{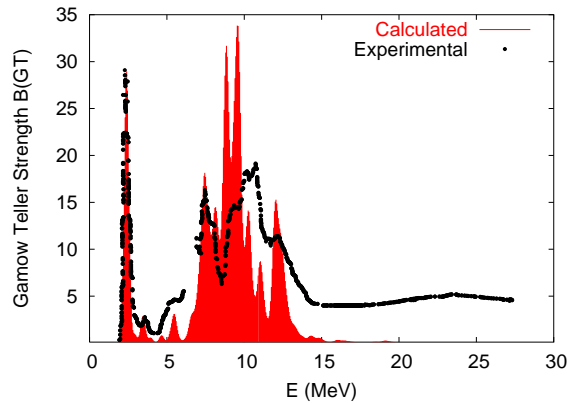}
    \caption{Gamow Teller strength in $^{48}$Ca$(p,n)^{48}$Sc
      from~\cite{Anderson.Lebo.ea:1990}---after elimination of the
      Fermi peak at around 6 MeV---compared with the calculated peaks
      (KB3 interaction) after 700 Lanczos
      iterations~\cite{Caurier.Poves.Zuker:1995}. The peaks have
      been smoothed by gaussians having the instrumental width of the
      first measured level.}
    \label{fig:gt48}
\end{figure}

Before moving to the explanation of the quenching of the GT
strength, it is convenient to recall the meaning of the valence space
as discussed in Section~\ref{sec:theory-calculations}. The
$^{48}$Ca$(p,n)^{48}$Sc reaction~\cite{Anderson.Lebo.ea:1990} provides
an excellent example. In Fig.~\ref{fig:gt48}, adapted from
\citet{Caurier.Poves.Zuker:1995}, the experimental data are compared
with the strength function produced by a calculation in the full
$pf$-shell using the interaction KB3.
The peaks have all $J=1$, $T=3$. In the $pf$ shell there are 8590 of
them and the calculation has been pushed to 700 iterations in the Lanczos strength
function 
 to ensure fully converged
eigenstates below 11 MeV. Of these eigenstates 30 are below 8
MeV: They are at the right energy and have the right strength profile.
At higher energies the peaks are much too narrow compared with
experiment. It means that they may well be eigenstates of the
effective Hamiltonian in the $pf$ shell, but not eigenstates of the
full system. Therefore, they should be viewed as doorways, subject to
further mixing with the background of intruders which dominates the
level density after 8~MeV, as corroborated by the experimental tail
that contains only intruders and can be made to start naturally at
that energy.

The KB3 effective interaction  is
doing a very good job, but it is certainly not decoupling 8590 $pf$
states from the rest of the space. If the fact is not explicitly
recognized we end up with the often
raised~\cite{Hjorth-Jensen.Kuo.Osnes:1995} ``intruder problem'':
Decoupling cannot be enforced perturbatively when intruders are
energetically close to model states.  Figure~\ref{fig:gt48} indicates
that although few eigenstates are well decoupled, it is possible to
make sense of many others if one interprets them as doorways.
The satisfactory description of the lowest states indicates that the
model space makes sense. It means that it ensures good decoupling at
the $S_2$ level, or simply in second order perturbation
theory, which guarantee a state-independent interaction.  

Energetically we are in good shape, and we concentrate on the
renormalization of the GT operator: the calculated strength has been
quenched by a factor $\approx (0.74)^2$. Why? Why this factor ensures
the right detailed strength for about 30 states below 8 MeV? 

\subsubsection{Quenching}
\label{sec:quenching:-no-core}
To understand the quenching problem it is best to start by the
(tentative) solution. The dressed states in Eq.~\eqref{eq:decij} are
normalized to unity in the model space. This trick is essential in the
formulation of linked cluster or $\exp{S}$ theories (hence
Eq.~\eqref{eqn:w0}). It makes possible the calculation of the
energy---and some transitions, such as the $E2$---without knowledge
of the norm of the exact wavefunction. In general though, we need an
expectation value between exact, normalized states: $\langle \hat
f\parallel{\cal T}\parallel\hat i\rangle^2$. If we write
\begin{gather}
  \label{eq:ifgt}
|\hat i\rangle=\alpha |0\hbar\omega\rangle+\sum_{n\ne
 0}\beta_n|n\hbar\omega\rangle,   
\end{gather}
and a similar expression in $\alpha',\, \beta'$ for $\langle \hat f|$,
we find
\begin{gather}
  \label{eq:gtran}
\langle \hat f\parallel{\cal T}\parallel\hat i\rangle^2=
\left(\alpha\alpha'\, T_0+\sum_{n\ne 0} \beta_n\beta'_n\,T_n \right)^2,   
\end{gather}
since the GT operator does not couple states with different number of
\hw\ excitations.
If we make two assumptions: (a) neglect the $n\ne 0$ contributions;
(b) $\alpha\approx\alpha'$; it follows that if the projection of the
physical wavefunction in the 0\hw\ space is $Q\approx \alpha^2$,
its contribution to the
transition will be quenched by $Q^2$.

\begin{figure}[htbp]
  \includegraphics[width=\linewidth]{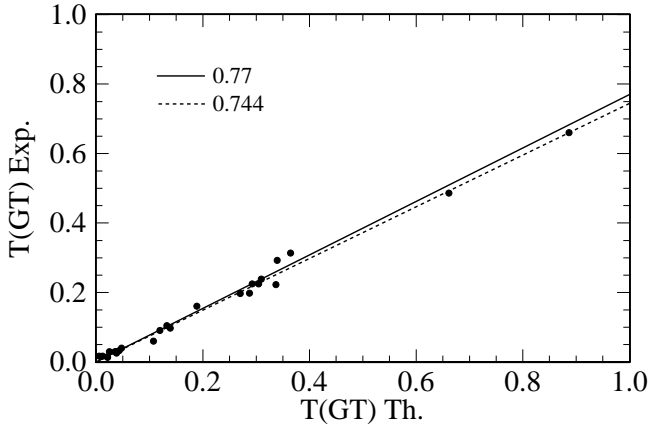}
  \caption{Comparison of the experimental and theoretical values of
    the quantity $T(GT)$ in the $pf$
    shell~\cite{Martinez-Pinedo.Poves.ea:1996a}. The $x$ and $y$
    coordinates correspond to theoretical and experimental values
    respectively. The dashed line shows the ``best-fit'' for
    $Q=0.744$. The solid line shows the result obtained in the
    $sd$-shell nuclei~\cite{Wildenthal.Curtin.Brown:1993}
      \label{fig:g-eff}} 

\end{figure}

Exactly the same arguments apply to transfer reactions---for which
${\cal T}=a_s$ (or $a_s^{\dagger}$)---but are simpler because $T_{n\ne 0}=0$.
The transition strength is given by the spectroscopic factor, which
can be identified with $Q$ when one particle is removed and $1-Q$ when
it is added. The assumption that the model amplitudes in the exact
wavefunctions are approximately constant is borne out by systematic
calculations of $Q$ in the $p$
shell~\cite[$Q=0.820(15)$]{Chou.Warburton.Brown:1993}, the $sd$
shell~\cite[$Q=0.77(2)$]{Wildenthal.Curtin.Brown:1993} and the $pf$
shell~\cite[illustrated in
Fig.~\ref{fig:g-eff},][$Q=0.744(15)$]{Martinez-Pinedo.Poves.ea:1996a}.

These numbers square well with the existing information on
spectroscopic factors from $(d,p)$~\cite[$Q\approx 0.7$]{Vold.ea:1978}
and $(e,e'p)$~\cite[$Q=0.7$]{Cavedon.ea.1982} data \citep[see
also][]{Pandharipande.Sick.ea:1997}. This consistency is significant
in that it backs assumption (a) above, which is trivially satisfied
for spectroscopic factors. It opens the perspective of accepting the
GT data as a measure of a very fundamental quantity that does not
depend on particular processes. 
The proposed ``solution'' to the quenching problem amounts to reading
data.

\begin{figure}[htb]
  \includegraphics[angle=90,width=0.9\linewidth]{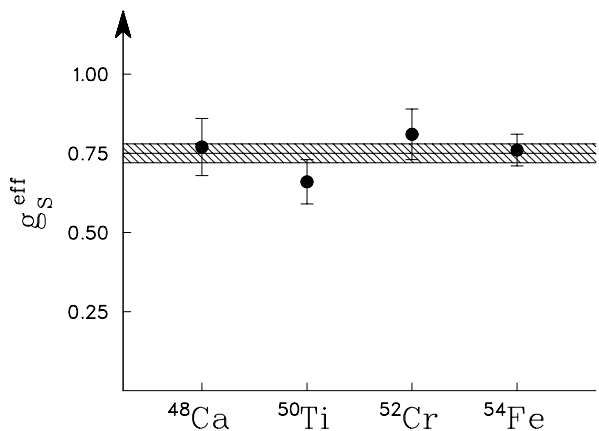}
  \caption{Effective spin g-factor of the M1 operator deduced from
    the comparison of shell model calculations and data for the total
    B(M1) strengths in the stable even mass N=28 isotones (from
    \citet{VonNeumann.Poves.ea:1998}) \label{fig:m1quen}}
\end{figure}

Experimentally, the challenge is
to locate all the strength, constrained by the Ikeda sum
rule that relates the direct and inverse processes.  The careful analysis
of~\citet{Anderson.Chittrakarn.ea:1985} suggests, but does not prove,
that the experimental tail in Fig.~\ref{fig:gt48} contains enough
strength to satisfy approximately the sum rule. A similar result is
obtained for $^{54}$Fe$(p,n)$~\cite{Anderson.Lebo.ea:1990}. More
recent experiments by the Tokyo group establish that the strength
located at accessible energies exhausts 90(5)\% of the sum rule in the
$^{90}$Zr$(n,p)$ and 84(5)\% in
$^{27}$Al$(p,n)$~\citep{Wakasa.Sakai.ea:1997,Wakasa.Sakai.ea:1998}.

The theoretical problem is to calculate $Q$. It has been compounded by
a sociological one: the full GT operator is $(g_A/g_V)\sigma\tau$,
where $g_A/g_V \approx -1.27$ is the ratio of weak axial and vector
coupling constants. The hotly debated question was whether $Q$ was due
to non nucleonic renormalization of $g_A$ or nuclear renormalization of
$\sigma\tau$~\cite{Osterfeld:1992,Arima:2001}. We have sketched above
the nuclear case, along the lines proposed
by~\citet*{Caurier.Poves.Zuker:1995}, but under a new guise that makes
it easier to understand.
The calculations of~\citet{Bertsch.Hamamoto:1982},
\citet{Drozdz.Klemt.ea:1986} and~\citet{Dang.Arima.ea:1997} manage to
place significant amounts of strength beyond the resonance region, but
they are based on 2p-2h doorways which fall somewhat short of giving
a satisfactory view of the strength functions. No-core calculations
are under way, that should be capable of clarifying the issue.

The purely nuclear origin of quenching is borne out by $(p,
p^{\prime})$, $(\gamma, \gamma^{\prime})$ and $(e, e^{\prime})$
experiments that determine the spin and convection currents in M1
transitions, in which $g_A/g_V$ play no role~\citep[see][for a
complete review]{Richter:1995}. An analysis of the data available for
the $N=28$ isotones in terms of full $pf$-shell calculations concluded
that agreement with experiment was achieved by quenching the
$\sigma\tau$ operator by a factor 0.75(2), fully consistent with the
value that explains the Gamow-Teller data
\cite{VonNeumann.Poves.ea:1998} (see Fig.~\ref{fig:m1quen}). 
These results rule out the hypothesis of a renormalization of
the axial-vector constant $g_A$: it is the $\sigma\tau$ operator that
is quenched.

\section{Spherical shell model description of nuclear rotations}
\label{sec:spher-shell-model}
Theoretical studies in the $pf$ shell came in layers. The first, by 
\citet*{McCullen.Bayman.Zamick:1964}, restricted to the $f_{7/2}$
space, was a success, but had some drawbacks: the spectra were not
always symmetric by interchange of particles and holes, and the
quadrupole moments had systematically the wrong sign. The first
diagonalizations in the full shell~[\citep{Pasquini:1976},
\citep{Pasquini.Zuker:1978}] solved these problems to a large extent,
but the very severe truncations necessary at the time made it
impossible to treat on the same footing the pairing and quadrupole
forces. The situation improved very much by dressing perturbatively
the pure two-body $f_{7/2}$ part of the Hamiltonian $H_2$, with a
three body term $H_{R1}$, mostly due to the quadrupole
force~\cite{Poves.Zuker:1981a}. The paper ended with the phrase:
  
``It may well happen, that in some cases, not in the $pf$ shell but
elsewhere, $H_{R1}$ will overwhelm $H_2$. Then, and we are only
speculating, we shall speak, perhaps, of the rotational coupling
scheme.''

Indeed, some nuclei were indicating a willingness to become rotational
but could not quite make it, simply because the perturbative treatment
was not enough for them. The authors missed prophecy by one extra
conditional: ``not in the $pf$ shell should have been ``not
necessarily...''

At the time it was thought impossible to describe rotational motion in
a spherical shell-model context. The glorious exception, discovered
by~\citet{Elliott:1958a,Elliott:1958b}, was (apparently) associated to
strict SU(3) symmetry, approximately realized only near $^{20}$Ne
and $^{24}$Mg.

\subsection{Rotors in the $pf$ shell}\label{sec:rotors-pf-shell} 

\begin{figure}[htb]
  \includegraphics[width=0.7\linewidth]{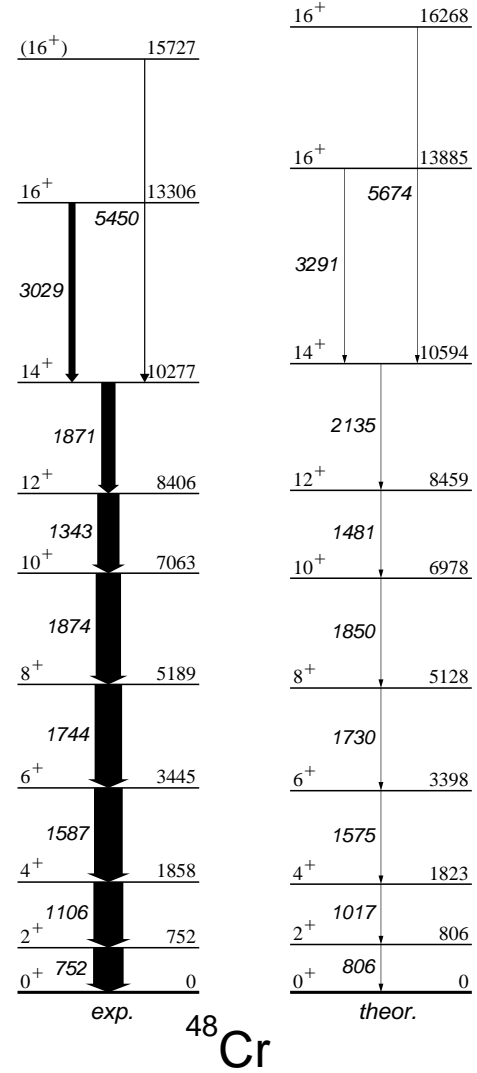}
  \caption{$^{48}$Cr level scheme; experiment $vs.$ theory
    \label{cr48exp}} 
\end{figure}

\begin{table}[htb]
\caption{$^{48}$Cr; quadrupole properties of the yrast band} 
 \begin{tabular*}{\linewidth}{@{\extracolsep{\fill}}cccccc}
   \hline\noalign{\smallskip} 
      J & B(E2)$_{exp}$ & B(E2)$_{th}$  & Q$_0$(t) &
       Q$_0$(s) &   Q$_0$(t)[f7/2,p3/2]\\
       \hline\noalign{\smallskip}
      2  &  321(41)  & 228 & 107 & 103 & 104 \\
      4  &  330(100) & 312 & 105 & 108 & 104 \\
      6  &  300(80)  & 311 & 100 &  99 & 103 \\
      8  &  220(60)  & 285 &  93 &  93 & 102 \\
     10  &  185(40)  & 201 &  77 &  52 &  98 \\
     12  &  170(25)  & 146 &  65 &  12 &  80 \\
     14  &  100(16)  & 115 &  55 &  13 &  50 \\
     16  &   37(6)   &  60 &  40  & 15 &  40 \\[5pt]
 \hline 
  \end{tabular*}
    \label{tab:cr48def}    
\end{table}

The fourth layer was started when the ANTOINE code
(Section~\ref{sec:antoine}) came into operation: $^{48}$Cr was
definitely a well deformed rotor~\cite{Caurier.Zuker.ea:1994}, as can
be surmised by comparing with the experimental spectrum from
~\citet{Lenzi.Napoli.ea:1996} in Fig.~\ref{cr48exp} ~\citep[see
also][]{Cameron.Bentley.ea:1993} and the transition properties in
Table~\ref{tab:cr48def} from ~\citet{Brandolini.Lenzi.ea:1998}
where we have used
\begin{gather}
\label{bmq}
 Q_0(s)=\frac{(J+1)\,(2J+3)}{3K^2-J(J+1)}\,Q_{spec}(J), \quad K\ne 1\\ 
 B(E2,J\;\rightarrow\;J-2)=\nonumber\\
\frac{5}{16\pi}\,e^2|\langle JK20|
 J-2,K\rangle |^2 \, Q_0(t)^2\quad K\ne 1/2,\, 1;
\label{bme2} 
\end{gather}
to establish the connection with the intrinsic frame descriptions: A
good rotor must have a nearly constant $Q_0$, which is the case up to
$J=10$, then $^{48}$Cr backbends. This is perhaps the most striking
result to come out of our $pf$ calculations, because such a behaviour
had been thought to occur only in much heavier nuclei.

\begin{figure}[htb]
  \includegraphics[width=0.9\linewidth]{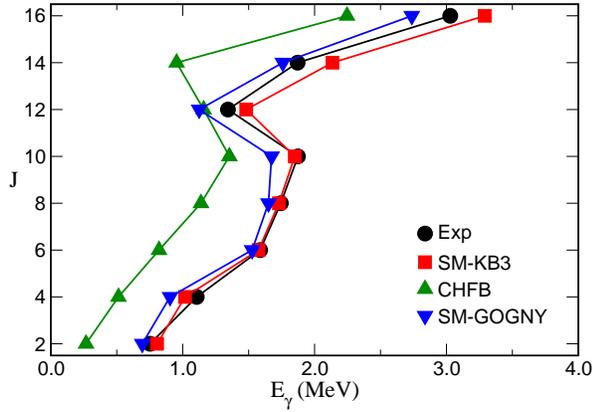}
  \caption{The yrast band of  $^{48}$Cr; experiment vs. the shell model
    calculations with KB3 and the Gogny force and the CHFB results
    also with the Gogny force.\label{fig:smgog}}
\end{figure}

\begin{figure}[htb]
  \includegraphics[width=0.9\linewidth]{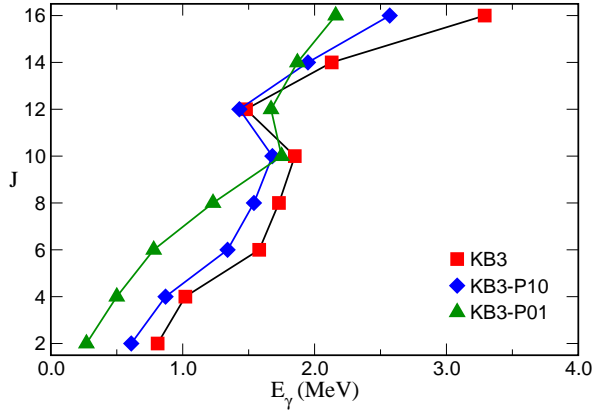}
  \caption{Gamma-ray energies along the yrast band of $^{48}$Cr (in
    MeV), full interaction (KB3), isoscalar pairing retired (KB3--P01)
    and isovector pairing retired (KB3--P01).\label{fig:cr48_2}}
\end{figure}

Fig.~\ref{fig:smgog} compares the experimental patterns with those
obtained with KB3, and with the Gogny force. The latter, when
diagonalized gives surprisingly good results. When treated in the
cranked Hartree Fock Bogoliubov (CHFB) approximation, the results are
not so good. The discrepancy is more apparent than real: The
predictions for the observables are very much the same in both cases
\citep[see][for the details]{Caurier.Egido.ea:1995}. The reason is
given in Fig~\ref{fig:cr48_2}, where exact KB3 diagonalizations are
done, subtracting either of the two pairing contributions, $JT=01$ and
10 \cite{Poves.Martinez-Pinedo:1998}.  It is apparent that
the $JT=01$-subtracted pattern is quite close to the CHFB one in
Fig.~\ref{fig:smgog}, especially in the rotational regime before the
backbend. The $JT=10$ subtraction goes in the same direction. The
interpretation is transparent: CHFB does not ``see'' proton-neutron
pairing at all, and it is not very efficient in the low pairing regime. As it
does everything else very well, the inevitable conclusion is that
pairing can be treated in first order perturbation theory, {\it i.e.}, the
energies are very sensitive to it, but not the wavefunctions.
Floods of ink have gone into ``neutron-proton'' pairing, which is a
problem for mean field theories but neither for the Gogny force nor
for the shell model. Furthermore, the results show that ordinary pairing
is also a mean field problem when nuclei are not superfluids.

\begin{figure}[htb]
  \includegraphics[width=0.9\linewidth]{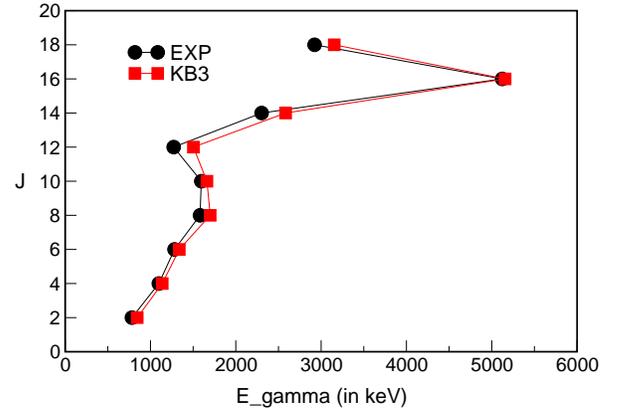}
  \caption{The yrast band of  $^{50}$Cr; experiment vs. the shell model
    calculations with the KB3 interaction\label{fig:cr50}}
\end{figure}

$^{48}$Cr has become a standard benchmark for models
(Cranked Hartree Fock Bogoliuvov~\cite{Caurier.Egido.ea:1995}, Cranked
Nilsson Strutinsky~\cite{Yugoaldavis.Aberg:1999}, Projected Shell
Model~\cite{Hara.Sun:1999}, Cluster Model~\cite{Descouvemont:2002},
etc.)
Nuclei in the vicinity also have strong rotational
features. The mirror pairs $^{47}$V-$^{47}$Cr and $^{49}$Cr-$^{49}$Mn
closely follow the semiclassical picture of a particle or hole hole
strongly coupled to a rotor ~\citep{Martinez-Pinedo.Zuker.ea:1997} in
full agreement with the experiments
\cite{Cameron.Bentley.ea:1991,Cameron.Bentley.ea:1994,OLeary.Bentley.ea:1997,
  Bentley.OLeary.ea:1998,Tonev.Petkov.ea:2002}.  $^{50}$Cr was
predicted to have a second backbending by
\citet{Martinez-Pinedo.Poves.ea:1996b}, confirmed
experimentally~\cite{Lenzi.Ur.ea:1997} (see figure~\ref{fig:cr50}).
When more particles or holes are added, the collective behaviour
fades, though even for $^{52}$Fe, a rotor like
band appears at low spin with an yrast trap at J=12$^+$, both are
accounted for by the shell model
calculations~\cite{Poves.Zuker:1981a}({\em sic}), \cite{Ur.Lenzi.ea:1998}.
For spectroscopic comparisons with the odd-odd nuclei, see 
\citet{Lenzi.Napoli.ea:1999,Brandolini.Medina.ea:2001} for $^{46}$V
and \citet{Svensson.Lenzi.ea:1998} for $^{50}$Mn.  Recently, a highly
deformed excited band has been discovered in $^{56}$Ni
\cite{Rudolf.Baktash.ea:1999}. It is dominated by the configuration
(1f$_{7/2}$)$^{12}$ (2p$_{3/2}$, 1f$_{5/2}$, 2p$_{1/2}$)$^{4}$. The
calculations reproduce the band, that starts at
about 5~MeV excitation energy and has a
deformation close to $\beta$=0.4.

\subsection{Quasi-SU(3).}
\label{sec:quasi-su3-pseudo}

To account for the appearance of backbending rotors, a theoretical
framework was developed by~\citet{Zuker.Retamosa.ea:1995}, and made
more precise by~\citet{Martinez-Pinedo.Zuker.ea:1997}. Here we give a
compact overview of the scheme that will be shown to apply even to the
classic examples of rotors in the rare-earth region.  

\begin{figure}[htb]
  \includegraphics[angle=270,width=0.9\linewidth]{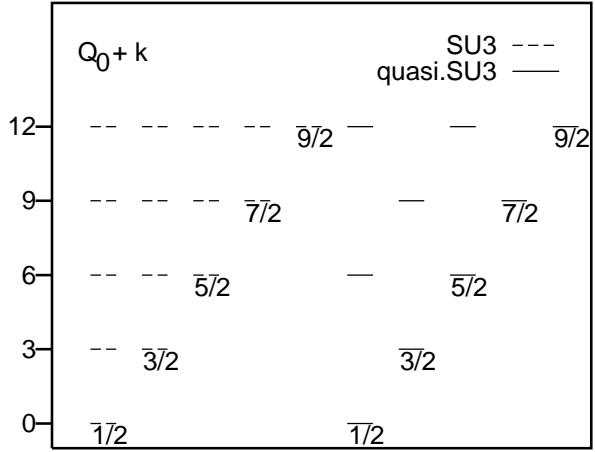}
  \caption{Nilsson orbits for SU(3) $(k= - 2p)$ and quasi-SU(3)
    ($k= - 2p + 1/2)$).\label{fig:su3}}
\end{figure}

Let us start by considering the quadrupole force alone, taken to act
in the $p$-th oscillator shell.  It will tend to maximize the
quadrupole moment, which means filling the lowest orbits obtained by
diagonalizing the operator $Q_0=2q_{20}=2z^2-x^2-y^2$. Using the
cartesian representation, $2q_{20}=2n_z-n_x-n_y$, we find eigenvalues
$2p$, $2p-3$,\ldots, etc., as shown in the left panel of
Fig.~\ref{fig:su3}, where spin has been included.  By filling the
orbits orderly we obtain the intrinsic states for the lowest
SU(3) representations~\cite{Elliott:1958a,Elliott:1958b}: $(\lambda
,0)$ if all states are occupied up to a given level and $(\lambda
,\mu)$ otherwise. For instance: putting two neutrons and two protons
in the $K=1/2$ level leads to the ($4p$,0) representation. For four
neutrons and four protons, the filling is not complete and we have the
(triaxial) ($8(p-1)$,4) representation for which we expect a low lying
$\gamma$ band.

In $jj$ coupling the angular part of the quadrupole operator
$q^{20}=r^2C^{20}$ has matrix elements 
\begin{gather}\label{eq:delta2}
\langle j\, m|C^2|j+2\,
m\rangle\approx \frac{3[(j+3/2)^2-m^2]}{2(2j+3)^2} ,\\\label{eq:delta1}  
\quad \langle j\,
m|C^2|j+1\, m\rangle=-\frac{3m[(j+1)^2-m^2]^{1/2}}{2j(2j+2)(2j+4)}    
\end{gather}
The $\Delta j=2$ numbers in Eq.~\eqref{eq:delta2} are---within the
approximation made---identical to those in $LS$ scheme, obtained by
replacing $j$ by $l$. The $\Delta j=1$ matrix elements
in~Eq.~\eqref{eq:delta2} are small, both for large and small $m$,
corresponding to the lowest oblate and prolate deformed orbits
respectively. If the spherical $j$-orbits are degenerate, the $\Delta
j=1$ couplings, though small, will mix strongly the two $\Delta j=2$
sequences (e.g., $(f_{7/2}p_{3/2})$ and $(f_{5/2}p_{1/2})$). The
spin-orbit splittings will break the degeneracies and favour the
decoupling of the two sequences. Hence the
idea~\citep{Zuker.Retamosa.ea:1995} of neglecting the $\Delta j=1$
matrix elements and exploit 
the correspondence
\[ l\longrightarrow j=l+1/2\quad m\longrightarrow m+1/2\times
{\rm sign}(m).\]
which is one-to-one except for $m=0$.
The resulting ``quasi SU(3)''
quadrupole operator respects SU(3) relationships, except for
$m=0$, where the correspondence breaks down.  The
resulting spectrum for quasi-$2q_{20}$ is shown in the right panel of
Fig~\ref{fig:su3}. The result is not exact for the $K=1/2$ orbits
but a very good approximation.

\begin{figure}[htb]
  \includegraphics[width=0.9\linewidth]{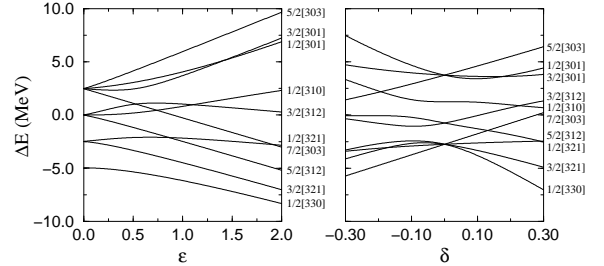}
  \caption{Nilsson diagrams in the $pf$ shell. Energy vs. single
    particle splitting $\varepsilon$ (left panel), energy vs.
    deformation $\delta$ (right panel). \label{fig:nilsson}}
\end{figure}

To check the validity of the decoupling, a Hartree calculation was
done for $H=\varepsilon H_{sp}+ H_q$, where $H_{sp}$ is the observed
single particle spectrum in $^{41}$Ca (essentially equidistant orbits
with 2MeV spacings) and $H_q$ is the quadrupole force in
Eq.~\eqref{Hq} with a properly renormalized coupling. The result is
exactly a~\citet{Nilsson:1955}
calculation~\cite{Martinez-Pinedo.Zuker.ea:1997},    

\begin{equation}
  \label{hmql}
  H_{Nilsson}=\hbar \omega \left( \varepsilon H_{sp}
  -\frac{\delta}{3}\, 2q_{20}\right),   
\end{equation}
where
\begin{equation}
  \label{beta}
  \frac{\delta}{3}=\frac{1}{4}\frac{\langle 2q_{20}\rangle}{\langle
  r^2\rangle}= 
  \frac{\langle 2q_{20}\rangle}{(p+3/2)^4}.
\end{equation}
In the right panel of Fig.~\ref{fig:nilsson} the results are given in
the usual form. 

In the left panel we have turned the representation around: since we
are interested in rotors, we start from perfect ones (SU(3)) and let
$\varepsilon$ increase. At a value of $\approx 0.8$ the four lowest
orbits are in the same sequence as the right side of
fig.~\ref{fig:su3} (Remember here that the real situation corresponds
to $\varepsilon \approx 1.0$). The agreement even extends to the next
group, although now there is an intruder (1/2[310] orbit). The
suggestion is confirmed by an analysis of the wavefunctions: For the
lowest two orbits, the overlaps between the pure quasi-SU(3)
wavefunctions calculated in the restricted $\Delta j=2$ space
(\emph{$fp$ from now on}) and the ones in the full $pf$ shell exceeds
0.95 throughout the interval \mbox{$0.5< \varepsilon <1$}. More
interesting still: the contributions to the quadrupole moments from
these two orbits vary very little, and remain close to the values
obtained at $\varepsilon=0$ \emph{i.e.}, from fig.~\ref{fig:su3}).

\begin{figure}[htb]
  \includegraphics[angle=270,width=0.9\linewidth]{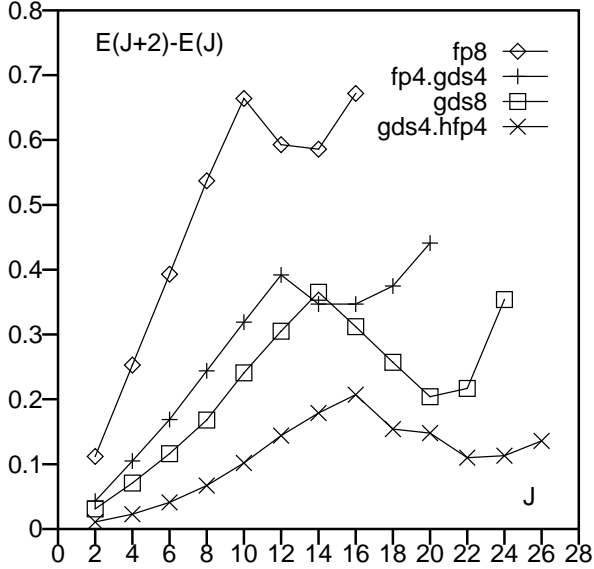}
  \caption{\label{fig:yrkls} Yrast transition energies
    $E_\gamma=E(J+2)-E(J)$ for different configurations, KLS
    interaction.} 
\end{figure}

\begin{figure}[htb]
  \includegraphics[angle=270,width=0.9\linewidth]{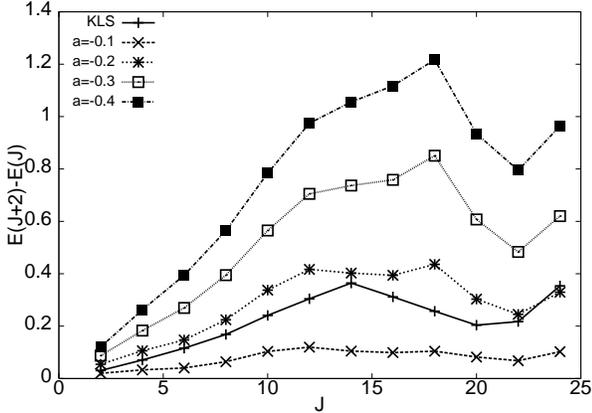}
  \caption{Backbending patterns in $(gds)^8$ $T=0$, with the KLS
    interaction and with a random interaction plus a constant
    $W_{rstu}^{JT}=-a$.  \label{fig:6}}
\end{figure}

We have learned that --for the rotational features--
calculations in the restricted $(fp)^n$ spaces
account remarkably well for the results in the full major shell
$(pf)^n$ (see last column of Table~\ref{tab:cr48def}).
 Let us move now to larger spaces. In Fig.~\ref{fig:yrkls}
we have yrast transition energies for different configurations of 8
particles in $\Delta j=2$ spaces. The force is KLS, $\hbar
\omega=9$~MeV, the single particle splittings uniform at
$\varepsilon=1$~MeV, and $gds$, say, is the lower sequence in the
$sdg$, $p=4$, shell.  Rotational behavior is fair to excellent at low
$J$.  As expected from the normalization property of the realistic
quadrupole force [Eq.~\eqref{q}] the moments of inertia in the
rotational region go as $(p+3/2)^2\,(p'+3/2)^2$, {\it i.e.} if we
multiply all the $E_\gamma$ values by this factor the lines become
parallel.  The intrinsic quadrupole moment $Q_0$ [Eq.~\eqref{bmq}]
remains constant to within 5\% up to a critical $J$ value at which the
bands backbend.

Why and how do the bands backbend? We have no simple answer, but
Fig.~\ref{fig:6} from~\citep{Velazquez.Zuker:2002} shows the behavior
of the $(gds)^8$ space under the influence of a symmetric random
interaction, gradually made more attractive by an amount $a$. There is
no need to stress the similarity with Fig.~\ref{fig:yrkls}. The
appearance of backbending rotors seems to be a general result of the
competition between deformation and alignment, characteristic of
nuclear processes.  

 The group theoretical aspects of quasi-SU3 have been recently discussed and
 applied to the description of the $sd$-shell nuclei
 by \citet*{Vargas.Hirsch.Draayer:2001,Vargas.Hirsch.Draayer:2002}.

\subsection{Heavier nuclei: quasi+pseudo SU(3)}\label{sec:heavier-nuclei}

We have seen that quasi-SU(3) is a variant of SU(3) that obtains for
moderate spin-orbit splittings. For other forms of single particle
spacings, the pseudo-SU(3)
scheme (\citet{Arima.Harvey:1969,Hecht.Adler:1969,Ratna.Draayer.ea:1973},
see also \citet{Vargas.Hirsch.Draayer:2002a} for more recent applications) 
will be favored (in which case we have to use the left panel of
Fig.~\ref{fig:su3}, with pseudo-$p=p-1$). Other variants of SU(3) may
be possible and are well worth exploring. In cases of truly large
deformation SU(3) itself may be valid in some blocks.

\begin{figure}
  \includegraphics[angle=270,width=0.9\linewidth]{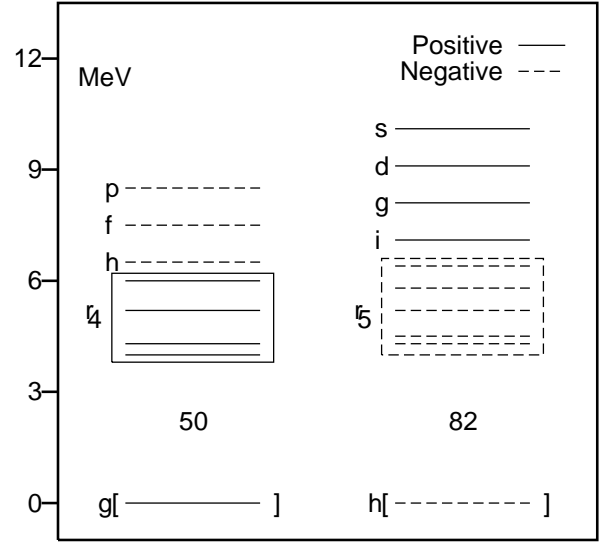}
  \caption{Schematic single particle spectrum above $^{132}$Sn. $r_p$ is
 the set of orbits in shell $p$ excluding the largest. For the upper
 shells the label $l$ is used for $j=l+1/2$\label{fig:sp}}
\end{figure}

To see how this works, consider Fig.~\ref{fig:sp} giving a schematic
view of the single particle energies in the space of two contiguous
major shells---in protons ($\pi$) and neutrons ($\nu$)---adequate for
a SM description of the rare earth region.
 
We want to estimate the quadrupole moments for nuclei at the onset of
deformation. We shall assume quasi-SU(3) operates in the upper shells,
and pseudo-SU(3) in the lower ones.  The number of particles in each
shell for which the energy will be lowest will depend on a balance of
monopole and quadrupole effects, but Nilsson diagrams
suggest that when nuclei acquire stable deformation, two orbits
$K$=1/2 and 3/2---originating in the upper shells of
Fig.~\ref{fig:su3}---become occupied, {\it i.e.}, the upper blocks are
precisely the 8-particle configurations we have studied at length.
Their contribution to the electric quadrupole moment is then
\begin{equation}
  \label{q0up}
  Q_0=8[e_\pi(p_\pi-1)+e_\nu(p_\nu-1)],
\end{equation}
with $p_\pi=5$, $p_\nu=6$; $e_\pi$ and $e_\nu$ are the effective
charges.

Consider even-even nuclei with Z=60-66 and N=92-98, corresponding to 6
to 10 protons with pseudo-$p=3$, and 6 to 10 neutrons with
pseudo-$p=4$ in the lower shells. From the left part of
Fig.~\ref{fig:su3} we obtain easily their contribution to $Q_0$, which
added to that of Eq.~(\ref{q0up}) yields a total
\begin{equation}
  \label{q0down}
Q_0=56e_{\pi}+(76+4n)e_{\nu},
\end{equation}
for $^{152+2n}$Nd, $^{154+2n}$Sm, $^{156+2n}$Gd and $^{158+2n}$Dy
respectively. At fixed $n$, the value is constant in the four cases
because the orbits of the triplet K=1/2, 3/2, 5/2 in
Fig.~\ref{fig:su3} have zero contribution for $p$=3.  $Q_0$ (given in
dimensionless oscillator coordinates, {\it i.e.}, $r \rightarrow r/b$ with
\mbox {$b^2\approx 1.01{\rm A}^{-1/3}fm^2$}), is related to the $E2$
transition probability from the ground state by
\mbox{$B(E2)\uparrow=10^{-5}{\rm A}^{2/3}Q_0^2$}.  The results, using
effective charges of $e_\pi=1.4$, $e_\nu=0.6$ calculated
in~\citep{Dufour.Zuker:1996} are compared in table~\ref{be2} with the
available experimental values . The agreement is quite remarkable and
no free parameters are involved. Note in particular the quality of the
prediction of constancy (or rather $A^{2/3}$ dependence) at fixed $n$,
which does not depend on the choice of effective charges.  The
discrepancy in $^{152}$Nd is likely to be of experimental origin,
since systematics indicate, with no exception, much larger rates for a
$2^+$ state at such low energy (72.6 keV).
\begin{table}
\caption{$B(E2)\uparrow$ in $e^2b^2$ compared with experiment 
\protect{\cite{Raman.Nestor.ea:1989}}}
\begin{ruledtabular}
\begin{tabular}{cllll}
N   & Nd & Sm & Gd & Dy\\
\colrule
92 & 4.47 & 4.51 & 4.55 & 4.58\\
   &  2.6(7) & 4.36(5) & 4.64(5) & 4.66(5)\\
\colrule
94 & 4.68 & 4.72 & 4.76 & 4.80\\
  &   &  & 5.02(5) & 5.06(4)\\
\colrule
96 & 4.90 & 4.95 & 4.99 & 5.03\\
  &   &  & 5.25(6) & 5.28(15)\\
\colrule
98 & 5.13 & 5.18 & 5.22 & 5.26\\
 &   &  &  & 5.60(5)\\
\end{tabular}
\end{ruledtabular}
\label{be2}
\end{table}
It is seen  that by careful analysis of exact results one
may come to very simple computational strategies. In the last example
on B(E2) rates, the simplicity is such that the computation reduces to
a couple of sums. 

\subsection{The  $^{36}$Ar
  and $^{40}$Ca super-deformed bands}\label{sec:36ar-40ca-super}
The arguments sketched above apply to the region around $^{16}$O,
where a famous 4-particles 4-holes (4p-4h) band starting at 6.05~MeV
was identified by~\citet{Carter.Mitchell.ea:1964}, followed by the
8p-8h band starting at 16.75~MeV~\cite{Chevalier.Scheibling.ea:1967}.
Shell model calculations in a very small space,
$p_{1/2}d_{5/2}s_{1/2}$, could account for the spectroscopy in
$^{16}$O, including the 4p-4h
band~\citep[ZBM]{Zuker.Buck.McGrory:1968}, but the 8p-8h one needs at
least three major shells and was tackled by an $\alpha$-cluster
model~\cite{Abgrall.Caurier.ea:1967,Abgrall.Baron.ea:1967}. It is
probably the first superdeformed band detected and explained.  

In $^{40}$Ca, the first excited $0^+$ state is the 4p-4h band-head. It
is only recently that another low-lying highly deformed band has been
found~\cite{Ideguchi.Sarantites.ea:2001}, following the discovery of a
similar structure in $^{36}$Ar~\cite{Svensson.Macchiavelli.ea:2000}.       

By applying the quasi+pseudo-SU(3) recipes of
Section~\ref{sec:heavier-nuclei} we find that the maximum deformations
attainable are of 8p-8h character in  $^{40}$Ca and 4p-4h in $^{36}$Ar
with $Q_0=180$~efm$^2$ and $Q_0=136$~efm$^2$ respectively.   

The natural generalization of the ZBM space consists in the
$d_{3/2}s_{1/2}fp$ orbits (remember: $fp\equiv f_{7/2}p_{3/2}$), which
is relatively free of center of mass spuriousness.
And indeed, it describes well the rotational
regime of the observed bands. However, to track them beyond backbend,
it is convenient to increase the space to $d_{3/2}s_{1/2}pf$. The
adopted interaction, \textsf{sdfp.sm} is the one originally constructed
by~\citet{Retamosa.Caurier.ea:1997}, and used
in~\cite{Caurier.Nowacki.ea:1998} (USD, KB3 for intra shell and KLS
for cross shell matrix elements), with minor monopole adjustments
dictated by new data on the single particle structure of
$^{35}$Si~\cite{Nummela.ea:2001a}.   

\begin{figure}
  \includegraphics[width=0.9\linewidth]{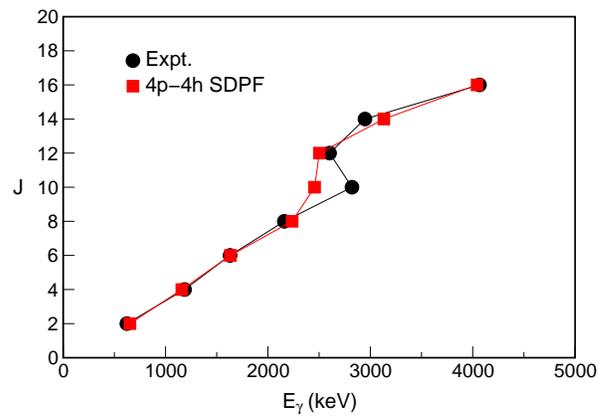}
  \caption{The superdeformed band of $^{36}$Ar, experiment
    vs. SM\label{fig:ar36}}
\end{figure}

The calculations are conducted in spaces of fixed number of particles
and holes.  In figure~\ref{fig:ar36} the calculated energy levels in
$^{36}$Ar are compared to the data. The agreement is excellent, except
at $J=12$ where the data show a clear backbending while the calculation
produces a much smoother upbending pattern.

In table~\ref{tab:ar36} the calculated spectroscopic quadrupole
moments ($Q_s$) and the B(E2)'s are used to compute the intrinsic $Q$
as in Eqs.~(\ref{bmq},\ref{bme2}), using standard effective charges
$\delta q_{\pi}$=$\delta q_{\nu}$=0.5.  As expected both Q$_0$(s) and
Q$_0$(t) are nearly equal and constant---and close to the quasi+pseudo
SU(3) estimate---up to the backbend. The calculated B(E2)'s agree well
with the experimental ones~\cite{Svensson.Macchiavelli.ea:2001}.  The
value of $Q_0$ corresponds to a deformation $\beta\approx 0.5$.

\begin{table}[htb]
\caption{\label{tab:ar36}
Quadrupole properties of the 4p-4h configuration's yrast-band in
 $^{36}$Ar (in e$^2$fm$^4$ and efm$^2$)}
 \begin{tabular*}{\linewidth}{@{\extracolsep{\fill}}cccccc} 
\noalign{\smallskip}\hline\noalign{\smallskip}
 & \multicolumn{2}{c}{ B(E2)(J$\rightarrow$J-2)} & & & \\
\noalign{\smallskip}
 J &  EXP &   TH &
 Q$_{spec}$ & Q$_0$(s) & Q$_0$(t) \\   
\hline\noalign{\smallskip}
 2 &        & 315 & -36.0 & 126 & 126 \\
 4 & 372(59) &435 & -45.9 & 126 & 124 \\
 6 & 454(67)& 453 & -50.7 & 127 & 120 \\
 8 & 440(70)& 429 & -52.8 & 125 & 114 \\
10 & 316(72)& 366 & -52.7 & 121 & 104  \\
12 & 275(72)& 315 & -53.0 & 119 &  96 \\
14 & 232(53)& 235 & -54.3 & 120 &  82  \\
16 & $>$84  & 131 & -56.0 & 122 &  61 \\
\noalign{\smallskip}\hline
\end{tabular*} 
\end{table}

Now we examine the 8p-8h band in
$^{40}$Ca~\cite{Ideguchi.Sarantites.ea:2001}.  The valence space
adopted for $^{36}$Ar is truncated by limiting the maximum number of
particles in the 1f$_{5/2}$ and 2p$_{1/2}$ orbits to two.

\begin{figure}[htb]
  \includegraphics[width=0.9\linewidth]{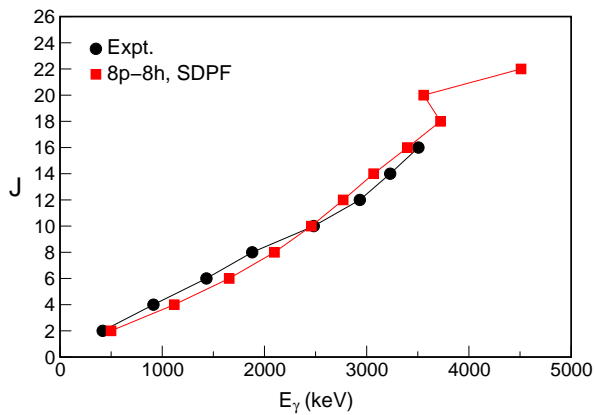}
  \caption{The superdeformed band in $^{40}$Ca; exp. vs. 8p-8h
    calculation\label{fig:ca40sd}}
\end{figure}

The experimental~\cite{Ideguchi.Sarantites.ea:2001} and calculated
yrast gamma-ray energies are compared in fig.~\ref{fig:ca40sd}. The
patterns agree reasonably well but the change of slope at $J=10$
---where the backbend in $^{48}$Cr starts (Fig.~\ref{fig:smgog})---is
missed by the calculation, which only backbends at $J=20$, the band
termination for the configuration $f_{7/2}^8 (d_{3/2}s_{1/2})^{-8}$.
The extra collectivity induced by the presence of $sd$
particles in pseudo-SU(3) orbitals is responsible or the
delay in the alignment, but apparently too strong to allow for the
change in slope at $J=10$.

\begin{table}[htb] 
\caption{{\label{tab:8p8h.sdpf}}
Quadrupole properties of the 8p-8h configuration's yrast-band in
$^{40}$Ca  (in e$^2$fm$^4$ and efm$^2$), calculated in the $sdpf$ valence space}
 \begin{tabular*}{\linewidth}{@{\extracolsep{\fill}}rccrr} 
 \hline\noalign{\smallskip}
 J & B(E2)(J $\rightarrow$ J-2)&
 Q$_{spec}$ &  Q$_0$(t) & Q$_0$(s) \\
 \noalign{\smallskip}\hline\noalign{\smallskip}
 2 & 589 &  -49.3 &  172 & 172 \\
 4 & 819 &  -62.4 &  170 & 172 \\
 6 & 869 &  -68.2 &  167 & 171 \\
 8 & 860 &  -70.9 &  162 & 168 \\
10 & 823 &  -71.6 &  157 & 164 \\
12 & 760 &  -71.3 &  160 & 160 \\
14 & 677 &  -71.1 &  149 & 157 \\
16 & 572 &  -72.2 &  128 & 158 \\
18 & 432 &  -75.0 &  111 & 162  \\
20 &  72 &  -85.1 & & \\
22 &   8 &  -79.1 & & \\
24 &   7 &  -81.5 & & \\
\noalign{\smallskip}\hline
\end{tabular*} 
\end{table}
The experimental $Q_0$(t)=180$^{+39}_{-29}$ obtained from the
fractional Doppler shifts corresponds to a deformation $\beta\approx
0.6$~\cite{Ideguchi.Sarantites.ea:2001}. It is extracted from an
overall fit that assumes constancy for all measured values, and
corresponds exactly to the quasi+pseudo SU(3) estimate. It also
squares well with the calculated 172~efm$^2$ in
Table~\ref{tab:8p8h.sdpf} where the steady decrease in collectivity
remains consistent with experiment within the quoted uncertainties.
A reanalysis of the experimental lifetimes in the superdeformed band
of $^{40}$Ca \cite{Chiara.Ideguchi.ea:2003} suggest that for low spins
the deformation would be smaller due to the mixing with less deformed states
of lower np-nh rank.

This calculation demonstrates that a detailed description of very
deformed bands is within reach of the shell model. The next step
consists in remembering that states at fixed number of particles are
doorways that will fragment in an exact calculation that remains to be
done.  

 The band in $^{36}$Ar has been also described by the projected shell model
 in~\cite{Long.Sun:2001}. Beyond mean field methods using the Skyrme
 interaction have been recently applied to both the  $^{36}$Ar
 and  $^{40}$Ca superdeformed bands by \citet*{Bender.Flocard.ea:2003}.

\subsection{Rotational bands of unnatural
  parity}\label{sec:rotat-bands-unnat}
 The occurrence of low-lying bands of opposite parity to the ground
 state band is very frequent in $pf$ shell nuclei. Their simplest
 characterization is as particle-hole bands 
 with the hole in the 1d$_{3/2}$ orbit. In a nucleus whose ground
 state is described by the configurations $(pf)^n$ the opposite parity
 intruders will be   (1d$_{3/2}$)$^{-1}$ $(pf)^{n+1}$. The
 promotion of a particle from the $sd$ to the $pf$ shell costs an
 energy equivalent to the local value of the gap, that in this region
 is about 7~MeV. On the other side, the presence of one extra particle
 in the $pf$-shell
 may produce an important increase of the correlation energy,
 that can compensate the energy lost by the particle hole jump.
 For instance, the very
 low-lying positive parity band of $^{47}$V, can be interpreted as         
  (1d$_{3/2}$)$^{-1}$ (a proton hole) coupled to $(pf^8)$~T=0. 
 Indeed, the correlation
 energy of this pseudo-$^{48}$Cr is larger than the correlation energy
 of the ground state of
 $^{47}$V and even larger than the correlation energy of the real
 $^{48}$Cr, explaining why the  band starts at
 only 260~keV of excitation energy~\cite{Poves.Sanchez-Solano:1998}. The most
 extreme case  is $^{45}$Sc, where the intruder band based
 in the configuration (1d$_{3/2}$)$^{-1}$ coupled to $^{46}$Ti, barely
 miss (just by 12~keV) to become the ground state. 
 Many bands of this
 type have been experimentally studied in recent years, mainly at
 the Gasp and Euroball detectors \cite{Brandolini.Medina.ea:1999}, and
 explained by shell model calculations.

\begin{figure}[htb]
  \includegraphics[width=0.9\linewidth]{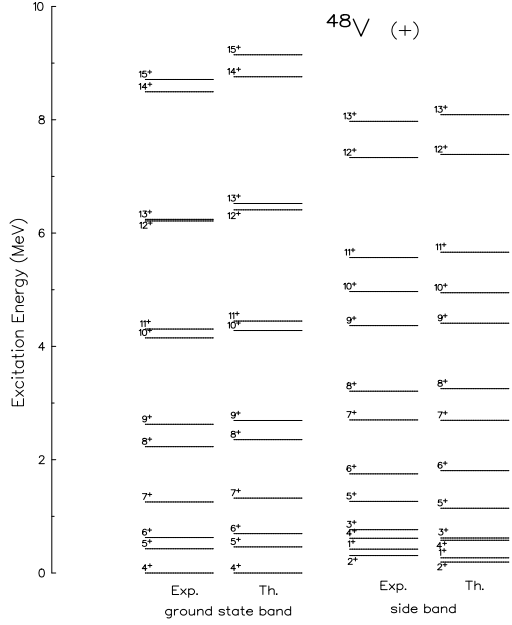}
  \caption{The ground state (K=4$^+$) and the side band (K=1$^+$) of
    $^{48}$V, experiment $vs.$ theory. \label{fig:v48p}}
\end{figure}
We present now some recent results in $^{48}$V
\cite{Brandolini.Marginean.ea:2002}.  The positive parity levels of
this odd-odd nucleus are very well reproduced by the shell model
calculation using the interactions KB3 (or KB3G) as can be seen in
Fig.~\ref{fig:v48p}, another example of good quality ``pure
spectroscopy''. The levels are grouped in two bands, one connected to
the 4$^+$ ground state, that would correspond to K=4$^+$ in a Nilsson
context (the aligned coupling of the last unpaired proton K=3/2$^-$
and the last neutron K=5/2$^-$) and a second one linked to the 1$^+$
member of the ground state multiplet that results of the anti-aligned
coupling.

\begin{figure}[htb]
  \includegraphics[width=1.0\linewidth]{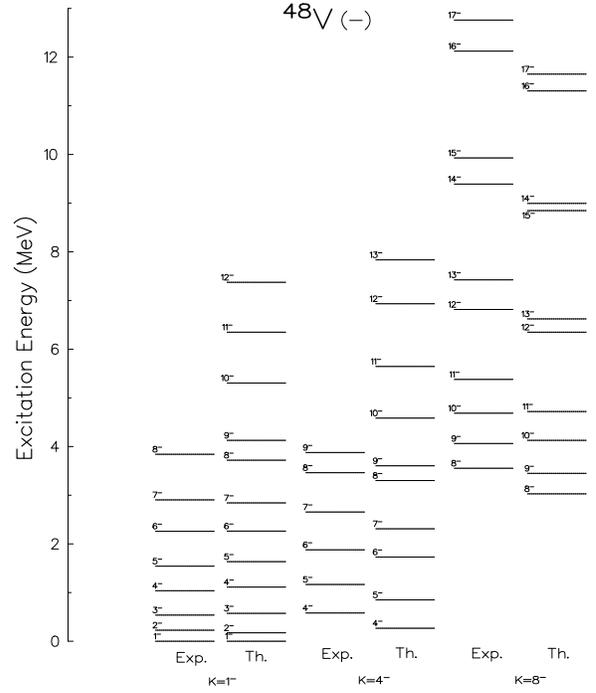}
  \caption{The negative parity  K=1$^-$ and  K=4$^-$ bands of
    $^{48}$V, experiment $vs.$ theory. \label{fig:v48n}}
\end{figure}

The negative parity states can be interpreted as the result of the
coupling of a hole in the 1d$_{3/2}$ orbit to $^{49}$Cr. As we have
mentioned earlier, the ground state band of $^{49}$Cr can be
understood in a particle plus rotor picture, and assigned K=5/2$^-$.
Now we have to couple the 1d$_{3/2}$ hole to our basic rotor
$^{48}$Cr. Therefore the structure of the negative parity states of
$^{48}$V is one particle and one hole coupled to the rotor core of
$^{48}$Cr.  Hence we expect that the lowest lying bands would have
K=1$^-$ and K=4$^-$ (aligned and anti-aligned coupling of the
quasiparticle and the quasihole). These two bands have been found
experimentally at about 500~keV of excitation energy.  Their structure
is in very good agreement with the SM predictions (see
Fig.~\ref{fig:v48n}) except for a small shift of the K=4$^-$ and
K=8$^-$ bandheads relative to the K=1$^-$ state.

\begin{figure}[htb]
  \includegraphics[width=1.0\linewidth]{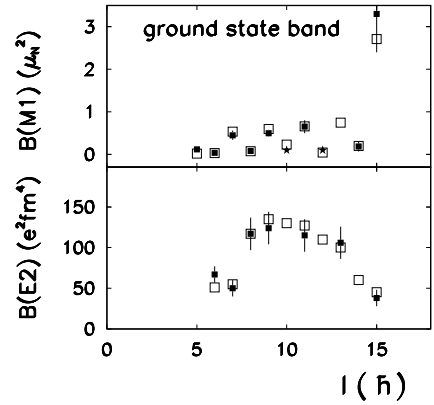}
  \caption{ Theoretical and experimental B(M1) and B(E2) transition
    probabilities in the ground state band of $^{48}$V
    \label{fig:v48trp}}
\end{figure}

A more stringent test of the theoretical predictions is provided by
the electromagnetic properties. In this particular nucleus the
experimental information is very rich and a detailed comparison can be
made for both the positive and negative bands and for E2 and M1
transitions.  The results for the ground state band are collected in
Fig.~\ref{fig:v48trp}.  Notice that the calculation reproduces even
the tiniest details of the experimental results, as, for instance the
odd-even staggering of the B(M1)'s. The B(E2)'s correspond to a
deformation $\beta$=0.2 that is clearly smaller than that of
$^{48}$Cr.

\begin{figure}[htb]
  \includegraphics[width=1.1\linewidth]{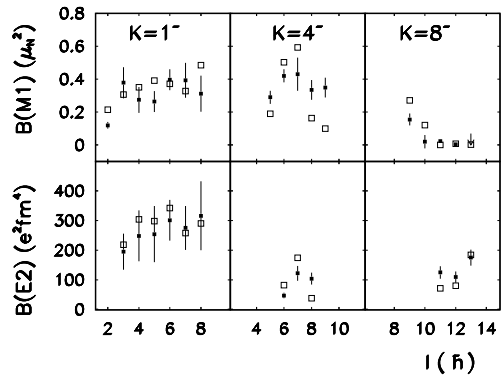}
  \caption{Theoretical and experimental B(M1) and B(E2) transition
    probabilities in the negative parity bands of $^{48}$V
    \label{fig:v48trn}} 
\end{figure}

The transition probabilities in the negative parity bands are
presented in Fig.~\ref{fig:v48trn}. Now the experimental results have
larger uncertainties. Despite that, the agreement in absolute values
and trends is very good for the lowest band (K=1$^-$). The
quantitative agreement is worse for the other two bands, nevertheless,
the main features of the data are well reproduced.  The deformation of
the K=1$^-$ band extracted from the B(E2) properties turns out to be
larger than the deformation of the ground state band and close to that
of $^{48}$Cr.

\section{Description of very neutron rich
  nuclei}\label{sec:descr-very-neutr}

The study of nuclei lying far from the valley of stability is one of
the most active fields in today's experimental nuclear physics. What
it specific about these nuclei, from a shell model point of view? As
everywhere else: the model space and the monopole behavior of the
interaction. They go together because the effective single particle
energies (ESPE, see Section~\ref{sec:bm0hb--calc}) depend on
occupancies, which in turn depend on the model space.

In broad terms, the specificity of light and medium neutron rich
nuclei is that the EI closures (corresponding to the filling of the $p+1/2$
orbit of each oscillator shell) take over as boundaries of the model
spaces. As discussed in Section~\ref{sec:shell-formation}, the
oscillator closures (HO) may be quite solid for double magic nuclei, but
they become vulnerable in the semi-magic cases.

For instance the $d_{3/2}$-$f_{7/2}$ neutron gap in $^{40}$Ca of
$\approx 7$~MeV goes down to $\approx 2.5$ MeV around $^{28}$O. Since
the $d_{5/2}$ orbit is well below its $sd$ partners, now quite close
to $f_{7/2}$, the natural model space is no longer the $sd$ shell, but
the EI space bounded by the $N=14$ and 28 closures, supplemented by the
$p_{3/2}$ subshell whenever the $p_{3/2}$-$f_{7/2}$ gap becomes
small: from $\approx 6$ MeV in $^{56}$Ni, it drops to $\approx 4.5$
MeV in $^{48}$Ca, then $\approx 2$ MeV in $^{40}$Ca, and finally
$\approx 0$ MeV in $^{28}$Si. As explained at the end of
Section~\ref{sec:shell-formation} this monopole drift provides direct
evidence for the need of three-body mechanisms.

For the $p$ shell, the situation is similar: The imposing 11.5 MeV
$p_{1/2}$-$d_{5/2}$ gap in $^{16}$O is down to some 3 MeV in $^{12}$C.
The monopole drift of the $s_{1/2}$-$d_{5/2}$ gap brings it from about
8 MeV in $^{28}$Si to nearly -1 MeV in $^{12}$C. According
to~\citet{Otsuka.Fujimoto.ea:2001} the drift may well continue: in
$^8$He, the $s_{1/2}$-$p_{1/2}$ bare gap is estimated at 0.8 MeV.

All the numbers above (except the last) are experimental. The bare
monopole values are smaller because correlations increase
substantially the value of the gaps, but they do not change
qualitatively the strong monopole drifts. Their main consequence is
that ``normal'' states, {\it i.e.}, those described by 0\hw\ $p$ or
$sd$ calculations often ``coexist'' with ``intruders'' that involve
promotion to the next oscillator shell. Let us examine how this
happens.

 \subsection{N=8; $^{11}$Li: halos}
\label{sec:n=8-11li-shell}

The 1/2$^+$ ground state in $^{11}$Be provided one of the first
examples of intrusion. The expected 0$\hbar \omega$ normal state lies
300~keV higher. The explanation of this behavior has varied with
time~\cite{Talmi.Unna:1960,Sagawa.Brown.ea:1993,Suzuki.Otsuka:1994,Aumann.ea:2001}, 
see also \citep{Brown:2001} for a recent review and
\cite{Suzuki.Fujimoto.ea:2003} for a
new multi-\hw calculation, but the idea has
remained unchanged. The normal state corresponds to a hole on the
$N=8$ closure. The monopole loss of promoting a particle to the $sd$
shell is compensated by a pairing gain for the $p_{1/2}^{-2}$ holes. A
quadrupole gain due to the interaction of the $sd$ neutron with the
$p^2$ protons is plausible, but becomes questionable when we note that
the same phenomenon occurs in $^9$He which has no $p^2$ particles,
suggesting the need for a further reduction of the monopole
loss~\cite{Otsuka.Fujimoto.ea:2001}.

The interest in $^{11}$Be---as a ``halo'' nucleus---was revived by the
discovery of the remarkable properties of $^{11}$Li, which sits at the
drip line ($\approx$~200~keV two-neutron
separation energy) and has a very large spatial
extension, due to a neutron halo
\cite{Tanihata.ea:1985,Hansen.Jonson:1987}.

The shell model has no particular problem with halo nuclei, whose
large size is readily attributed to the large size of the $s_{1/2}$
orbit. As shown by~\cite{Kahana.Lee.Scott:1969a} the use of Woods
Saxon wavefunctions affect the matrix elements involving this
orbit, but the uncertainties involved are easily absorbed by the
monopole field. As a consequence, the whole issue hinges on the 
$s_{1/2}$ contribution to the wavefunctions, which is sensitively
detected by the $\beta$ decay to the first excited 1/2$^{-}$ state in
$^{11}$Be. In \citet{Borge.ea:1997} it was shown that calculations
producing a 50\% split between the neutron closed shell and the
$s_{1/2}^2\, p_{1/2}^{-2}$ configuration lead to the right lifetime.    
This result was confirmed by~\citet{Simon.ea:1999}. Further
confirmation comes from~\citet{Navin.ea:2000}, with solid indications
that the supposedly semi-magic $^{12}$Be ground state is dominated by
the same $s_{1/2}^2\, p_{1/2}^{-2}$ configuration.

\subsection{N=20; $^{32}$Mg, deformed intruders}
\label{sec:n=20-32mg-deformed}
In the mid 1970's it was the $sd$ shell that attracted the most
attention. Nobody seemed to remember $^{11}$Be, and everybody
(including the authors of this review active at the time) were
enormously surprised when a classic
experiment~\citet{Thibault.ea:1975} established that the mass and
$\beta$-decay properties of the $^{31}$Na ground state---expected to
be semimagic at $N=20$---could not possibly be that of a normal
state~\cite{Chung.Wildenthal:1979}. The next example of a frustrated
semi-magic was $^{32}$Mg ~\cite{Detraz.ea:1979}.  
Early mean field calculations had interpreted the discrepancies
as due to deformation~\cite{Campi.Flocard.ea:1975} but the experimental
confirmation took some time~\cite{Guillemaud.ea:1984,Klotz.ea:1993,Motobayashi.Ikeda.ea:1995}.

Exploratory shell model calculations
by~\citet{Storm.Watt.Whitehead:1983}, including  the 1f$_{7/2}$ orbit in
the valence space, was able to improve the mass predictions, however, 
deformation was still absent. To obtain deformed solutions demanded the inclusion of the 2p$_{3/2}$
orbit as demonstrated by~\citet{Poves.Retamosa:1987}.  These
calculations were followed by many
others~\cite{Warburton.Becker.Brown:1990,Heyde.Woods:1991,
  Fukunishi.Otsuka.Sebe:1992,Poves.Retamosa:1994,Otsuka.Fukunishi:1996,
  Siiskonen.Lippas.Rikowska:1999}, that mapped an ``island of
inversion'', {\it i.e.} the region where the intruder configurations
are dominant in the ground states.  The detailed contour of this ``island''
depends strongly on the behavior of the  effective single particle
energies (ESPE),
in turn dictated by the monopole hamiltonian.
Let's recall, that according to
what we have learned in Section~\ref{sec:quasi-su3-pseudo}, the
configurations $(d_{5/2}s_{1/2})^{2-4}_p\,
(f_{7/2}p_{3/2})^{2-4}_n$,corresponding to $N=20,\, Z=10$-12, have a
``quasi-SU(3)'' quadrupole coherence close to that of SU(3) ({\it i.e.},
maximal). 

\begin{figure}[htb]
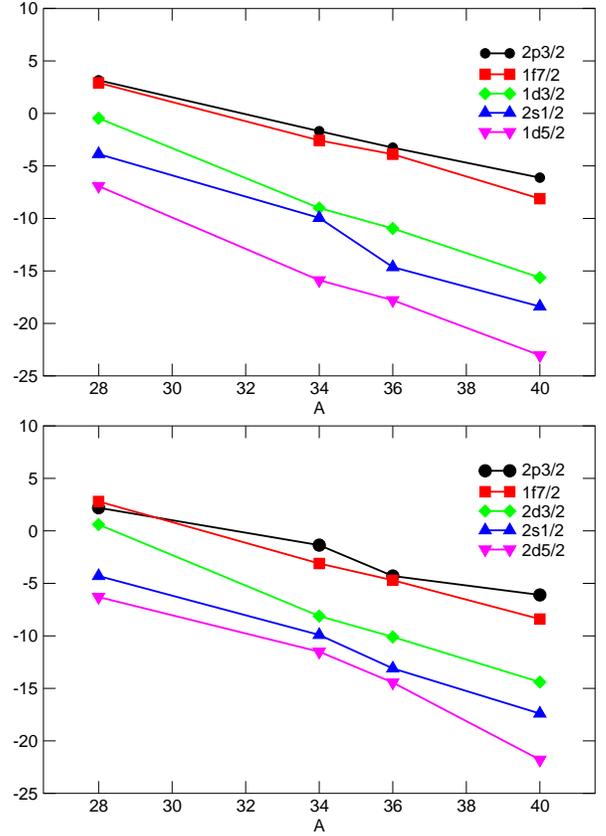

  \includegraphics[width=0.9\linewidth]{rmpgif/poves_fig47a.eps}\\
  \includegraphics[width=0.9\linewidth]{rmpgif/poves_fig47b.eps}
  \caption{Effective single particle energies (in MeV) at N=20 with the
    $sdpf.sm$ interaction (upper panel) and the Tokyo group
    interaction (bottom panel)\label{fig:espe_n20}}       
\end{figure}

The ESPE in Fig.~\ref{fig:espe_n20} represent $H_m$ for the
\textsf{sdfp.sm} and Tokyo group
interactions~\cite{Utsuno.Otsuka.ea:1999}. Within details they are
quite close, and such as efficiently to favor deformation. Both
interactions lead to an ``island of inversion'' for $Z$=10, 11 and 12;
$N$=19, 20 and 21.

\begin{figure}[htb]
  \includegraphics[width=0.9\linewidth]{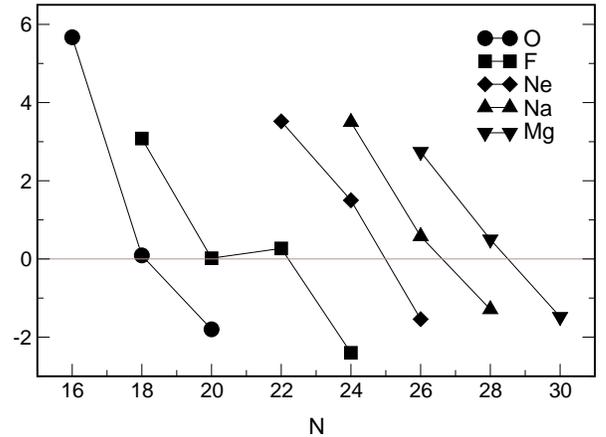}
  \caption{Two neutron separation energies (in MeV) calculated with the
    $sdpf.sm$ interaction.\label{fig:s2n}}
\end{figure}

Consider now some detailed information obtained with $sdpf.sm$ by
\citet{Caurier.Nowacki.Poves:2001}.  The $S_{2N}$ values of
Fig.~\ref{fig:s2n} locate the neutron drip line, consistent with what
is known for oxygen and fluorine, where the last bound isotopes are
$^{24}$O and $^{31}$F~\cite{Sakurai.ea:1999}. Note the kink due to
deformed correlations in the latter.  For the other chains, the
behavior is smoother and the last predicted bound isotopes are
$^{34}$Ne, $^{37}$Na and $^{40}$Mg.

\begin{table*}
\caption{Properties of the even magnesium isotopes. N stands for normal
  and I for intruder. Energies in MeV, B(E2)'s in
  e$^2$fm$^4$ and Q's in efm$^2$}
\label{tab:mg}
\vspace{0.1cm}
 \begin{tabular*}{\textwidth}{@{\extracolsep{\fill}}cccccccccc} 
\hline\noalign{\smallskip}
   &  & $^{30}$Mg  &   &   & $^{32}$Mg   &  &     & $^{34}$Mg   &  \\
\hline\noalign{\smallskip}
    & N  & I  & EXP  &  N  & I  & EXP   & N  & I  & EXP \\
\hline\noalign{\smallskip}
$\Delta$E(0$^+_{\textrm{I}}$) &      & +3.1  &     &    & -1.4  &   
&    & +1.1   & \\

0$^+$ &    0.0 & 0.0  &      & 0.0 & 0.0  &   
& 0.0  & 0.0   & \\

2$^+$     &1.69   & 0.88  &1.48      &1.69 & 0.93  & 0.89  
  &1.09  & 0.66   &0.67 \\
4$^+$     &4.01   & 2.27  &      &2.93 & 2.33  & (2.29)  
  &2.41  & 1.86   &2.13 \\
6$^+$     &6.82   &3.75   &      &9.98   &3.81   &    
  &3.52  & 3.50   & \\
B(E2)     &   &   &   &   &   &   &   &      & \\
2$^+$ $\rightarrow$ 0$^+$  &   53   &112   & 59(5)     &36 & 98
  & 90(16)    &75  & 131   & 126(25)\\
4$^+$ $\rightarrow$ 2$^+$  &   35   &144   &      &17   &123   
  &  &88   & 175   & \\
6$^+$ $\rightarrow$ 4$^+$  &   23   &140   &      &2   &115   &   
  &76  & 176    & \\
Q$_{spec}$(2$^+$)  &    -12.4  &-19.9   &      & -11.4  &-18.1   &
  &-15.4  & -22.7   & \\
\hline
\end{tabular*}
\end{table*}

Some results for the even Mg isotopes (N=18, N=20 and N=22) are
gathered in table~\ref{tab:mg}. In $^{30}$Mg the normal configuration
is the one that agrees with the existing experimental data
\cite{Pritychenko.ea:1999}.  In $^{32}$Mg the situation is the
opposite as the experimental data (the 2$^+$ excitation energy
\cite{Guillemaud.ea:1984} and the 0$^+$~$\rightarrow$~2$^+$ B(E2)
\cite{Motobayashi.Ikeda.ea:1995}) clearly prefer the intruder.  A
preliminary measure of the 4$^{+}$ excitation energy reported by
\citet{Azaiez.ea:1999} goes in the same direction. Data and
calculations suggest prolate deformation with $\beta \approx$~0.5.  In
$^{34}$Mg the normal configuration that contains two $pf$ neutrons, is
already quite collective. It can be seen in the table that it
resembles the $^{32}$Mg ground state. The 4p-2h intruder is even more
deformed ($\beta \approx$~0.6) and a better rotor. Results from the
Riken experiments of \citet{Yoneda.ea:2000} and
\citet{Iwasaki.ea:2001} seem to favor the intruder option. In the
QMCD calculations of \citet{Utsuno.Otsuka.ea:1999}, the ground state
band is dominantly 4p-2h; the 2$^+$ comes at the right place, but the
4$^{+}$ is too high. Clearly, $^{34}$Mg is at the edge of the ``island
of inversion''. Another manifestation of the intruder presence in the
region has been found at Isolde \cite{Nummela.ea:2001b}: The decay of
$^{33}$Na indicates that the ground state of $^{33}$Mg has
J$^{\pi}$=3/2$^+$ instead of the expected J$^{\pi}$=3/2$^-$ or
J$^{\pi}$=7/2$^-$. This inversion is nicely reproduced by the
$sdpf.sm$ calculation.

\subsection{N=28: Vulnerability}
\label{sec:N=28}

Let us return for a while on Fig.~\ref{fig:espe_n20}. The scale does
not do justice to a fundamental feature: the drift of the
$p_{3/2}-f_{7/2}$ gap, which decreases as protons are removed. As
explained at the end of Section~\ref{sec:shell-formation}, this
behavior is contrary to a very general trend in heavier nuclei, and
demands a three-body mechanism to resolve the contradiction. Next we
remember that the oscillator closures are quite vulnerable even at the
strict monopole level. As we have seen, quadrupole coherence takes full
advantage of this vulnerability. On the contrary, the EI closures are
very robust. However, because of the drift of the $p_{3/2}-f_{7/2}$
gap even the $N=28$ closure becomes vulnerable. The $sdpf.sm$
interaction leads to a remarkable result summarized in  
Table~\ref{tab:n28gap}: the decrease of the gap combined with the gain
of correlation energy of the 2p-2h lead to a breakdown of the $N=28$
closure for $^{44}$S and $^{40}$Mg. The double magic $^{42}$Si resists. 
Towards N=Z, $^{52}$Cr
and $^{54}$Fe exhibit large correlation energies  associated to the prolate deformed character of their
2p-2h neutron configurations ($\beta \sim$~0.3).

\begin{figure}[htb]
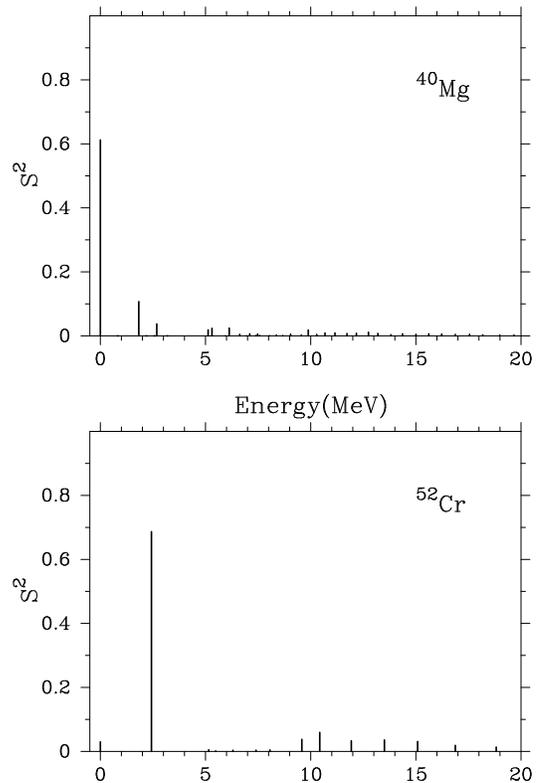

  \includegraphics[angle=270,width=0.8\linewidth]{rmpgif/poves_fig49a.ps}\\
  \includegraphics[angle=270,width=0.8\linewidth]{rmpgif/poves_fig49b.ps} 
  \caption{LSF of the 2p-2h 0$^+$ bandheads of $^{40}$Mg (upper panel)
    and $^{52}$Cr (bottom panel) in the full 0$\hbar \omega$
    space\label{fig:mg40}}
\end{figure}

When the 2p-2h bandheads of  $^{40}$Mg and $^{52}$Cr are allowed to
mix in the full 0$\hbar \omega$ space, using them as pivots in the LSF
procedure, they keep their identity to a large extent. This can be
seen in Fig.~\ref{fig:mg40}, where we have plotted
the strength functions of the 2p-2h 0$^+$ states in the full space. In $^{40}$Mg
the 2p-2h state represents the 60\%  of the ground state while in $^{52}$Cr it
is the dominant component (70\%) of the first excited 0$^+$.
This is a very interesting illustration  of the mechanism of
intrusion; the intruder state is present in both nuclei, but it is
only in the very neutron rich one that it becomes the ground state. 

\begin{table}
\caption{N=28 isotones: quasiparticle neutron gaps, difference in
  correlation energies between the 2p-2h and the 0p-0h configurations
  and their relative position}
\label{tab:n28gap}       % Give a unique label
 \begin{tabular*}{\linewidth}{@{\extracolsep{\fill}}lccccccccc} 
\hline\noalign{\smallskip}
& $^{40}$Mg 
& $^{42}$Si 
& $^{44}$S
& $^{46}$Ar 
& $^{48}$Ca
& $^{50}$Ti
& $^{52}$Cr
& $^{54}$Fe 
& $^{56}$Ni \\
\noalign{\smallskip}\hline\noalign{\smallskip}
gap         & 3.35    & 3.50     & 3.23  &  3.84  & 4.73 & 5.33 &
5.92 & 6.40 & 7.12 \\
$\Delta$E$_{\mathcal{C}orr}$   & 8.45    & 6.0      & 6.66  &  5.98  &
4.08 & 7.59 & 10.34 & 10.41 & 6.19 \\
E$^*_{2p-2h}$ & -1.75   & 1.0      & -0.2  &  1.7   & 5.38 &
3.07 & 1.50 & 2.39 & 8.05 \\
\noalign{\smallskip}\hline
 \end{tabular*}
\end{table}

Table~\ref{tab:n28} gives an idea of the properties of the isotones in
which configuration mixing is appreciable, but no clearcut deviation
from N=28 magicity can be detected. 
\begin{figure*}[hbt]
\resizebox{0.8\textwidth}{!}{%
  \includegraphics{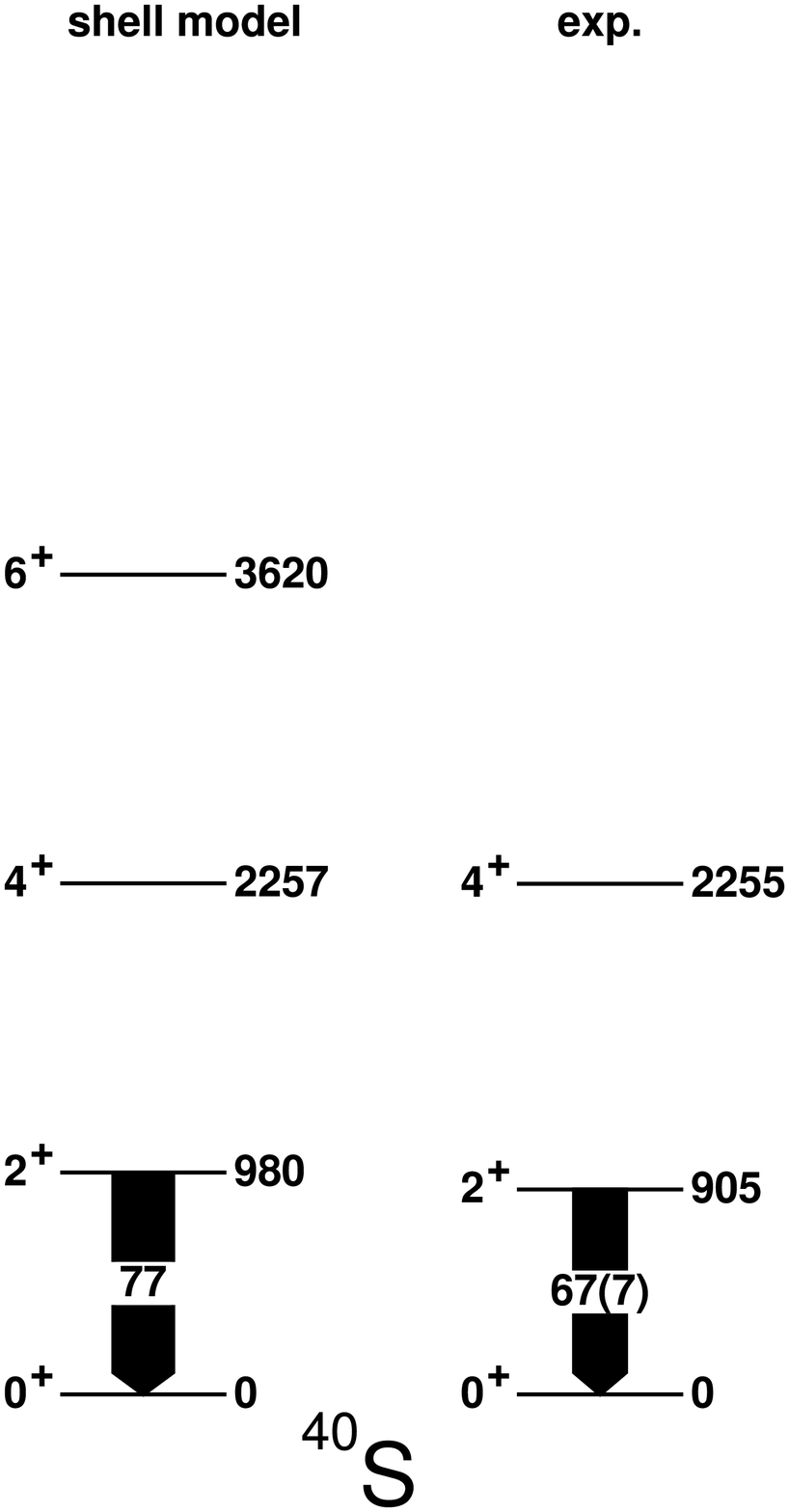}
  \includegraphics{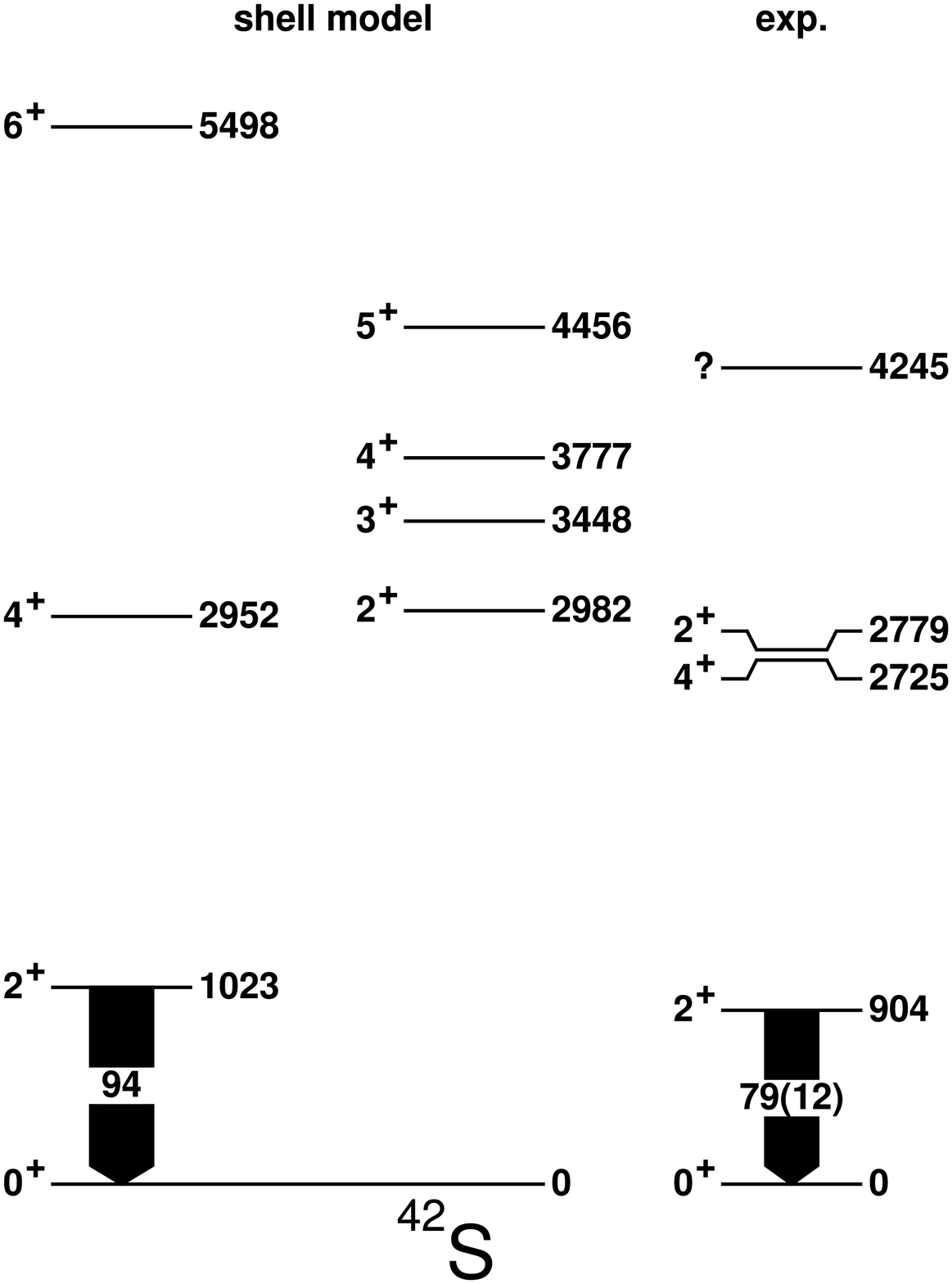}
  \includegraphics{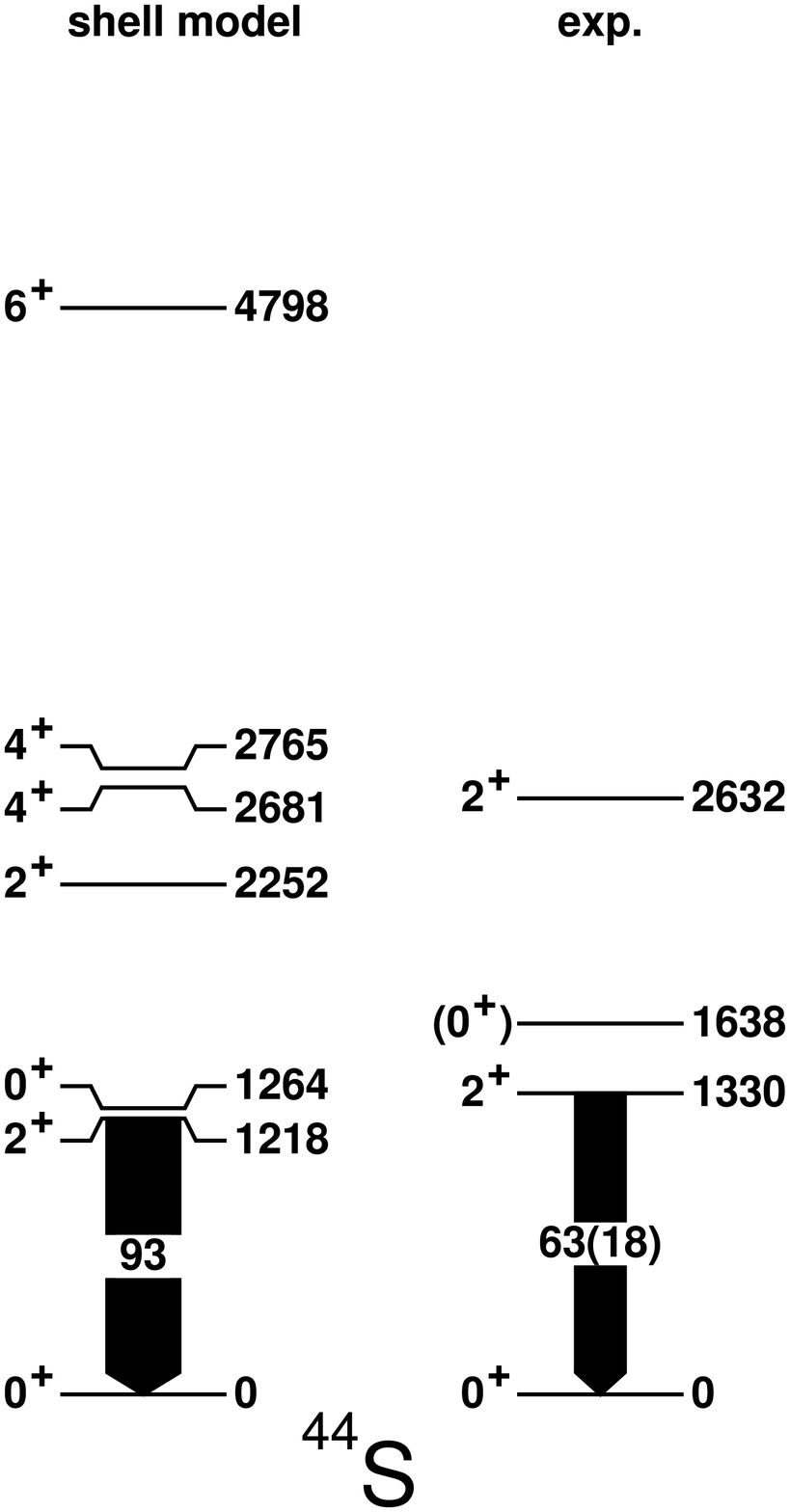}}
\caption{Predicted level schemes of the heavy sulphur isotopes,
         compared with the experiment. Energies in keV, B(E2)'s in
         e$^2$~fm$^4$}
\label{fig:sulphurs}       % Give a unique label
\end{figure*}
In Fig.~\ref{fig:sulphurs} we have plotted the low energy spectra of
the heaviest known sulphur isotopes. The agreement with the
accumulated experimental results \citep[see][for a recent
review]{Glasmacher:1998} is excellent and extends to the new data of
\citet{Sohler.ea:2002}.  Analyzing  their  proton occupancies
we conclude that the rise in
collectivity along the chain is correlated with the equal filling of
the $d_{3/2}$ and $s_{1/2}$ orbitals, (the $d_{5/2}$ orbital
remains always nearly closed).  The maximum proton collectivity is
achieved when both orbitals are degenerate, which corresponds to the
Pseudo-SU(3) limit. For the neutrons, the maximum collectivity occurs
at $N=24$, the $f_{7/2}$ mid-shell. According to the calculation,
$^{42}$S is a prolate rotor, with an incipient $\gamma$ band. In
$^{44}$S the spherical and deformed configurations mix equally.

The N=27 isotones also reflect the regular transition from sphericity
to deformation in their low-lying spectrum.  The excitation energy of
the $3/2^-$ state should be sensitive to the correlations and to the
neutron gap.  While in $^{47}$Ca, it lies quite high (at around 2~MeV)
due to the strong $f_{7/2}$ closure, in $^{45}$Ar it appears at about
0.4 MeV.  Concerning $^{43}$S, the information comes from two recent
experiments: the mass measurements at Ganil by
\citet{Sarazin.Savajols.ea:2000} that observed a low-lying isomer
around 400 keV excitation energy and the MSU Coulex experiment of
\citet{Ibbotson.Glasmacher.ea:1999} that detected a strong E2
transition from the ground state to an excited state around 940 keV.
According to our calculations the ground state corresponds to the
deformed configuration and has spin 3/2$^-$.  The spherical single
hole state 7/2$^-$ would be the first excited state and his lifetime
is consistent with that of the experimental isomer.  The third known
state is short lived and it should correspond to the 7/2$^-$ member of
the ground state band.
\begin{table}
\caption{N=28 isotones: Spectra, quadrupole properties and occupancies}
\label{tab:n28}       % Give a unique label
 \begin{tabular*}{\linewidth}{@{\extracolsep{\fill}}lcccc} 
\hline\noalign{\smallskip}
& $^{40}$Mg 
& $^{42}$Si 
& $^{44}$S
& $^{46}$Ar  \\
\noalign{\smallskip}\hline\noalign{\smallskip}
E$^*$(2$^+$)(MeV)   & 0.81   & 1.49     & 1.22 &  1.51     \\
E$^*$(4$^+$)   & 2.17   & 2.68     & 2.25 &  3.46      \\
E$^*$(0$^+_2$) & 1.83   & 1.57     & 1.26 &  2.93    \\
$Q(2^+$)(e~fm$^2$)   & -21   & 16      & -17    & 20  \\
B(E2)(e$^2$~fm$^4$) & 108    & 71     & 93   & 93    \\
$\langle {n}_{7/2}\rangle$  & 5.54 & 6.16 &  6.16  & 6.91  \\
$({f}_{7/2})^8$ \% & 3  & 28   & 24   & 45  \\
\noalign{\smallskip}\hline
 \end{tabular*}
\end{table}

\subsection{N=40; from ``magic'' $^{68}$Ni to deformed
  $^{64}$Cr}  \label{sec:n=40-from-magic}

Systematics is a most useful guide, but it has to be used with care.
The $p$ and $sd$ shells can be said to be ``full'' 0\hw\ spaces, in
the sense that the region boundaries are well defined by the $N=4$, 8
and $20$ closures, with the exception for the very neutron-rich halo
nuclei and the island of inversion discussed previously. In the $pf$-shell
it is already at $N=Z\approx 36$ that the 0\hw\ model space
collapses under the invasion of deformed intruders. Naturally, we
expect something similar in the neighborhood of 
$N=40$ isotones. Recent
experiments involving the Coulomb excitation of $^{66-68}$Ni
~\cite{Sorlin.ea:2002}, the $\beta$ decay of $^{60-63}$V
~\cite{Sorlin.ea:2003}, and the decay and the spectroscopy of the
$^{67m}$Fe ~\cite{Sawicka.Daugas.ea:2003}, support this hypothesis. In
addition, the $\beta$ decays of the neutron-rich isotopes $^{64}$Mn and
$^{66}$Mn, investigated at Isolde ~\cite{Hannawald.ea:1999}, that
show  a sudden drop of the $2^+$ energies in the daughter nuclei
$^{64}$Fe and $^{66}$Fe, point  also in the same direction.

In defining the new model spaces, the full $pf$ results are quite
useful: $^{56}$Ni turns out to be an extraordinarily good pivot (refer
to Fig.~\ref{fig:ni56.con}), and as soon as a few particles are added
it becomes a good core. Therefore, for the first time we are faced
with a genuine EI space $r_3\, g_{9/2}$ (remember $r_3$ is the
``rest'' of the $p=3$ shell). Its validity is restricted to nearly
spherical states. Deformation will demand the addition of the  
 $d_{5/2}$ subshell. 
\begin{table}[t]
\caption{2$^+_1$ energies and  
$g_{9/2}$ intruder occupation in the  Nickel isotopic chain, from
~\cite{Sorlin.ea:2002}}  
\begin{tabular*}{\linewidth}{@{\extracolsep{\fill}}lccccccc}
\hline\hline\noalign{\smallskip}
& $^{62}$Ni
& $^{64}$Ni 
& $^{66}$Ni
& $^{68}$Ni
& $^{70}$Ni
& $^{72}$Ni
& $^{74}$Ni  \\
\noalign{\smallskip}\hline\noalign{\smallskip}
E(2$^+$)$_{calc.}$   & 1.11  & 1.24     & 1.49 &  1.73 & 1.50 & 1.42 &
1.33  \\ [5pt] 
E(2$^+$)$_{exp.}$   & 1.173  & 1.346     & 1.425 &  2.033 & 1.259 &  &
\\ [5pt] 
B(E2$\uparrow$)$_{calc.}$ & 775    & 755     & 520   & 265 & 410  &
505 & 690\\ [5pt] 
%B(E2$\uparrow$)$_{exp.}$  & 915    & 590     & 650   & 255 &      &
%&   \\ [5pt] 
$\langle {n}_{9/2}\rangle$  & 0.24 & 0.43 &  0.67  & 1.07 & 0.84 &
0.55 & 0.45 \\ 
\noalign{\smallskip}\hline\hline
 \end{tabular*}
\label{tab:ni68}
\end{table}

\begin{figure}[htb]
  \includegraphics[width=0.9\linewidth]{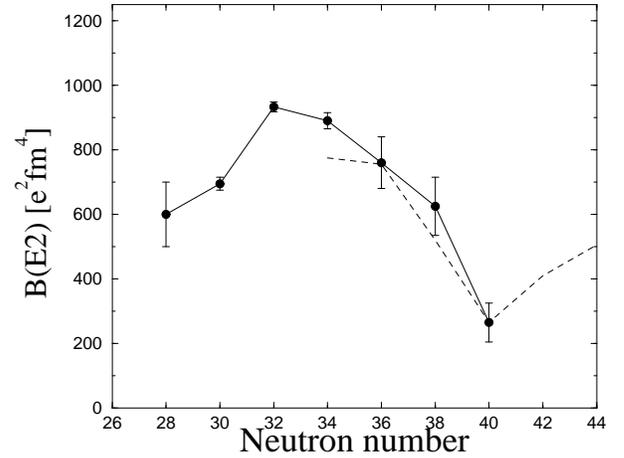}
  \caption{B(E2)  (0$^+$ $\rightarrow$ 2$^+$) in
    e$^2$~fm$^4$ for the Nickel isotopes, from
    ~\cite{Sorlin.ea:2002}\label{fig:nibe2}} 
\end{figure}

The  effective interaction is based on three
blocks: i) the TBME from KB3G effective
interaction~\cite{Poves.Sanchez-Solano.ea:2001}, ii) the G-matrix of
ref.~\cite{Hjorth-Jensen.Kuo.Osnes:1995} with the modifications of
ref.~\cite{Nowacki:1996} and iii) the Kahana, Lee and Scott G-matrix
\cite{Kahana.Lee.Scott:1969b} for the remaining matrix elements.  We
have computed all the nickel isotopes and followed their behavior at and
beyond N=40.  Let's first focus on $^{68}$Ni. The calculated low
energy spectrum, an excited 0$^+$ at 1.85~MeV, 2$^+$ at 2.15~MeV,
5$^-$ at 2.72~MeV and 4$^+$ at 3.10~MeV, fit remarkably well with the
experimental results. The ground state wave function is 50\% closed
shell. This number is very sensitive to modifications of the $pf$-$g$
gap, and smaller values of the closed shell probability can be obtained
without altering drastically the rest of the
properties. The SM results have been compared with those of other
methods in~\cite{Langanke.Terasaki.ea:2003}.  
\begin{table}[htb]
\centering
\caption{Spectroscopic properties of $^{60-64}$Cr in the $fpgd$ valence
  space}
\begin{tabular*}{\linewidth}{@{\extracolsep{\fill}}lccc}
\hline \hline
 \noalign{\smallskip} 
 & $^{60}$Cr
 & $^{62}$Cr
 & $^{64}$Cr \\
 \noalign{\smallskip}\hline\noalign{\smallskip}
E$^*$(2$^+$) (MeV)               &  0.67  & 0.65   & 0.51 \\
Q$_s$(e.fm$^2$)                  &  -23   & -27    & -31  \\
B(E2)$\downarrow$(e$^2$.fm$^4$)    &  288   & 302    & 318  \\
Q$_i$(e.fm$^2$) from Q$_s$       &   82   & 76     & 109  \\
Q$_i$(e.fm$^2$) from B(E2)       &  101   & 103    & 106  \\[10pt]
E$^*$(4$^+$) (MeV)               & 1.43   & 1.35   & 1.15 \\
Q$_s$(e.fm$^2$)                  & -37    & -30    & -43  \\
B(E2)$\downarrow$(e$^2$.fm$^4$)    & 426    & 428    & 471  \\
Q$_i$(e.fm$^2$) from Q$_s$       & 102    &  84    & 119  \\
Q$_i$(e.fm$^2$) from B(E2)       & 117    & 117    & 123  \\
\noalign{\smallskip} \hline \hline
\end{tabular*}
\label{tab:crfpgd}
\end{table}
In table~\ref{tab:ni68} we compare the excitation
energy of the 2$^+$ states in the Nickel chain with the available
experimental data (including the very recent $^{70}$Ni value
\cite{Sorlin.ea:2002}). The agreement is quite good, although the
experiment gives a larger peak at N=40. Similarly the B(E2)'s in
Fig.~\ref{fig:nibe2} follow the experimental trends including the drop at
$^{68}$Ni recently measured~\cite{Sorlin.ea:2002}. Note that for a
strict closure the transition would vanish.

\begin{figure}[htb]
  \includegraphics[angle=270,width=0.9\linewidth]{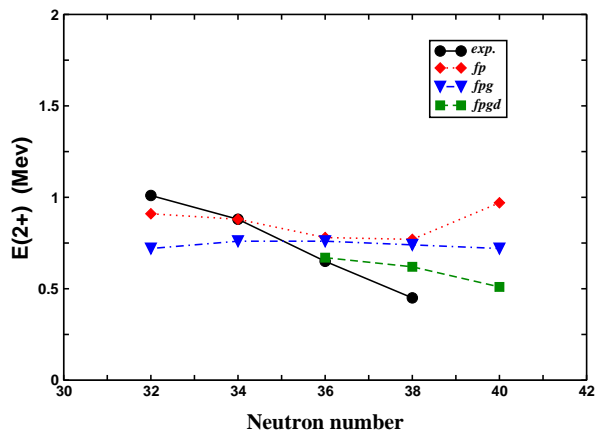} 
  \caption{Experimental and calculated 2$^+$ energies in the $pf$,
    $pfg$ and $pfgd$ spaces. \label{fig:crlev}}
\end{figure}

The $\beta$ decays of  $^{60}$V and $^{62}$V indicate very low energies
for the 2$^+_1$ states in $^{60}$Cr and $^{62}$Cr
\cite{Sorlin.ea:2000,Sorlin.ea:2003}.  If N=40 were a magic number,
one would expect higher 2$^+$ energies as the shell closure
approaches, particularly in $^{64}$Cr. As seen in
Fig.~\ref{fig:crlev}, exactly the opposite seems to happen
experimentally. The three possible model spaces tell an interesting
story: $pf\equiv r_3$ is acceptable at $N=32,\, 34$. The addition of
$g_{9/2}$ does more harm than good because the gaps have been
arbitrarily reduced to reproduce the $N=36$ experimental point by
allowing 2p-2h jumps including $d_{5/2}$. But then, in $N=38$ the
2$^+$ is too high, and the experimental trend promises no improvement
at $N=40$.

Properties of the 2$^+_1$ energies given by the $pfgd$ calculation are
collected in  table~\ref{tab:crfpgd}. They
point to prolate structures with deformation $\beta\approx 0.3$.

\section{Other regions and themes}
\label{sec:other}

\subsection{Astrophysical applications}
\label{sec:astro}
Astrophysical environments involve conditions of temperature and
densities that are normally not accessible in laboratory experiments.
The description of the different nuclear processes requires
then theoretical estimates.  As discussed in previous sections,
shell-model calculations are able to satisfactorily reproduce many
experimental results so that it should be possible to obtain reliable
predictions for nuclei and/or conditions not yet accessible
experimentally.

As a first example, consider the decay properties of nuclei totally
stripped of the atomic electrons at typical cosmic rays energies of
300~MeV/A. A nucleus such as $^{53}$Mn, unstable in normal conditions,
becomes stable.  Another isotope, $^{54}$Mn, could be used as a
chronometer to study the propagation of iron group nuclei on
cosmic-rays~\cite{Duvernois:1997} provided its half-life under
cosmic-rays conditions is known.  The necessary calculation involves
decay by second-forbidden unique transitions to the ground states of
$^{54}$Cr and $^{54}$Fe. Due to phase space arguments the $\beta^-$
decay to $^{54}$Fe is expected to dominate. In two difficult and
elegant experiments the very small branching ratio for the $\beta^+$
decay to the ground state of $^{54}$Cr has been measured: $(1.8 \pm
0.8)\times 10^{-9}$ by~\citet{Zaerpoor.Chan.ea:1997} and
$(1.20\pm0.26)\times 10^{-9}$ by~\citet{Wuosmaa.Ahmad.ea:1998}. Taking
the weighted mean of this values, $(1.26\pm0.25) \times 10^{-9}$, and
knowing that of $^{54}$Mn, 312.3(4)~d, a partial $\beta^+$ half-life
of $(6.8\pm 1.3)\times 10^8$~yr is obtained compared with $5.6\times
10^8$~yr from the shell model calculation
of~\citet{Martinez-Pinedo.Vogel:1998}, which yields $5.0\times
10^5$~yr for the dominant $\beta^-$ branch. Both theoretical results
are sensitive to uncertainties in the renormalization of the unique
second-forbidden operators that should be removed taking the ratio of
the $\beta^-$ and $\beta^+$ half-lives.  Multiplying the theoretical
ratio by the experimental value of the $\beta^+$ half-life yields a
value of $(6.0\pm 1.2)\times 10^5$~yr for the $\beta^-$ branch. Using
a similar argument \citet{Wuosmaa.Ahmad.ea:1998} estimate the partial
$\beta^-$ half-life to be $(6.0\pm 1.3[\mathrm{{stat}]\pm 1.1
  [\mathrm{theor}]})\times 10^5$. The 
influence of this half-life on the age of galactic cosmic-rays has been
discussed recently by~\citet{Yanasak.Wiedenbeck.ea:2001}. Another
possible cosmic ray chronometer, $^{56}$Ni, could measure the time
between production of iron group nuclei in supernovae and the
accelaration of part of this matterial to form cosmic
rays~\cite{Fisker.Martinez-Pinedo.Langanke:1999}. Before acceleration,
the decay of $^{56}$Ni proceeds by electron capture to the 1$^+$ state
in $^{56}$Co with a half-live of 6.075(20)~days. After acceleration
$^{56}$Ni is stripped of its electrons, the transition to the 1$^+$
state is no longer energetically allowed and the decay proceeds to the
3$^+$ state at 158~keV via a second forbidden unique transition.
Currently only a lower limit for the half-life of totally ionized
$^{56}$Ni could be established ($2.9\times
10^4$~y)~\cite{Sur.Norman.ea:1990}. A recent shell-model calculation
by~\cite{Fisker.Martinez-Pinedo.Langanke:1999} predict a half-live of
$4\times 10^4$~y that is too short for $^{56}$Ni to serve as a cosmic
ray chronometer.

Nuclear beta-decay and electron capture are important during the
late stages of stellar evolution~\citep[see][for a recent
review]{Langanke.Martinez-Pinedo:2003}.  At the relevant conditions in
the star electron capture and $\beta$ decay are dominated by
Gamow-Teller (and Fermi) transitions. Earlier determinations of the
appropriate weak interaction rates were based in the phenomenological
work of \citet*{Fuller.Fowler.Newman:1980,Fuller.Fowler.Newman:1982b,%
  Fuller.Fowler.Newman:1982a,Fuller.Fowler.Newman:1985}. The shell
model makes it possible to refine these estimates. For the $sd$ shell
nuclei, important in stellar oxygen and silicon burning, we refer
to~\citet{Oda.Hino.ea:1994}. More recently, it has been possible to
extend these studies to $pf$ shell nuclei relevant for the
pre-supernova evolution and
collapse~\cite{Langanke.Martinez-Pinedo:2001,
  Langanke.Martinez-Pinedo:2000,Caurier.Langanke.ea:1999}. The
astrophysical impact of the shell-model based weak interaction rates
have been recently studied
by~\citet{Heger.Langanke.ea:2001,Heger.Woosley.ea:2001}

\begin{figure}[htbp]
  \includegraphics[width=0.90\linewidth]{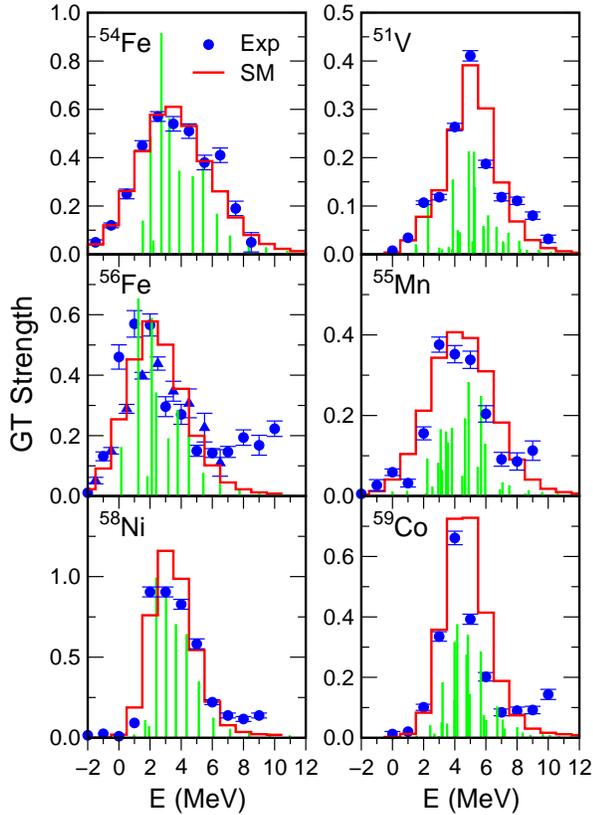}
  \caption{Comparison of shell model GT$_+$ distributions with
    experimental data \cite{Roennqvist.Conde.ea:1993,%
      El-Kateb.Jackson.ea:1994,Alford.Brown.ea:1993} for selected
    nuclei. The shell model results (discrete lines) have been folded
    with the experimental resolution (histograms) and include a
    quenching factor of $(0.74)^2$ \citep[adapted
    from][]{Caurier.Langanke.ea:1999}.
     \label{fig:GTplus}}
\end{figure}

\begin{figure}[htbp]
  \includegraphics[width=0.90\linewidth]{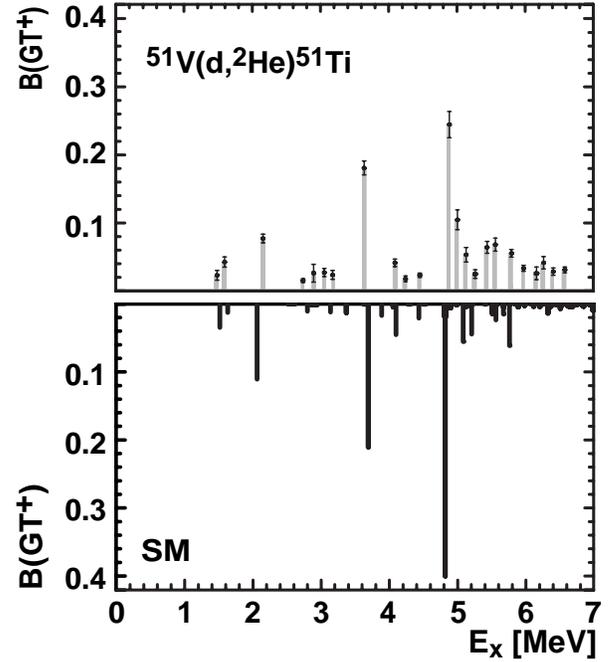}
  \caption{Comparison of the shell-model GT$_+$ distribution (lower
    panel) for $^{51}$V with the high resolution $(d,{}^2\text{He})$
    data \citep[from][]{Baeumer.Berg.ea:2003}. The shell-model
    distribution includes a quenching factor of $(0.74)^2$.
    \label{fig:v51comp}}
\end{figure}

The basic ingredient in the calculation of the different weak
interaction rates is the Gamow-Teller strength distribution. The
GT$_+$ sector directly determines the electron capture
rate and also contributes to the beta-decay rate through the thermal
population of excited states~\cite{Fuller.Fowler.Newman:1982b}. The
GT$_-$ strength contributes to the determination of the
$\beta$-decay rate. To be applicable to calculating stellar weak
interaction rates the shell-model calculations should reproduce the
available GT$_+$ (measured by $(n,p)$-type reactions) and GT$_-$
(measured in $(p,n)$-type reactions). Figure~\ref{fig:GTplus} compares
the shell-model $GT_+$ distributions with the pioneering measurements
performed at TRIUMF. These measurements had a typical energy
resolution of $\approx 1$~MeV. Recently developed techniques,
involving $(d,{}^2\text{He})$ charge-exchange reactions at
intermediate energies~\cite{Rakers.Baeumer.ea:2002}, have improved
the energy resolution by an order of magnitude or
more. Figure~\ref{fig:v51comp} compares the shell-model GT$_+$
distribution computed using the KB3G
interaction~\cite{Poves.Sanchez-Solano.ea:2001} with a recent
experimental measurement of the $^{51}$V$(d,{}^2\text{He})$ performed
at KVI~\cite{Baeumer.Berg.ea:2003}.

\begin{figure}[htbp]
  \includegraphics[width=0.90\linewidth]{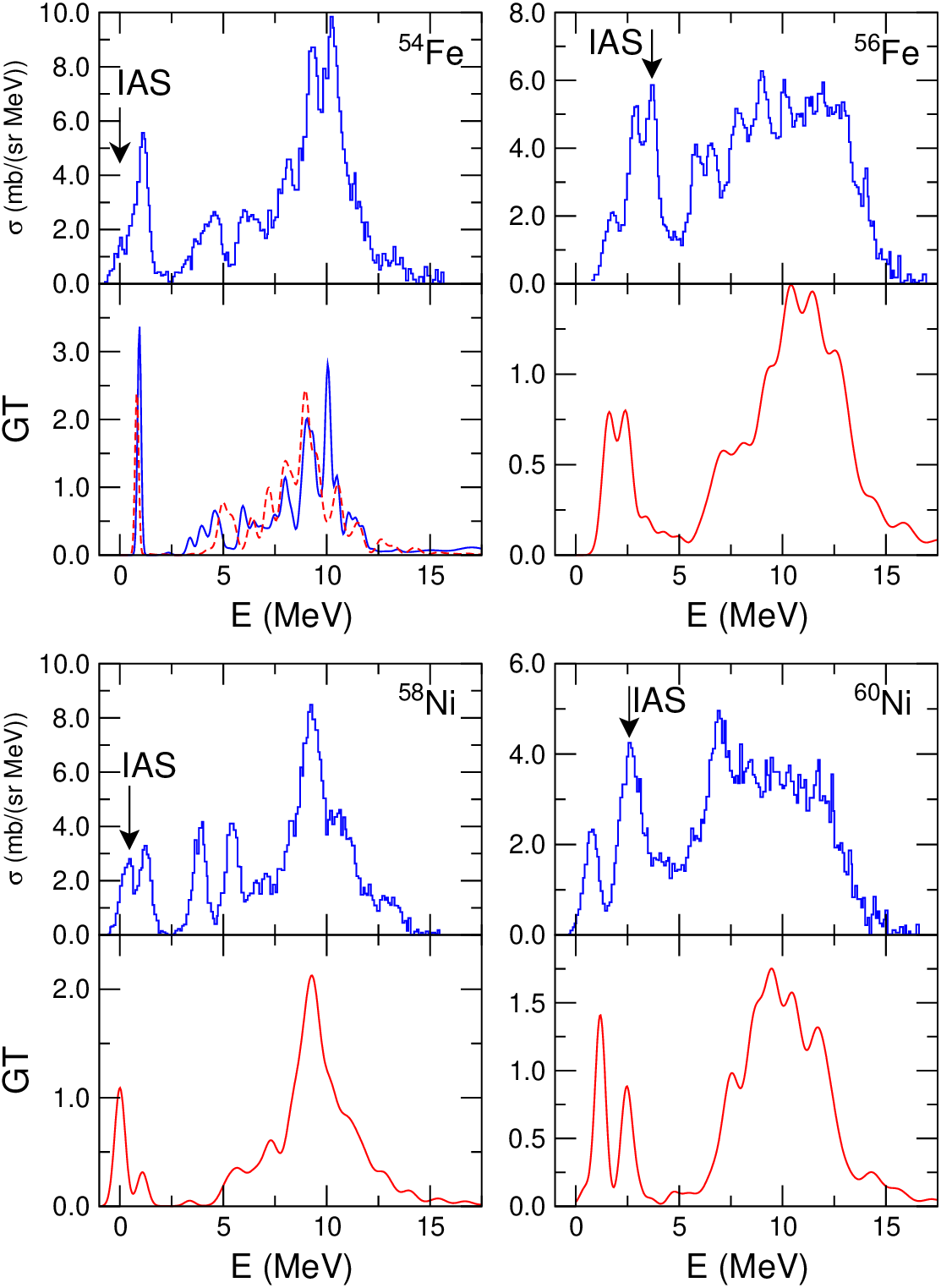}
  \caption{Comparison of $(p,n)$ $L=0$ forward angle cross section
    data~\cite{Rapaport.Taddeucci.ea:1983} (upper panels) with the
    calculated GT$_-$ strength distributions (lower panels). For
    $^{54}$Fe the solid curve in the lower panel shows the
    experimental GT$_-$ data from~\cite{Anderson.Lebo.ea:1990}, while
    the shell model results are given by the dashed line. The
    Fermi transition to the isobaric analog state (IAS) is included in
    the data (upper panels) but not in the calculations. A quenching
    factor of $(0.74)^2$ is included in the calculations.
    \citep[from][]{Caurier.Langanke.ea:1999}. \label{fig:GTminus}}
\end{figure}

Figure~\ref{fig:GTminus} compares the shell-model GT$_-$ distributions
with the data obtained in $(p,n)$ charge-exchange reaction
measurements for $^{54,56}$Fe and
$^{58,60}$Ni~\cite{Rapaport.Taddeucci.ea:1983,Anderson.Lebo.ea:1990}.
The GT$_-$ operator acting on a nucleus with neutron excess and ground
state isospin $T$ can lead to states in the daughter nucleus with
three different isospin values $(T-1,T,T+1)$. As a consequence, the
GT$_-$ strength distributions have significantly more structure and
extend over a larger excitation energy interval than the GT$_+$
distributions, making their theoretical reproduction more challenging.
Nevertheless, the agreement with the experimental data is quite
satisfactory. The shell-model results for $^{58}$Ni have been recently
compared with high-resolution data (50~keV) obtained using the
$({}^3\text{He},t)$ reaction~\cite{Fujita.Fujita.ea:2002}.

Shell-model diagonalization techniques have been used to determine
astrophysically relevant weak interaction rates for nuclei with $A\leq
65$. Nuclei with higher masses are relevant to study the collapse
phase of core-collapse
supernovae~\cite{Langanke.Martinez-Pinedo:2003}. The calculation of
the relevant electron-capture rates is currently beyond the
possibilities of shell-model diagonalization calculations due to the
enormous dimensions of the valence space. However, this dimensionality
problem does not apply to Shell-Model Monte-Carlo methods (SMMC, see
section~\ref{sec:monte-carlo-methods}). Moreover, the high
temperatures present in the astrophysical environment makes necessary
a finite temperature treatment of the nucleus, this makes SMMC methods
the natural choice for this type of calculations. Initial studies
by~\citet{Langanke.Kolbe.Dean:2001} showed that the combined effect of
nuclear correlations and finite temperature was rather efficient in
unblocking Gamow-Teller transitions on neutron rich germanium
isotopes. More recently this calculations have been extended to cover
all the relevant nuclei in the range $A=65$--112
by~\citet{Langanke.Martinez-Pinedo.ea:2003}. The resulting
electron-capture rates have a very strong influence in the
collapse~\cite{Langanke.Martinez-Pinedo.ea:2003} and
post-bounce~\cite{Hix.Messer.ea:2003}

The astrophysical r-process is responsible for the synthesis of at
least half of the elements heavier than $A\approx
60$~\cite{Wallerstein.Iben.ea:1997}. Simulations of the r-process
require the knowledge of nuclear properties far from the valley of
stability~\cite{Kratz.Pfeiffer.Thielemann:1998,Pfeiffer.Kratz.ea:2001}.
As the relevant nuclei are not experimentally accessible, theoretical
predictions for the relevant quantities (i.e., neutron separation
energies and half-lives) are needed. The calculation of $\beta$ decay
half-lives usually requires two ingredients: the Gamow-Teller strength
distribution in the daughter nucleus and the relative energy scale
between parent and daughter (i.e. the $Q_\beta$ value). Due to the
huge number of nuclei relevant for the r process, the estimates of the
half-lives are so far based on a combination of global mass models and
the quasi particle random-phase approximation \citep[see][for a
description of the different models]{Langanke.Martinez-Pinedo:2003}.
However, recently shell-model calculations have become available for
some key nuclei with a magic neutron number
$N=50$~\cite{Langanke.Martinez-Pinedo:2003},
$N=82$~\cite{Brown.Clement.ea:2003,Martinez-Pinedo.Langanke:1999}, and
$N=126$~\cite{Martinez-Pinedo:2001}. All this calculations suffer from
the lack of spectroscopic information on the regions of interest that
is necessary to fine tune the effective interactions. This situation
is improving at least for $N=82$ thanks to the recent spectroscopic
data on $^{130}$Cd~\cite{Dillmann.Kratz.ea:2003}.

Nuclear reaction rates are the key input data for simulations of
stellar burning processes. Experiment-based reaction rates for the
simulation of explosive processes such as novae, supernovae, x-ray
bursts, x-ray pulsars, and merging neutron stars are scarce because of
the experimental difficulties associated with radioactive beam
measurements~\cite{Kaeppeler.Thielemann.Wiescher:1998}. Most of the
reaction rate tables are therefore based on global model predictions.
The most frequently used model is the statistical Hauser Feshbach
approach~\cite{Rauscher.Thielemann:2000}. For nuclei near the
drip-lines or near closed shell configurations, the density of levels
is not high enough for the Hauser Feshbach approach to be applicable.
For these cases alternative theoretical approaches such as the nuclear
shell model need to be applied. Shell-model calculations were used for
the determination of the relevant proton capture reaction rates for
$sd$-shell nuclei necessary for rp process
studies~\cite{Herndl.Goerres.ea:1995}. This calculations have been
recently extended to include $pf$-shell
nuclei~\cite{Fisker.Barnard.ea:2001}.

Knowledge of neutrino nucleus reactions is necessary for many
applications, e.g. the neutrino oscillation studies, detection of
supernova neutrinos, description of the neutrino transport in
supernovae and nucleosynthesis studies. Most of the relevant neutrino
reactions have not been studied experimentally so far and their cross
sections are typically based on nuclear theory~\citep[see][for a
recent review]{Kolbe.Langanke.ea:2003}. The model of choice for the
theoretical description of neutrino reactions depends of the energy of
the neutrinos that participate in the reaction. 

For low neutrino energies, comparable to the nuclear excitation
energy, neutrino-nucleus reactions are very sensitive to the
appropriate description of the nuclear response that is very sensitive
to correlations among nucleons. The model of choice is then the
nuclear shell-model. $0\hbar\omega$ calculations have been used for
the calculation of neutrino absorption cross
sections~\cite{Sampaio.Langanke.Martinez-Pinedo:2001} and scattering
cross sections~\cite{Sampaio.Langanke.ea:2002} for selected $pf$ shell
nuclei relevant for supernovae evolution. For lighter nuclei complete
diagonalizations can be performed in larger model spaces, e.g.
$4\hbar\omega$ calculations for
$^{16}$O~\cite{Haxton:1987,Haxton.Johnson:1990} and $6\hbar\omega$
calculations for
$^{12}$C~\cite{Hayes.Towner:2000,Volpe.Auerbach.ea:2000}. Other
examples of shell-model calculations of neutrino cross sections are
the neutrino absorption cross sections on $^{40}$Ar
of~\citet{Ormand.Pizzochero.ea:1995} for solar
neutrinos~\citep[see][for an experimental evaluation of the same cross
section]{Bhattacharya.Garcia.ea:1998}, this cross section have been
recently evaluated by~\cite{Kolbe.Langanke.ea:2003} for supernova
neutrinos. And the evaluation by~\citet{Haxton:1998} of the solar neutrino
absorption cross section on $^{71}$Ga relevant for the GALLEX and SAGE
solar neutrino experiments. 

For higher neutrino energies the standard method of choice is the
random phase approximation as the neutrino reactions are sensitive
mainly to the total strength and energy centroids of the different
multipoles contributing to the cross section. In some selected cases,
the Fermi and Gamow-Teller contribution to the cross section could be
determined from shell-model calculation that is supplemented by RPA
calculations for higher multipoles. This type mix calculation has been
carried out for several iron
isotopes~\cite{Kolbe.Langanke.Martinez-Pinedo:1999,Toivanen.Kolbe.ea:2001}
and for $^{20}$Ne~\cite{Heger.Kolbe.ea:2003}.

\subsection{$\beta\beta$-decays} 
\label{sec:betabeta-decays}

The double beta decay is the rarest nuclear weak process.  It takes place
between two even-even isobars, when the decay to the intermediate
nucleus is energetically forbidden or hindered by the large spin
difference between the parent ground state and the available states in
the intermediate nuclei.  It comes in three forms: The two-neutrino
decay $\beta\beta_{2\nu}$:
$$\displaystyle ^A_Z X _N \longrightarrow _{Z+2}^A X_{N-2} + e^-_1 +
e^-_2 + \bar{\nu}_1 + \bar{\nu}_2 $$
is just a second order process
mediated by the Standard model weak interaction It conserves the
lepton number and has been already observed in a few nuclei.

The second mode, the neutrinoless decay $\beta\beta_{0\nu}$:
$$\displaystyle ^A_Z X _N \longrightarrow ^A _{Z+2} X_{N-2} + e^-_1 +
e^-_2$$
needs an extension of the standard model of electroweak
interactions as it violates lepton number.  A third mode,
$\beta\beta_{0\nu,\chi}$ is also possible
$$\displaystyle ^A_Z X _N \longrightarrow ^A _{Z+2} X_{N-2} + e^-_1 +
e^-_2 + \chi $$
in some extensions of the standard model and proceeds
via emission of a light neutral boson, a Majoron $\chi$.  The last two
modes, not yet experimentally observed, require massive neutrinos --an
issue already settled by the recent measures by
Super-Kamiokande~\cite{Fukuda.Hayakawa.ea:1998b},
SNO~\cite{Ahmad.Allen.ea:2002a}
 and KamLAND~\cite{Eguchi.Enomoto.ea:2003}.
Interestingly, the double beta decay without emission of neutrinos
would be the only way to sign the Majorana character of the neutrino
and to distinguish between the different scenarios for the neutrino
mass differences.  Experimentally, the three modes show different
electron energy spectra (\citep[see figure 2 in][]{Zdesenko:2002}),
the $\beta\beta_{2\nu}$ and $\beta\beta_{0\nu,\chi}$ are characterized
by a continuous spectrum ending at the maximum available energy
$Q_{\beta\beta}$, while the $\beta\beta_{0\nu}$ spectrum consists in a
sharp peak at the end of the $Q_{\beta\beta}$ spectrum. This should,
in principle, make it easier the signature of this mode.  In what
follows we shall concentrate in the $\beta\beta_{2\nu}$ and
$\beta\beta_{0\nu}$ modes.

The theoretical expression of the  half-life of the  2$\nu$ mode can
be  written as:
\begin{equation} 
 [T^{2\nu}_{1/2}]^{-1} = G_{2\nu} |M^{2\nu}_{GT}|^2,
\end{equation}
with
\begin{equation} 
M^{2\nu}_{GT}(J)= \sum_{m} \displaystyle{
 \frac{ \langle J^+ || \vec{\bm{\sigma}}t_- || 1^+_m \rangle 
        \langle 1^+_m || \vec{\bm{\sigma}}t_- || 0^+ \rangle}    
      {E_m + E_0(J)}}
\end{equation}
(there is an implicit sum over all the nucleons). 
$G_{2\nu}$ contains the phase space factors and the axial coupling constant
$g_A$. The calculation of $M^{2\nu}_{GT}$ 
requires the precise knowledge of the ground state of the parent nuclei 
and the ground state and occasionally a few excited states of the 
grand daughter, both even-even. Besides, it is necessary to
have a good description of the Gamow- Teller strength functions of
both of them, which implies a detailed description of the odd-odd
intermediate nucleus. This is why this calculation is a challenge for
the nuclear models, and why agreement with the experiment in this channel
is taken as a quality factor to be applied to the predictions  of the 
models for  the neutrinoless mode.

\begin{figure}[htb]
  \includegraphics[width=0.9\linewidth]{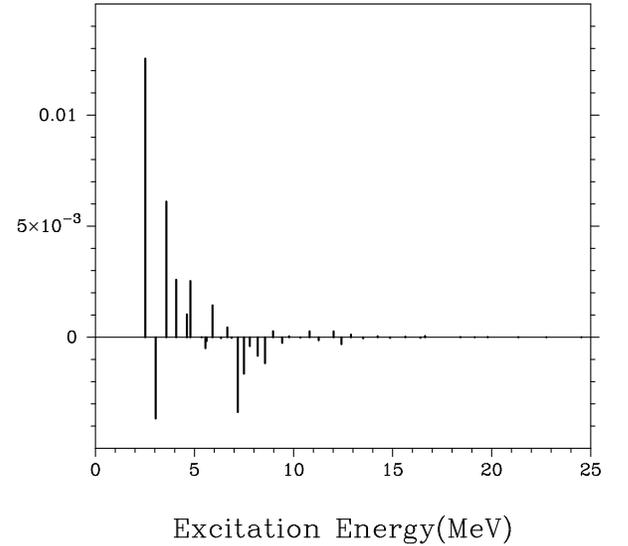}
  \caption{LSF for $^{48}$Ca $\rightarrow$ $^{48}$Ti 2$\nu$ decay. 
    Each bar corresponds to a contribution to the matrix element.
    Notice the interfering positive and negative contributions.
    \label{fig:a48bb2nu}}
\end{figure}

It is also a show-case of the use of the Lanczos strength function (LSF) method. It works as
follows: Once the relevant wave functions of parent and grand daughter
are obtained, it is straightforward to build the doorway states
$\vec{\sigma}t_- | 0^+_{initial} \rangle$ and $\vec{\sigma}t_+ |
J^+_{final} \rangle$. In a second step, one of them is fragmented
using LSF, producing, at iteration N, N 1$^+$ states in the
intermediate nucleus, with excitation energies E$_m$. Overlapping
these vectors with the other doorway, entering the appropriate energy
denominators and adding up the N contributions gives an approximation
to the exact value of $M^{2\nu}_{GT}$ (N=1 is just the closure
approximation).  Finally, the number of iterations is increased until
full convergence is reached.  The method is very efficient, for
instance, in the A=48 case, 20 iterations suffice largely. The
contributions of the different intermediate states to the final matrix
element are  plotted in  Fig.~\ref{fig:a48bb2nu}.

 In the $pf$-shell there is only one double
 beta emitter, namely $^{48}$Ca. For many years, the experimental
 information was limited to a lower limit on the 2$\nu$-$\beta \beta$
 half-life T$^{2\nu}_{1/2}$$>$~3.6$\times$10$^{19}$~$yr$
 \cite{Bardin.Gollon.ea:1970}. 
 The calculation of the   $^{48}$Ca half-life was one of the first
 published results stemming from the full $pf$-shell calculations
 using the code ANTOINE. 
  The resulting matrix elements are:
$M^{2\nu}_{GT}(0^+)$=0.083 and $M^{2\nu}_{GT}(2^+)$=0.051.
For the ground state to ground state decay 
$^{48}$Ca$\rightarrow$$^{48}$Ti, G$_{2\nu}$=1.1$\times$10$^{-17}$~$yr^{-1}$
\cite{Tsuboi.Muto.Horie:1994}.
The phase space factor hinders the transition to the 2$^+$ that
represents only about 3\% of the total probability. Putting everything
together, and using the Gamow-Teller quenching factor, already
 discussed,  the resulting half-life is 
 T$^{2\nu}_{1/2}$$=$~3.7$\times$10$^{19}$~$yr$,
\cite{Caurier.Poves.Zuker:1990} \citep[see also erratum
in][]{Caurier.Zuker.ea:1994}. The prediction was a success, because
 a later measure gave
T$^{2\nu}_{1/2}$$=$~4.3$^{+2.4}_{-1.1}$[stat]$\pm$1.4[syst] 
 $\times$10$^{19}$~$yr$
\cite{Balysh.DeSilva.ea:1996}. 

Among the other $\beta\beta$ emitters in nature (around 30), only a few  
are potentially interesting for experiment since they have a
$Q_{\beta\beta}$ value, sufficiently large ($\ge 2.5 MeV$) for the
$0\nu$  
signal not to be drowned in the surrounding natural radioactivity.
With the exception of $^{150}$Nd, all of them can be described 
within a shell model approach. The results for the
$2\nu$ mode of the lightest emitters  $^{48}$Ca,  $^{76}$Ge
and  $^{82}$Se, for which  full space calculations are doable, are
gathered 
in table~\ref{tab:2nu} \cite{Caurier.Nowacki.ea:1996}. The SMMC method
has also been applied to the calculation of 2$\nu$ double beta decays
in~\cite{Radha.Dean.ea:1996}.

\begin{table}[htb]
\caption{Calculated T$_{1/2}^{2\nu}$ half-lives for several nuclei and
 $0^+ \rightarrow 0^+$ transitions}
    \begin{tabular*}{\linewidth}{@{\extracolsep{\fill}}cccc}
\hline\hline \noalign{\smallskip}  
% \begin{tabular}{cccc} 
Parent                  &  $^{48}Ca$       &  $^{76}Ge$     &
$^{82}Se$    \\
$T_{1/2}^{2\nu}$ th.(y) &  $ 3.7 \times 10^{19}$  & $ 2.6 \times 10^{21}$
& $3.7 \times 10^{19}$ \\
$T_{1/2}^{2\nu}$ exp.(y)&  $ 4.3 \times 10^{19}$  & $ 1.8 \times
10^{21}$ & $8.0 \times 10^{19}$ \\
\hline \hline
\label{tab:2nu}
\end{tabular*}
 \end{table}

The expression for the neutrinoless beta decay half-life, in the
$0^{+}\rightarrow 0^{+}$ case,  can be brought
to the following form 
\cite{Takasugi:1981,Doi.Kotani.Takasugi:85}:
\begin{equation}
 [T_{1/2}^{(0\nu)}(0^{+}->0^{+}]^{-1}=\nonumber 
\end{equation}
\begin{equation}
  G_{0\nu}\left(M_{GT}^{(0\nu)} - \left(\frac{g_V}{g_A}\right)^{2}
M_{F}^{(0\nu)}\right)^2 \left(\frac{\langle m_{\nu}
  \rangle}{m_{e}}\right)^{2}\nonumber 
\label{eq:0vh1}
\end{equation}
where $\langle m_{\nu} \rangle$ is the effective neutrino mass,
$G_{0\nu}$ the 
kinematic space factor and M$_{GT}^{(0\nu)}$ and  M$_{F}^{(0\nu)}$ the
following matrix elements ($m$ and $n$ sum over nucleons):
\begin{equation}
M_{GT}^{(0\nu)}=\left<0_{f}^{+}\right\|\displaystyle\sum_{n,m}h(r)
(\vec{\bm{\sigma}}_{n}\cdot
\vec{\bm{\sigma}}_{m}) t_{n\_}t_{m\_}\left\|0_{i}^{+}\right>
\end{equation}
\begin{equation}
M_{F}^{(0\nu)}=\left<0_{f}^{+}\right\|\displaystyle\sum_{n,m}h(r) 
t_{n\_}t_{m\_}
\left\|0_{i}^{+}\right>,
\end{equation}
 where, due to the presence of the neutrino propagator, the ``neutrino 
potential'' $h(r)$ is introduced.  
In this case, the matrix elements are just expectation values of two
body operators, without a sum over intermediate states. This was
believed to make the results less dependent of the nuclear model
employed to obtain the wave function. An assumption that has not
survived to the actual calculations. Full details of the calculations
as well as predictions for other 0$\nu$ and 2$\nu$ decays in heavier
double beta emitters can be found in
\citet{Retamosa.Caurier.Nowacki:1995} and
\citet{Caurier.Nowacki.ea:1996} \citep[see also][for a recent and very
comprehensive review of the nuclear aspects of the double beta
decay]{Suhonen.Civitarese:1998}. The upper bounds on the neutrino mass
resulting of our SM calculations, assuming a reference half-life
T$_{1/2}^{0\nu} \ge 10^{25} \; y.$ are collected in
table~\ref{tab:0nu}.
 
\begin{figure}[htb]
  \includegraphics[width=0.8\linewidth]{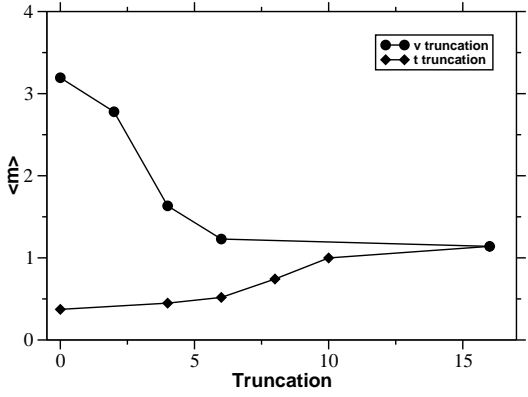}
  \caption{Variation of the $\langle m_{\nu} \rangle$ limit as a
    function of seniority truncations (circles) and configuration
    truncations (diamonds) \label{fig:bbconv}}
\end{figure}

\begin{table}[htb]
 \caption{0$\nu$ matrix elements and upper bounds on the neutrino mass
   for 
 T$_{1/2}^{0\nu} \ge 10^{25}$  y. $\langle m_{\nu} \rangle$ in eV.}
    \begin{tabular*}{\linewidth}{@{\extracolsep{\fill}}cccc}
\hline\hline \noalign{\smallskip}  
Parent             &   $^{48}Ca$ & $^{76}Ge$  & $^{82}Se$ \\ 
$M_{GT}^{0\nu}$    &        0.63 & 1.58       & 1.97     \\
$M_{F}^{0\nu}$     &       -0.09 & 0.19       & -0.22    \\
$\langle m_{\nu} \rangle$ & 0.94 & 1.33       & 0.49     \\ 
\hline \hline
\label{tab:0nu}
\end{tabular*}
\end{table}
For the heavier emitters, some truncation scheme has to be employed and
the seniority truncation seem to be the best, since the dimensions
are strongly reduced in particular for 0$^+$ states.  An interesting
feature in the calculation of the double beta decay matrix elements is
shown in Fig.~\ref{fig:bbconv} for the $^{76}$Ge case : the
convergence of the $M_{GT}^{0\nu}$ matrix element is displayed as a
function of the truncation of the valence space, either with the
seniority $v$ or with the configurations $t$ (t is the maximum number
of particles in the $g_\frac{9}{2}$ orbital).  The $t=16$ and $v=16$
values correspond to the full space calculation and are consequently
equal. The two truncation schemes show very distinct patterns, with
the seniority truncation being more efficient.  Such patterns have
been also observed in the Tellurium and Xenon isotopes.  This seems to
indicate that the most favorable nuclei for the theoretical
calculation of the $0\nu$ mode would be the spherical emitters, where
seniority is an efficient truncation scheme.

\smallskip

\subsection{Radii isotope shifts in the Calcium isotopes}
\label{sec:is}

We have already seen in section~\ref{sec:rotat-bands-unnat} that
deformed np-nh configurations can appear at very low excitation energy
around shell closures and even to become yrast in the case of neutron
rich nuclei. This situation simply reflects the limitation of the
spherical mean field
description of the nucleus, and shows that even in magic
cases, the correlations produce a sizeable erosion of the Fermi
surface. This is the case in the Calcium isotopic chain.  Experiments
based on optical isotope shifts or muonic atom data, reveal that the
nuclear charge radii $\langle r_c^2 \rangle$ follow a characteristic
parabolic shape with a pronounced odd-even staggering
~\cite{Fricke.Bernhardt.ea:1995,Palmer.ea:1984}.

For the description of the nuclei around N=Z=20, a model space
comprising the orbits 1d$_\frac{3}{2}$, 2s$_\frac{1}{2}$,
1f$_\frac{7}{2}$ and 2p$_\frac{3}{2}$ is a judicious choice
\citep[see][]{Caurier.Langanke.ea:2001}.  The interaction is the same
used to describe the neutron rich nuclei in the $sd$-$pf$ valence
space, called sdpf.sm in section~\ref{sec:descr-very-neutr}, with
modified single particle energies to reproduce the spectrum of
$^{29}$Si.
   
\begin{figure}[htb]
   \includegraphics[width=0.9\linewidth]{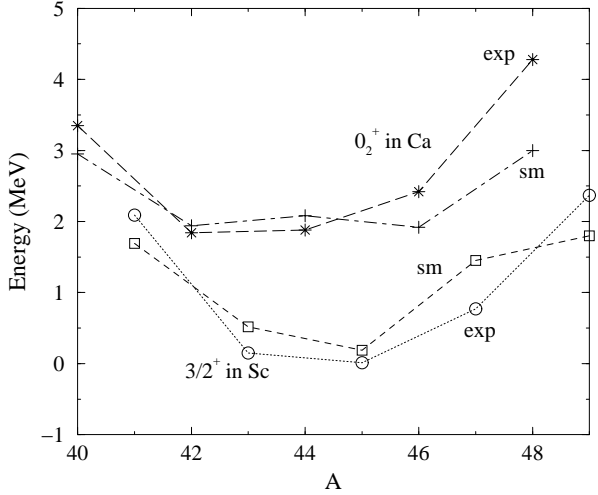}
   \caption{Comparison between calculated and experimental intruder
     excitation energies in the Calcium and Scandium isotopic chains
    \label{fig:zbm2_intr}}
\end{figure}

Several important features of the nuclei at the $sd$/$pf$ interface
are reproduced by the calculation, among others, the excitation
energies of the intruder 0$^+_2$ states in the Calcium isotopes as
well as the location of the $\frac{3}{2}^+$ states in the Scandium
isotopes (see Fig.~\ref{fig:zbm2_intr}), the excitation energies of
the 2$^+$ and 3$^-$ and the B(E2)'s between the 2$^+_1$ and the
0$^+_1$ states in the Calcium isotopes.

\begin{figure}[htb]
  \includegraphics[width=0.9\linewidth]{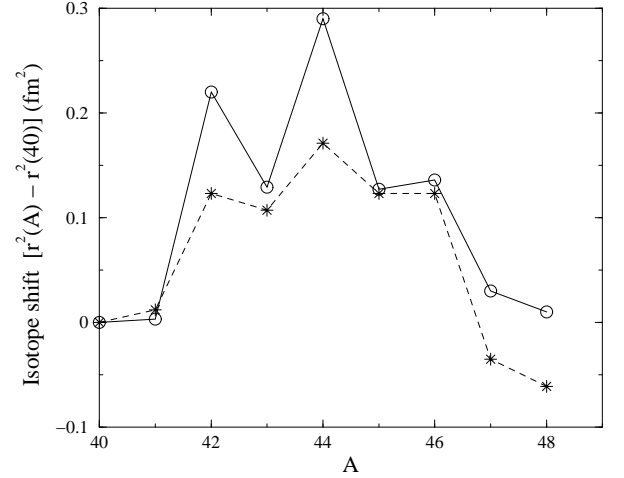}
   \caption{Isotopes shifts in Calcium chain: experimental (circles)
     versus shell model calculations (stars)
    \label{fig:zbm2_is}}
\end{figure}

Due to the cross shell pairing interaction, protons and neutrons
are lifted from the $sd$ to the $fp$ orbitals. The former  produce  an
increase of   
$\langle r^2_c \rangle$ that, using HO wave functions,  can be expressed as: 
\begin{equation}
\delta r^2_c(A) = \displaystyle\frac{1}{Z}n^\pi_{pf}(A)b^2
\end{equation}
where Z=20 for the calcium chain, $b$ is the oscillator parameter, and
$n^\pi_{pf}$ 
is extracted from the calculated wave functions. 
The charge radii isotope shifts of the Calcium isotopes relative
to $^{40}$Ca are shown in Fig.~\ref{fig:zbm2_is}, together with the
experimental values. 
The global trends are very well reproduced, although 
the calculated shifts are a bit smaller than 
the experimental ones. This is probably due to the
limitations in the  valence space, that excludes  the
1d$_\frac{5}{2}$, the 
1f$_\frac{5}{2}$ and the 2p$_\frac{1}{2}$ orbits.

\subsection{Shell model calculations in heavier nuclei} 
\label{sec:n126}

 There are some regions of heavy nuclei in which physically sound valence
 spaces can be designed that are at the same time tractable. A good
 example is the space comprising the neutron orbits between N=50 and
 N=82 for the tin isotopes
 \cite{Hjorth-Jensen.Kuo.Osnes:1995,Nowacki:1996}. When protons are
 allowed the dimensions grow rapidly and the calculations have been
 limited until now to nuclei with few particle or holes on the top of
 the closed 
 shells~\cite[see][for a review of the work of the Napoli
 group]{Covello.Andreozzi.ea:1997}.
 
 The SMMC, that can overcome these limitations, has been applied in
 this region to $^{128}$Te and $^{128}$Xe that are candidates to
 $\gamma$ soft nuclei by \citet{Alhassid.Bertch.ea:1996} and also to
 several Dysprosium isotopes A=152-162 in the Kumar-Baranger space by
 \citet{White.Koonin.ea:2000}.  The QMCD method has been also applied
 to the study of the spherical to deformed transition in the even
 Barium isotopes with A=138-150 by \citet{Shimizu.Otsuka.ea:2001}.

\begin{figure}[hbt]
  \includegraphics[width=0.9\linewidth]{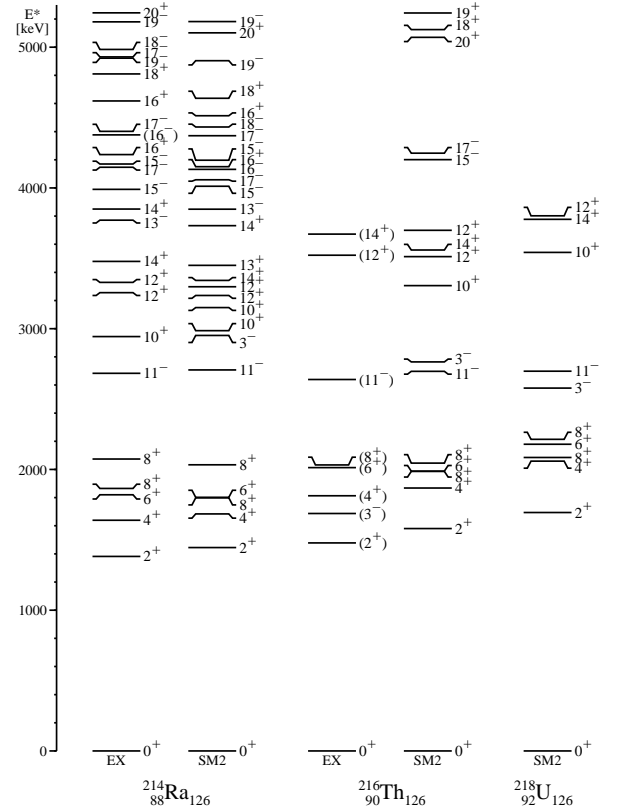}
  \caption{Experimental vs theoretical spectrum of $^{214}$Ra,
    $^{216}$Th, and $^{218}$U
    \label{fig:th216}}
\end{figure}

Hints on the location of the hypothetical islands of super heavy
elements are usually inspired in mean-field calculations of the single
particle structure.  The predictions for nuclei far from stability can
be tested also in nuclei much closer to stability, where shell model
calculations are now feasible. In particular, recent systematic mean
field calculations suggest a substantial gap for $Z=92$ and $N=126$
($^{218}$U) corresponding to the 1h$_\frac{9}{2}$ shell closure. On
the other hand, the single quasiparticle energies extrapolated from
lighter N=126 isotones up to $^{215}$Ac, do not support this
conclusion. To shed light on these controversial predictions, shell
model calculations (as well as experimental spectroscopic studies of
$^{216}$Th ~\cite{Hauschild.Rejmund.ea:2001}) were undertaken in the
$N=126$ isotones up to $^{218}$U.  Shell model calculations were
performed in the 1h$_{\frac{9}{2}}$, 2f$_{\frac{7}{2}}$,
1i$_{\frac{13}{2}}$, 3p$_{\frac{3}{2}}$, 2f$_{\frac{5}{2}}$,
3p$_{\frac{1}{2}}$ proton valence space, using the realistic
Kuo-Herling interaction~\cite{Kuo.Herling:1971} as modified by Brown
and Warburton~\cite{Warburton.Brown:1991}.  The calculation reproduces
nicely the ground state energies, the 2$^+$ energy systematics as well
as the high-spin trends. The cases of $^{214}$Ra,  $^{216}$Th and
$^{218}$U are shown in
figure~\ref{fig:th216}. The only deviations between theory and
experiment are in the 3$^-$ energy and reflect the particular nature
of theses states which are known to be very collective and
corresponding to particle hole excitations of the $^{208}$Pb core.

No shell gap for Z=92, corresponding to the 1h$_\frac{9}{2}$ closure,
is predicted.  On the contrary, the ground state of the N=126 isotones
is characteristic of a superfluid regime with seniority zero
components representing more than 95\% of the wave functions for all
nuclei.

\subsection{Random Hamiltonians}\label{sec:random-hamiltonians}
The study of random Hamiltonians is a vast interdisciplinary subject
that falls outside the scope of this review~\citep[see][for a
collection of the pioneering papers]{Porter:65}. Therefore here we
shall only give a bibliographical guide to work that has recently
attracted wide attention and may have consequences in future shell
model studies.
\citet{Johnson.Bertsch.ea:1998,Johnson.Bertsch.ea:1999} noticed that
random interactions had a strong tendency to produce $J=0$ ground
states. This is an empirical fact that was hitherto attributed to the
pairing force.
\citet{Bijker.Frank.ea:1999,Bijker.Frank:2000a,Bijker.Frank:2000b}
showed that in an interacting boson context this also occurred, and
was associated to the typical forms of collectivity found in the IBM.
For the fermion problem no collectivity occurs for purely random
interactions~\cite{Horoi.Brown.ea:2001}. The origin of $J=0$ ground
state-dominance was attributed to ``geometric
chaoticity''~\cite{Mulhall.Volya.ea:2000}. The geometric aspects of
the phenomenon were investigated in some simple cases
by~\citet{Zhao.Arima:2001,Zhao.Arima.ea:2001}
and~\citet{Huu-Tai.Frank.ea:2003}. \citet{Velazquez.Zuker:2002} argued
that the general cause of $J=0$ ground state-dominance was to be found
in time-reversal invariance, and showed that when the random matrix
elements were displaced to have a negative centroid, well developed
rotational motion appeared in the valence spaces where the realistic
interactions would also produce it (as in Fig.~\ref{fig:6}). This is
very much in line with what was found
by~\citet{Cortes.Haq.Zuker:1982}, and the interesting point is that
the collective ingredient induced by the displacement of the matrix
elements is the quadrupole force.

It is unlikely that we shall learn much more from purely random
Hamiltonians. However, the interplay of random and collective
interactions may deserve further study. In particular: We know that a
monopole plus pairing Hamiltonian is a good approximation. Would it be
a good idea to replace the rest of the interaction by a random one,
instead of neglecting it?

%\cite{Molina.Gomez.ea:2000,Relano.Gomez.ea:2002}

\section{Conclusion}
\label{sec:concl-persp}
Nuclei are idiosyncratic, especially the lighter ones, accessible to
shell model treatment. Energy scales that are well separated in other
systems (as vibrational and rotational states in molecules) do overlap
here, leading to strong interplay of collective and single particle
modes. Nonetheless, secular behavior---both in the masses and the
spectra---eventually emerges and the $pf$ shell is the boundary region
were rapid variation from nucleus to nucleus is replaced by smoother
trends. As a consequence, larger calculations become associated with
more transparent physics, and give hints on how to extend the shell
model philosophy into heavier regions where exact diagonalizations
become prohibitive. 
In this review we have not hesitated to advance some ideas on how this
could be achieved, by suggesting some final solution to the monopole
problem, and exploiting the formal properties of the Lanczos construction.

The shell model has been craft and science: one invented model spaces and
interactions and forced them on the spectra. Sometimes it worked very
well. Then one wondered why such a phenomenology succeeded, to discover
that there was not so much phenomenology after all. It is to be hoped
that this state of affairs will persist.

\acknowledgments

This review owes much to many colleagues; in the first place to
Joaqu\'{\i}n Retamosa who has been a member of the Strasbourg-Madrid
Shell Model collaboration for many years, then to Guy Walter, Mike
Bentley, Franco Brandolini, David Dean, Jean Duflo, Marianne Dufour,
Luis Egido, Jos\'e Mar\'{\i}a G\'omez, Hubert Grawe, Mar\'{\i}a Jos\'e
Garc\'{\i}a Borge, Morten Hjorth-Jensen, Karlheinz Langanke, Silvia
Lenzi, Peter Navr\'atil, Peter von Neumann-Cosel, Achim Richter, Luis
Miguel Robledo, Dirk Rudolph, Jorge S\'anchez Solano, Olivier Sorlin,
Carl Svensson, Petr Vogel to name only a few.  This work has been
partly supported by the IN2P3-France, CICyT-Spain agreements and by a
grant of the MCyT-Spain, BFM2000-030. Many thanks are given to Benoit
Speckel for his continuous computational support.

\appendix

\section{Basic definitions and results}
\label{sec:basics}

\begin{itemize}
\item $p$ is the principal oscillator quantum number

\item $D_p=(p+1)(p+2)$ is the (single fluid) degeneracy of shell $p$ 

\item Orbits are called $r$, $s$, etc. $D_r=2j_r+1$

\item $\hat\delta_{rs}=\delta_{j_rj_s}$ but $r\ne s$, and both have the same
parity. 

\item $m_r$ is the number of particles in orbit $r$, $T_r$ is used for both
the isospin and the isospin operator. In neutron-proton ($np$) scheme,
$m_{rx}$ specifies the fluid $x$. Alternatively we use $n_r$ and $z_r$.

\item $V^{\Gamma}_{rstu}$ or $\mathcal{V}^{\Gamma}_{rstu}$ are two
 body matrix elements.
 $W^{\Gamma}_{rstu}$
is used after the monopole part has been subtracted.

\end{itemize}

 A few equations have to exhibit explicitly angular
momentum ($J$), and isospin ($T$) conservation. We use Bruce
French's  notations~\cite{French:1966}:
$\Gamma$ stands for $JT$. Then $(-1)^\Gamma=(-1)^{J+T}$,
$[\Gamma]=(2J+1)(2T+1)$, and in general $F(\Gamma)=F(J)F(T)$.
Also$(-1)^r=(-1)^{j_r+1/2}$,
$[r]=2(2j_r+1)$. Expressions carry to neutron-proton formalism simply
by dropping the isospin factor.

The one particle creation and anhilation operators
\begin{equation}
A_{rr_z}=a_{rr_z}^{\dagger}\qquad
B_{rr_z}=\tilde{a}_{rr_z}=(-1)^{r+r_z}a_{r-r_z}\,. 
\label{(I.21)}
 \end{equation}
can be coupled to quadratics in $A$ and $B$,
\begin{eqnarray}
X_{\Gamma\Gamma_z}^{\dagger}(rs)&=&(A_rA_s)^\Gamma_{\Gamma_z},\quad
X_{\Gamma\Gamma_z}(rs)=(B_rB_s)^\Gamma_{\Gamma_z},
\nonumber\\
\quad S^\gamma_{\gamma_z}(rt)&=&(A_rB_t)^\gamma_{\gamma_z}.
\label{(I.24)}
\end{eqnarray}
$Z_{\Gamma\Gamma_z}^{\dagger}(rs)=(1+\delta_{rs})^{-1/2}
X_{\Gamma\Gamma_z}^{\dagger}(rs)$
is the normalized pair operator and $Z_{\Gamma\Gamma_z}(rs)$ its
Hermitian conjugate.

For reduced matrix elements we use Racah's definition
\begin{gather}
  <\alpha\alpha_z| P_{\gamma_z}^\gamma| \beta\beta_z>=\label{(I.26)} \\
  (-1)^{\alpha-\alpha_z}\begin{pmatrix}\alpha&\gamma&\beta\\
    -\alpha_z&\gamma_z&\beta_z\end{pmatrix}\,< \alpha\parallel
  P^\gamma\parallel \beta>\,.  \nonumber
\end{gather}

The normal and multipole
 representations of $H$ are obtained through the basic recoupling

\begin{multline}
-(X_\Gamma^{\dagger}(rs)X_\Gamma(tu))^0=
-(-1)^{u+t-\Gamma}\left\lbrack \frac{\Gamma}{ r}\right\rbrack^{1/2}
\delta_{st} S_{ru}^0+\\
\sum_\gamma [\Gamma\gamma]^{1/2}
(-1)^{s+t-\gamma-\Gamma}
\begin{Bmatrix}r&s&\Gamma\\ u&t&\gamma\end{Bmatrix}
(S_{rt}^\gamma S_{su}^\gamma)^0,
\label{(I.36)} 
\end{multline}

whose inverse is
\begin{gather}
\label{(I.37)}
(S_{rt}^\gamma S_{su}^\gamma)^0=
(-1)^{u-t+\gamma}\left\lbrack \frac{\gamma}{r}\right\rbrack^{1/2}
\delta_{st} S_{ru}^0-\\
\sum_\Gamma [\Gamma\gamma]^{1/2}
(-1)^{s+t-\gamma-\Gamma}
\begin{Bmatrix}r&s&\Gamma\\ u&t&\gamma\end{Bmatrix}
(X_\Gamma^{\dagger}(rs)X_\Gamma(tu))^0.\nonumber
\end{gather}

The ``normal'' representation of ${\cal V}$ is then
\begin{multline}
{\cal V}=\sum_{  r\leq s\, t\leq u,\Gamma} {\cal V}_{rstu}^\Gamma
Z_{rs\Gamma}^{\dagger}\cdot  Z_{tu\Gamma}=
\\
-\sum_{(rstu)\Gamma}\xi _{rs}\xi_{tu}
[\Gamma]^{1/2} {\cal V}_{rstu}^\Gamma
(X_{rs\Gamma}^{\dagger}  X_{tu\Gamma})^0,
\label{(I.40)}
\end{multline}
where we have used
\begin{equation}\xi _{r s}=
\begin{cases}
(1+\delta_{rs})^{-1/2}&\text{if r$\leq$ s},\\
(1+\delta_{rs})^{1/2}/2&\text{if no restriction},
\end{cases}
\label{I.39}
\end{equation}
so as to have complete flexibility in the sums.
According to (\ref{(I.36)}), ${\cal V}$ can be transformed into the
``multipole'' representation
\begin{multline}
{\cal V}=
\sum_{(rstu)\gamma}\xi _{rs}\xi_{tu}\left [
[\gamma]^{1/2} \omega_{rtsu}^\gamma
(S_{rt}^\gamma S_{su}^\gamma)^0\right. \\
\left. +\delta_{st}\hat\delta_{ru}
[s]^{1/2} \omega_{russ}^0 S_{ru}^0\right ],
\label{(I.41)}
  \end{multline}

where (sum only over Pauli-allowed $\Gamma$)

\begin{equation}
\omega_{rtsu}^\gamma =\sum_\Gamma(-1)^{s+t-\gamma-\Gamma}
\left\{\begin{array}{ccc}
   r&s&\Gamma\\  u&t&\gamma
   \end{array}\right\}
W_{rstu}^\Gamma[\Gamma],\label{3A}
\end{equation}
\begin{equation}
W_{rstu}^\Gamma =\sum_\gamma(-1)^{s+t-\gamma-\Gamma}
\left\{\begin{array}{ccc}
   r&s&\Gamma\\  u&t&\gamma
   \end{array}\right\}
\omega_{rtsu}^\gamma[\gamma].\label{3B}
\end{equation}

\medskip
Eq.(\ref{(I.37)}) suggests an alternative to (\ref{(I.41)})
\begin{eqnarray}
{\cal V}&=&\sum_{(rstu)\gamma}\xi _{rs}\xi_{tu}
[\gamma]^{1/2} \omega_{rstu}^\gamma\label{(I.43)} \\
&&\times\left\lbrack
(S_{rt}^\gamma S_{su}^\gamma)^0-(-1)^{\gamma+r-s}\left\lbrack
\frac{\gamma}{r}\right\rbrack^{1/2}\delta_{st}\hat\delta_{ru}S_{ru}^0
\right\rbrack,
\nonumber
\end{eqnarray}

where each term is associated with a pure two body operator.

\section{Full form of  ${\cal H}_m$}\label{sec:hm}
If we were interested only in extracting the diagonal parts of the
monopole Hamiltonian in $jt$ scheme, ${\cal H}_{mT}$, the solution
would consist in calculating traces of ${\cal H}$, and showing that
they can be written solely in terms of number and isospin operators.
Before we describe the technique for dealing with the non diagonal parts
of ${\cal H}_{mT}$ {\it i.e.} for generalizing to them the notion of
centroid, some remarks on the interest of such an operation may be in
order. 

The full  ${\cal H}_m$ contains all that is required for Hartree Fock
(HF) 
variation, but it goes beyond. Minimizing the energy with
respect to a determinantal state will invariably lead to an isospin
violation because neutron and proton radii tend to
equalize~\cite{Duflo.Zuker:2002}, which demands different orbits of
neutrons and protons. Therefore, to assess  accurately the amount of
isospin violation in the presence of isospin-breaking forces we must
ensure its conservation in their absence. More generally, a full
diagonalization of ${\cal H}_m$ is of great intrinsic interest.
Now, the technical details. 

Define the generalized number and isospin operators       
\begin{equation}
S_{rs}=\hat\delta_{rs}[r]^{1/2}S_{rs}^{00}\,,\qquad
T_{rs}=\frac{1}{2}\hat\delta_{rs}[r]^{1/2}S_{rs}^{01}
\label{(II.1)}
 \end{equation}
which, for $\delta_{rs}=1$, reduce to $S_{rr}=n_r,\;T_{rr}=T_r$.
By definition  ${\cal H}_m$ contains the two body quadratic forms in
$S_{rs}$ and $T_{rs}$:
\begin{subequations}
\begin{equation}
S_{rtsu}=\zeta_{rs}\zeta_{tu}(S_{rt}S_{su}-\delta_{st}S_{ru}),
\label{(II.2a)}
\end{equation}
\begin{equation}
T_{rtsu}=\zeta_{rs}\zeta_{tu}(T_{rt}\cdot  T_{su} - \frac{3}{ 4}
\delta_{st}S_{ru}),
\label{(II.2b)}
\end{equation}
\end{subequations}
which in turn become for $\delta_{rt}=\delta_{su}=1$,  $m_{rs}$ and
$T_{rs}$ in Eqs.~(\ref{eq:nrs}) and~(\ref{eq:trs})

It follows that the form of ${\cal H}_{mT}$ must be

\begin{equation}
{\cal H}_{mT}= {\cal K}+
\sum_{all}
(a_{rtsu}S_{rtsu}-b_{rtsu}T_{rtsu})
\hat\delta_{rt}\hat\delta_{su},
\label{hm}
  \end{equation}
where the sum is over all possible contributions. This is a special
Hamiltonian containing only $\lambda=0$ terms. 
Transforming to the
normal representation through Eq.~(\ref{3B})
\begin{eqnarray*}
\omega^{00} & = & \delta_{\lambda 0}\delta_{\tau 0}\Longrightarrow
{\cal V}_{rstu}^{J 0}={\cal V}_{rstu}^{J 1} =[rs]^{-1/2}\\
\omega^{01} & = & \delta_{\lambda 0}\delta_{\tau 1}\Longrightarrow
{\cal V}_{rstu}^{J 0}=3[rs]^{-1/2}\,,
{\cal V}_{rstu}^{J 1} =-[rs]^{-1/2}
\end{eqnarray*}
and therefore
\begin{subequations}
\begin{eqnarray}
S_{rtsu}&=&\sum_{\Gamma}Z_{rs\Gamma}^{\dagger}\cdot  Z_{tu\Gamma}=
\label{(II.11a)}\\
&=&\sum_JZ_{rsJ 0}^{\dagger}\cdot  Z_{tuJ 0}+
\sum_JZ_{rsJ 1}^{\dagger}\cdot  Z_{tuJ 1}\nonumber
\end{eqnarray}
\begin{equation}
-T_{rtsu}=\frac{3}{ 4}\sum Z_{rsJ 0}^{\dagger}\cdot  Z_{tuJ 0}-\frac{1}{ 4}
\sum Z_{rsJ 1}^{\dagger}\cdot  Z_{tuJ 1}
\label{(II.11b)}
\end{equation}
\end{subequations}
and inverting
\begin{subequations}
\begin{equation}
\sum_JZ_{rsJ 0}^{\dagger}\cdot Z_{tuJ 0}=\frac{1}{ 4}
(S_{rtsu}-4T_{rtsu})
\label{(II.12a)}
\end{equation}
\begin{equation}
\sum_JZ_{rsJ 1}^{\dagger}\cdot  Z_{tuJ 1}=\frac{1}{ 4}
(3S_{rtsu}+4T_{rtsu}).
\label{(II.12b)}
\end{equation}
\end{subequations}

When $j_r=j_s=j_t=j_u$ and $r\ne s$ and $t\ne u$, both $S_{rust}$,
$T_{rust}$ and $S_{rtsu}$, $T_{rtsu}$ are present. 
They can be calculated from (\ref{(II.12a)}) and (\ref{(II.12b)}) by
exchanging $t$ and $u$,
\begin{subequations}
\begin{eqnarray}
\sum Z_{rsJ 0}^{\dagger}\cdot  Z_{utJ 0}&=&
-\sum (-1)^J Z_{rsJ 0}^{\dagger}\cdot  Z_{tuJ 0}=\nonumber\\
&=&\frac{1}{ 4}(S_{rust}-4T_{rust})\label{(II.13a)}
\end{eqnarray}
\begin{eqnarray}
\sum Z_{rsJ 1}^{\dagger}\cdot  Z_{utJ 1}&=&
\sum (-1)^J Z_{rsJ 1}^{\dagger}\cdot  Z_{tuJ 1}=\nonumber\\
&=&\frac{1}{ 4}(3S_{rust}+4T_{rust})
\label{(II.13b)}
\end{eqnarray}
\end{subequations}
and by combining (\ref{(II.12a)}), (\ref{(II.12b)}),
(\ref{(II.13a)}) and (\ref{(II.13b)})
\begin{subequations}
\begin{eqnarray}
&&\sum_J Z_{rsJ 0}^{\dagger}\cdot  Z_{tuJ 0}\frac{(1\pm
(-1)^J)}{2}=\label{(II.14a)}\\
&&\frac{1}{ 8}( (S_{rtsu}-4T_{rtsu})\mp
(S_{rust}-4T_{rust}))\nonumber
\end{eqnarray}
\begin{eqnarray}
&&\sum_J Z_{rsJ 1}^{\dagger}\cdot  Z_{tuJ 1}\frac{(1\pm(-1)^J)}{2}=\label{(II.14b)}\\
&&\frac{1}{ 8}( (3S_{rtsu}+4T_{rtsu})\pm
(3S_{rust}+4T_{rust}))\nonumber
\end{eqnarray}
\end{subequations}

To write ${\cal H}_{mT}$ we introduce the notations 
\begin{gather}
\Phi (P)=1-(1-\delta_{rs})(1-\delta_{tu}),\\
\Phi (e)\equiv (\hat\delta_{rs}-\delta_{rs})
(\hat\delta_{tu}-\delta_{tu}),
\label{(II.5)}
\end{gather}
So for $r=s$ or $t=u$, $\Phi (P)=1$ and for $j_r=j_s=j_t=j_u$ and
$\Phi (P)=0$, $\Phi (e)=1$. Then
\begin{gather}
  {\cal H}_{mT} ={\cal K}+\nonumber\\
  \sum_{\substack{ r\leq s\\ t\leq u\\T,\rho=\pm}}
  \hat\delta_{rt}\hat\delta_{su}\left\lbrack (1-\Phi (e))\overline
    {\cal V}_{rstu}^T \Omega_{rstu}^T +\Phi (e) \overline {\cal
      V}_{rstu}^{\rho T}\Omega_{rstu}^{\rho T}
  \right\rbrack\nonumber\\
  \rho={\rm sign~}(-1)^J, \quad  \Omega_{rstu}^T =\sum_J
  Z_{rsJT}^{\dagger}\cdot  Z_{tuJT}\nonumber\\ 
  \Omega_{rstu}^{\pm T} =\sum_J Z_{rsJT}^{\dagger}\cdot Z_{tuJT}\,
  \frac{(1\pm(-1)^J)}{2}.
\label{(II.16)}
\end{gather}
The values of the generalized centroids $\overline {\cal
  V}_{rstu}^T$ and $\overline {\cal V}_{\rho T}$ are
determined by demanding that ${\cal H}-{\cal H}_{mT}={\cal
  H}_M$ contain no contributions with $\lambda=0$. In other words
\begin{gather}
W_{rstu}^{JT}= {\cal V}_{rstu}^{JT}-\nonumber\\
  -\hat\delta_{rt}\hat\delta_{su}\left\lbrack (1-\Phi (e)) \overline
    {\cal V}_{rstu}^T+\Phi (e) \overline {\cal V}_{rstu}^{\rho
      T}\right\rbrack\,,
\label{(II.15)}
\end{gather}
must be such that $\omega_{rtsu}^{0\tau}=0$, and  from Eq.~(\ref{3B})
\begin{equation}
\sum_{(J)}[J]
W_{rstu}^{JT}=0\therefore\overline  {\cal V}_{rstu}^T=
\sum_{(J)}[J]{\cal V}_{rstu}^{JT} / \sum_{(J)}[J].
\label{(II.9)}
\end{equation}
Applying this prescription to all the terms leads to (obviously
$\hat\delta_{rt}\hat\delta_{su}=1$ in all cases)
\begin{eqnarray}
&&\overline  {\cal V}_{rstu}^T=\sum_{(J)} {\cal V}_{rstu}^{JT}[J]
/ \sum_{(J)}[J]
\nonumber\\
&&\sum_{(J)}[J] =\frac{1}{ 4}\;\frac{D_r(D_s+2
\Phi (P)(-1)^T)}{1+\Phi (P)}
\nonumber\\
&&\overline  {\cal V}_{rstu}^{\pm T}=\sum_J {\cal V}_{rstu}^{JT}[J]
(1\pm(-1)^J)/ \sum_J[J] (1\pm(-1)^J)
\nonumber\\
&&\sum_J[J](1\pm (-1)^J) =\frac{1}{ 4}\,D_r(D_r\mp 2)
\nonumber\\
&&D_r=[r],\quad \quad
\Phi (e)=1,\;{\rm for~}
\overline{\cal V}_{rstu}^{\pm T}\,.\label{(II.18)}
\end{eqnarray}

Through eqs.(\ref{(II.12a)}), (\ref{(II.12b)}), (\ref{(II.14a)}), and
(\ref{(II.14b)}) we can obtain the form of ${\cal H}_{mT}$ in terms of
the monopole operators by regrouping the coefficients affecting each
of them.  To simplify the presentation we adopt the following
convention
\[\left\{
\begin{array}{llll}
\alpha\equiv rstu&r\leq s,t\leq u,&
\hat\delta_{rt}\hat\delta_{su}=1,& {\rm BUT}\\
S_\alpha=S_{rtsu},&S_{\bar\alpha}=S_{rust},&T_\alpha=T_{rtsu},&
T_{\bar\alpha}=T_{rust},
\end {array}
\right.\]
 then
\begin{subequations}
\label{(II.17)}
\begin{eqnarray}
{\cal H}_{mT} ={\cal K}+\sum_\alpha
(1-\Phi (e))(a_\alpha S_\alpha+b_\alpha T_\alpha)\nonumber\\
+\Phi (e) (a_\alpha^d S_\alpha+b_\alpha^d T_\alpha
+a_\alpha^e S_{\bar\alpha}+b_\alpha^e T_{\bar\alpha}),\,{\rm with}\\
a_\alpha =\frac{1}{ 4}(3\bar {\cal V}^1_\alpha+\bar {\cal
  V}_\alpha^0)\nonumber\\ 
b_\alpha =\frac{1}{ 4}( \bar {\cal V}^1_\alpha-\bar {\cal V}_\alpha^0)\\
a_\alpha^d =\frac{1}{ 8}(3\bar {\cal V}_\alpha^{+1}+3\bar {\cal V}_\alpha^{-1}
+\bar {\cal V}_\alpha^{+0}+\bar {\cal V}_\alpha^{-0})\nonumber\\
a_\alpha^e =\frac{1}{ 8}(3\bar {\cal V}_\alpha^{+1}-3\bar {\cal V}_\alpha^{-1}
-\bar {\cal V}_\alpha^{+0}+\bar {\cal V}_\alpha^{-0})\nonumber\\
b_\alpha^d =\frac{1}{2}({3}\bar {\cal V}_\alpha^{+1}+{3}\bar {\cal
  V}_\alpha^{-1} 
-\bar {\cal V}_\alpha^{+0}-\bar {\cal V}_\alpha^{-0})\nonumber\\
b_\alpha^e =\frac{1}{2}({3}\bar {\cal V}_\alpha^{+1}-\phantom{3}\bar
{\cal V}_\alpha^{-1} 
+\bar {\cal V}_\alpha^{+0}-\bar {\cal V}_\alpha^{-0})
\end{eqnarray}
\end{subequations}

\subsection{Separation of ${\cal H}_{mnp}$ and ${\cal H}_{m0}$}

In $j$ formalism ${\cal H}_{mnp}$ is ${\cal H}_{mT}$ under another guise:
neutron and
proton shells are differentiated and the  operators $T_{rs}$ and $S_{rs}$
are written in terms of four scalars $S_{r_xs_y};\;\; x,y=n \;{\rm or} \;p$.

We may also be interested in extracting only the
purely isoscalar contribution to ${\cal H}_{mT}$, which we call
${\cal H}_{m0}$.
The power of French's product notation becomes particularly evident here,
because the form of both terms is identical. It demands some algebraic
manipulation  to
find
\begin{gather}
{\cal H}_{mnp}\;{\text or}\;{\cal H}_{m0} ={\cal K}+\sum_\alpha \biggl[
\bar {\cal V}_\alpha S_\alpha(1-\Phi (e))+ \nonumber\\
+\frac{1}{2}
((\bar {\cal V}_\alpha^++\bar {\cal V}_\alpha^-)S_\alpha+
(\bar {\cal V}_\alpha^--\bar {\cal V}_\alpha^+)S_{\bar\alpha})\Phi
(e)\biggr]
\label{(II.19)}
 \end{gather}
with
\begin{eqnarray}
\overline  {\cal V}_{rstu}=\sum_{(\Gamma)} {\cal V}_{rstu}^\Gamma[\Gamma]/
\sum_{(\Gamma)}[\Gamma] \nonumber\\
\sum_{(\Gamma)}[\Gamma] =D_r(D_s-\Phi (P))/
( 1+\Phi (P))\nonumber\\
 \overline  {\cal V}_{rstu}^\pm=\sum_{\Gamma} {\cal V}_{rstu} [\Gamma]
(1\pm(-1)^{\Gamma+2r})\quad\nonumber\\
\sum_{\Gamma}[\Gamma](1\pm(-1)^{\Gamma+2r})=D_r(D_r\mp(-1)^{2r})
\label{(II.20)}
\end{eqnarray}
Of course we must remember that for ${\cal H}_{mnp}$,

$D_r=2j_r+1,\;(-1)^{2r}=-1,\;\Gamma\equiv J,$ etc., while for ${\cal H}_{m0}$

$D_r=2(2j_r+1),\;(-1)^{2r}=+1,\;\Gamma\equiv JT$; etc.

It should be noted that ${\cal H}_{m0}$
is not obtained by simply discarding the $b$ coefficients in
 eqs.(\ref{(II.17)}), because we can extract some $\gamma=$00
 contribution from the $T_\alpha$ operators. The point will become
 quite clear when considering the diagonal contributions.

\subsection{Diagonal forms of $  {\cal H}_m$}

We are going to specialize to the diagonal terms of ${\cal H}_{mT}$ and
${\cal H}_{mnp}$, which involve only ${\cal V}_{rsrs}^\Gamma$ matrix
elements whose centroids will be called simply ${\cal V}_{rs}$ and
${\cal V}_{rs}^T$ (the overline in 
$\overline  {\cal V}_{rstu}^T$ was  meant to avoid confusion with  possible
matrix elements ${\cal V}_{rstu}^1$ or ${\cal V}_{rstu}^0$, it can be safely
dropped now). 
 Then (\ref{(II.17)}) becomes Eq.~(\ref{eq:hmjtd}) and (\ref{(II.19)}) becomes

\begin{equation}
{\cal H}_{mnp}^d\;or\;{\cal H}_{m0}^d =
{\cal K}+\sum_{r\leq s}   {\cal V}_{rs}\,n_r(n_s-\delta_{rs}) /(1+\delta_{rs})
\label{(II.22b)}
\end{equation}

We rewrite the relevant centroids incorporating explicitly the Pauli
restrictions
\begin{subequations}
\begin{eqnarray}\label{(II.22)}
{\cal V}_{rs}&=&\frac{\sum_\Gamma {\cal V}_{rsrs}^\Gamma[\Gamma]
( 1-(-1)^\Gamma\delta_{rs})}{ D_r(D_s-\delta_{rs})}\\
{\cal V}_{rs}^T&=&\frac{4\sum_J {\cal V}_{rsrs}^{JT}[J]
( 1-(-1)^{J+T}\delta_{rs})}{D_r(D_s+2\delta_{rs}(-1)^T)}
\label{(II.23)}
\end{eqnarray}
\begin{eqnarray}
a_{rs}&=&\frac{1}{ 4}(3{\cal V}^1_{rs}+{\cal V}^0_{rs})={\cal
  V}_{rs}+\frac{3}{ 4}\, 
\frac{\delta_{rs}}{D_r-1}\,b_{rs}\nonumber\\
b_{rs}&=&{\cal V}^1_{rs}-{\cal V}^0_{rs}\label{(II.24)}
\end{eqnarray}
\end{subequations}
The relationship between $a_{rs}$ and ${\cal V}_{rs}$ 
makes it possible to combine eqs.~(\ref{eq:hmjtd}) and (\ref{(II.22b)})
in a single form
\begin{eqnarray}
&&{\cal H}_m^d ={\cal K}+
\sum \frac{1}{ (1+\delta_{rs})}\biggl[{\cal V}_{rs}\,n_r(n_s-\delta_{rs})+
\nonumber\\
&& +b_{rs}\left(T_r\cdot  T_s-\frac{3
n_r\bar n_r}{ 4(D_r-1)}\delta_{rs}\right)\biggr]
\label{(II.25)}
 \end{eqnarray}

in which {\it now} the $b_{rs}$ term can be dropped to obtain
${\cal H}_{m 0}^d$ {\it or} ${\cal H}_{mnp}^d$.

In the np scheme each orbit $r$ goes into two $r_n$ and $r_p$
and the centroids can be obtained through
($x,y=n$ or $p$, $x\not=y$)
\begin{eqnarray}
{\cal V}_{r_xs_y}&=&\frac{1}{2}\left\lbrack {\cal
    V}_{rs}^1\left(1-\frac{2\delta_{rs}}{D_r} 
\right)+ {\cal V}_{rs}^0\left(1+\frac{2\delta_{rs}}{ D_r}
\right)\right\rbrack\nonumber\\
{\cal V}_{r_xs_x}&=&{\cal V}_{rs}^1
\label{(II.27}
 \end{eqnarray}

Note that the diagonal terms depend on the representation:
${\cal H}_{mnp}^d\not ={\cal H}_{mT}^d$ in general.

\section{The center of mass problem}\label{sec:center-mass-problem}
What do you do about center of mass? is probably the standard question
most SM practitioners prefer to ignore or dismiss. Even if we may be
tempted to do so, there is no excuse for ignoring what the problem is,
and here we shall explain it in sufficient detail, so as to dispel
some common misconceptions.

The center of mass (CM)  problem arises because in a many body treatment it is
most convenient  to work with $A$ coordinates and momenta while only $A-1$
of them can be linearly independent since the solutions cannot depend
on the center of mass coordinate $R=(\sum_i r_i)/\sqrt{A}$ or momentum
$P=(\sum_i p_i)/\sqrt{A}$. The way out is to impose a factorization of
the wavefunctions into relative and CM parts: $\Phi(r_1\, r_2\, \ldots
r_A)=\Phi_{rel}\, \phi_{CM}$. The potential energy is naturally given
in terms of relative values, and for the kinetic energy we should do
the same by referring to the CM momentum,
\begin{equation}
  \label{eq:k2b}
  \sum_i
  (p_i-\frac{P}{\sqrt{A}})^2=\sum_ip_i^2-P^2=\frac{1}{A}\sum_{ij}(p_i-p_j)^2,
\end{equation}
and change accordingly ${\cal K}_{ij}$ in Eq.~\eqref{Hec}.
As we are only interested in wavefunctions in which the center of mass
is at rest (or in its lowest possible state), we can add a CM operator
to the Hamiltonian ${\cal H}\Longrightarrow {\cal
  H}+\lambda(R^2+P^2)$, and calling $r_{ij}=r_i-r_j$ we have
\begin{equation}
  \label{eq:CM}
  R^2+P^2=\sum_i(r_i^2+p_i^2)-\frac{1}{A}\sum_{ij}(r_{ij}^2+p_{ij}^2),
\end{equation}
so that upon diagonalization the eigenvalues will be of the form
$E=E_{rel}+\lambda (N_{CM} + 3/2)$, and it only remains to select the states
with $N_{CM}$=0. We are taking for granted separation of CM and relative
coordinates. Unfortunately, this happens only for spaces that are {\em
  closed} under the CM operator~\eqref{eq:CM}: for $N_{CM}$ to be a good
quantum number the space must include all possible states with $N_{CM}$
oscillator quanta. The problem was raised by
\citet{Elliott.Skyrme:1955} who initiated the study of
``particle-hole'' (ph) excitations on closed shells. They noted that
acting with $R-iP$ on the IPM ground state of $^{16}$O ($|0\rangle$)
leads to
\begin{equation}
  \label{eq:ES}
 \sqrt{\frac{1}{18}}(\bar p_1s_1-\sqrt{2}\bar p_3s_1-\sqrt{5}\bar
 p_1d_3+ \bar p_3d_3+3\bar p_3d_5)|0\rangle,
\end{equation}
where $\bar p_{2j}$ removes a particle and $s_1$ or $d_{2j}$ add a
particle on $|0\rangle$. As $R-iP$ has tensorial rank $J^{\pi}T=1^-0$,
Eq.~\eqref{eq:ES} is telling us that out of five possible $1^-0$
excitations, one is ``spurious'' and has to be discarded.  

Assume now that we are interested in 2p2p excitations. They involve
jumps of two oscillator quanta (2\hw) and the CM eigenstates
$(R-iP)^2|0\rangle$ involve the operator in Eq.~\eqref{eq:ES}, but also
jumps to other shells, of the type $\bar s\, p\bar p\, (sd)\equiv \bar
s\, (sd)$ or $(\bar{sd})\, (pf) \bar p\, (sd)\equiv \bar p\, (pf)$.
Therefore, as anticipated, relative-CM factorization can be achieved
only by including all states involving a given number of oscillator
quanta.  The clean way to proceed is through complete $N$\hw spaces,
discussed in Sections~\ref{sec:theory-calculations}
and~\ref{sec:no-core}.  The 0\hw and EI spaces are also free of
problems (the latter because no 1\hw $1^-0$ states exist).  It remains
to analyze the EEI valence spaces, where CM spuriousness is always
present but strongly suppressed because the main contributors to
$R-iP$ --of the type $pj\Longrightarrow p+1 j\pm 1$, with the largest
$j$-- are always excluded.  Consider EEI(1)=$p_{1/2},\, d_{5/2},\,
s_{1/2}$. The only possible 1\hw $1^-0$ state is $\bar p_1s_1$, which
according to Eq.~\eqref{eq:ES} accounts for $(1/18)\%=5.6\%$ of the
spurious state.  This apparently minor problem was unduly transformed
into a serious one through a proposal
by~\citet{Gloeckner.Lawson:1974}.  It amounts to project CM spuriousness
through the ${\cal H}\Longrightarrow {\cal H}+\lambda(R^2+P^2)$
prescription in the EEI(1) space by identifying $R-iP\equiv \bar
p_1s_1$. The procedure is manifestly incorrect, as was repeatedly
pointed out~\cite[see for instance][]{Whitehead.Watt.ea:1977} but the
misconception persists. An interesting and viable alternative was put
forward by~\citet{Dean.Russel.ea:1999} in calculations with two
contiguous major shells. Further work on the subject would be welcome.

\end{document}